\documentclass[twocolumn,twocolappendix]{aastex631}
\usepackage{amsmath}
\usepackage{amssymb}
\usepackage[utf8]{inputenc}
\usepackage[english]{babel}
\usepackage{amsfonts}
\newcommand{\gdet}{\sqrt{-g}}
\newcommand{\p}{\partial}
\usepackage{graphics}
\usepackage{graphicx}
\usepackage{epstopdf}
\usepackage{footnote}
\usepackage{txfonts}
\usepackage{lipsum}
\usepackage{float}
\usepackage{gensymb}

\makesavenoteenv{tabular}

\DeclareSymbolFont{cmletters}{OML}{cmm}{m}{it}
\DeclareMathSymbol{v}{\mathalpha}{cmletters}{"76}

\definecolor{darkblue}{rgb}{0.0,0.0,0.3}
\hypersetup{colorlinks,breaklinks,
            linkcolor=darkblue,urlcolor=darkblue,
            anchorcolor=darkblue,citecolor=darkblue}

\newcommand{\rg}{r_\mathrm{g}}
\newcommand{\rbhi}{r_\mathrm{BH,0}}
\newcommand{\rbh}{r_\mathrm{BH}}

\begin{document}
  

\title{Black Hole-Disk Interactions in Magnetically Arrested Active Galactic Nuclei: 
\\General Relativistic Magnetohydrodynamic Simulations Using A Time-Dependent, Binary Metric}

\author{Sean M. Ressler}
\affiliation{Canadian Institute for Theoretical Astrophysics, University of Toronto, Toronto, On, Canada M5S 3H8}

\author{Luciano Combi}
\affiliation{Perimeter Institute for Theoretical Physics, Waterloo, Ontario N2L 2Y5, Canada}
\affiliation{Department of Physics, University of Guelph, Guelph, Ontario N1G 2W1, Canada}

\author{Xinyu Li}
\affiliation{Canadian Institute for Theoretical Astrophysics, University of Toronto, Toronto, On, Canada M5S 3H8}
\affiliation{Department of Astronomy, Tsinghua University, 30 Shuangqing Rd, Haidian District, Beijing, 100084, China}
\affiliation{David A. Dunlap Department of Astronomy, University of Toronto, 50 St. George Street, Toronto, ON M5S 3H4}

\author{Bart Ripperda}
\affiliation{Canadian Institute for Theoretical Astrophysics, University of Toronto, Toronto, On, Canada M5S 3H8}
\affiliation{David A. Dunlap Department of Astronomy, University of Toronto, 50 St. George Street, Toronto, ON M5S 3H4}
\affiliation{Department of Physics, University of Toronto, 60 St. George Street, Toronto, ON M5S 1A7}
\affiliation{Perimeter Institute for Theoretical Physics, Waterloo, Ontario N2L 2Y5, Canada}

\author{Huan Yang}
\affiliation{Department of Astronomy, Tsinghua University, 30 Shuangqing Rd, Haidian District, Beijing, 100084, China}
\affiliation{Perimeter Institute for Theoretical Physics, Waterloo, Ontario N2L 2Y5, Canada}
\affiliation{Department of Physics, University of Guelph, Guelph, Ontario N1G 2W1, Canada}



\begin{abstract}
Perturber objects interacting with supermassive black hole accretion disks are often invoked to explain observed quasi-periodic behavior in active galactic nuclei (AGN).
We present global, 3D general relativistic magnetohydrodynamic (GRMHD) simulations of black holes on inclined orbits colliding with magnetically arrested thick AGN disks using a binary black hole spacetime with mass ratio $0.1$.  
We do this by implementing an approximate time-dependent binary black hole metric into the GRMHD {\tt Athena++} code.
The secondary enhances the unbound mass outflow rate 2--4 times above that provided by the disk in quasi-periodic outbursts, eventually merging into a more continuous outflow at larger distances.  
We present a simple analytic model that qualitatively agrees well with this result and can be used to extrapolate to unexplored regions of parameter space.
We show self-consistently for the first time that spin-orbit coupling between the primary black hole spin and the binary orbital angular momentum causes the accretion disk and jet directions to precess significantly (by 60$\degree$--80$\degree$) on long time-scales (e.g., $\sim$ 20 times the binary orbital period).
Because this effect may be the only way for thick AGN disks to consistently precess, it could provide strong evidence of a secondary black hole companion if observed in such a system.
Besides this new phenomenology, the time-average properties of the disk and accretion rates onto the primary are only marginally altered by the presence of the secondary, consistent with our estimate for a perturbed thick disk. 
This situation might drastically change in cooled thin disks.

\end{abstract}

\keywords{Active galactic nuclei, Supermassive black holes,Accretion, Magnetohydrodynamical simulations, General relativity}

\section{Introduction}

{Active galactic nuclei (AGN) produce copious amounts of electromagnetic radiation with strong variabilities occurring on timescales of minutes to years  at all observed frequencies \citep{ulrich1997variability, czerny2004role, ricci2023changing, sartori2018model}. Although this variation is mainly stochastic (dominated by red noise) some of these sources seem to exhibit (quasi-)periodic emission \citep{graham2015possible, charisi2016population,d2023observational}.} {A natural mechanism to explain possible periodic behavior in AGN is the presence of a binary system in the central engine, e.g., a supermassive binary black hole \citep{begelman1980massive}.  In this scenario, (magneto-)hydrodynamical accretion processes \citep{noble2021mass} and kinematic effects such as Doppler boosting \citep{d2015relativistic} might link the AGN periodicity with the binary properties.}
{In particular, repeated collisions between a smaller binary companion and the accretion disk surrounding the central supermassive black hole has long been discussed as a possible mechanism for quasi-periodic behavior \citep{Zentsova1983,Lehto1996,Ivanov1998,Komossa2006,Dai2010,Franchini2023,Linial2023,Linial2024,Pasham2024}.} 
Typically it is proposed that the secondary object is either a star or a smaller mass black hole, which impacts the disk, ejects matter, and/or heats up the disk material via bow shocks.
For instance, the source OJ287 shows consistent quasi-periodic outbursts at optical wavelengths and is supposed to host a supermassive binary black hole system (though the masses of the black holes are uncertain, \citealt{Sillanpaa1988,Dey2018,Laine2020,Komossa2023}).
{Moreover, one possible explanation for recent observations of quasi-periodic X-ray eruptions in AGN \citep{miniutti2019nine} is an association with flaring emission due to disk collisions \citep{Linial2023, franchini2023quasi,tagawa2023flares,xian2021x,Zhou:2024bjt}.} 

The case where the secondary object is a black hole is particularly interesting. For a given effective size, black holes are much more massive than stars and so they can have a stronger influence on the primary black hole {at close separations} through effects like spin-orbit coupling, where the orbital angular momentum and black hole spin directions will oscillate as a function of time. {In fact, this has been proposed as a possible mechanism for periodic radio variability in AGNs \citep{vonFellenberg2023} and optical variability in blazars \citep{abraham1999beaming,romero2000beaming, britzen2023precession} }

{If this model is correct, then it would mean that these quasi-periodic observational features are also the electromagnetic counterparts to low frequency gravitational wave sources expected to be detected by the Laser Interferometer Space Antenna (LISA, \citealt{Flanagan1998,Amaro2007,Berry2013}) at merger, and pulsar timing arrays during the inspiral regimes.
Modelling these systems in detail could then have important implications for multi-messenger astronomy.}

Because of the strong gravitational field and non-linear magnetohydrodynamics involved, self-consistent models of black hole-disk interaction can only be accurately built using 3D General-Relativistic Magnetohydrodynamical (GRMHD) simulations.
There are essentially two different approaches to simulate binary black hole accretion systems {in General Relativity} (for detailed reviews, see \citealt{gold2019relativistic, Cattorini2024}).
The most accurate approach is to couple MHD to numerical relativity (e.g., \citealt{Farris2012,Giacomazzo2012,Gold2014,Paschalidis2021}); this method, however, is computationally expensive and numerically challenging. In practice {these simulations can only be evolved for a short amount of time (typically $\sim$ tens of orbits) close to merger and are usually far from steady-state.}
A computationally cheaper and numerically simpler approach is to approximate the dynamical spacetime by some semi-analytic expression for use in existing GRMHD codes (e.g., \citealt{Noble2012,bowen2018quasi,Gold2014,mundim2014approximate,Lopez2021,Avara2023}).
The approximate metric must be constructed with care to be globally applicable and an accurate representation of the evolving spacetime.
The accuracy can also become poor at times close to merger.
\citet{Lopez2021} and \citet{Combi2021} recently proposed an approximate metric constructed by superimposing two boosted Kerr black holes, demonstrating it to be accurate during the inspiral regime leading up to merger and well-defined for the entire domain.
The robustness of the metric in GRMHD was exhibited in a follow-up work, where it was applied to an equal mass binary system with spinning black holes \citep{Combi2022}.  

Using semi-analytical approximations for the binary black hole spacetime as in \citet{Combi2021} allows for a much larger exploration of the vast parameter space in binary black hole accretion.  
This is not only because avoiding numerical relativity makes the computation faster, but it will allow the community to take full advantage of the numerous advances made in single black hole GRMHD codes if these are adapted for time-dependent metrics.
For instance, in the past couple of decades, the GRMHD accretion community has been able to explore effects such as varying black hole spin \citep{EHT5,EHT_SGRA_5}, varying magnetic flux supply \citep{Sasha2011,Narayan2012}, varying disk tilt \citep{McKinney2013,Liska2018,White2019b,Chatterjee2020}, including radiative effects \citep{BHLIGHT,White2023}, including non-ideal physics \citep{Ressler2015,Foucart2017,grim,Ripperda2019}, and studying a variety of initial conditions informed by larger scales \citep{Ressler2020b,Ressler2021,Kaaz2023,Cho2023,Lalakos2023}. 
Furthermore, since the user base for GRMHD is currently much larger than that for numerical relativity (e.g., \citealt{Porth2019}), it could encourage more researchers to study binary black hole systems. 

Recently, \citet{Sukova2021} investigated the collisions of spherical objects with AGN disks using GRMHD simulations.  
The simulations were essentially agnostic to the type of secondary object (e.g., a star or black hole), and simply enforced that a region with a particular ``influence radius'' be comoving with the object. 
The authors found that the presence of the secondary could significantly modify the structure of the accretion flow and produce strong outbursts of relativistic outflow 1-2 orders of magnitude larger than the ``background'' outflow rate.
Such strong features in the accretion and outflow properties would seem more than enough to explain some of the observed quasi-periodic behavior in AGN.
However, the simulations of \citet{Sukova2021} were primarily performed in 2D axisymmetry, with only one 3D perturber simulation with a limited runtime (5000 primary black hole light crossing times) and initialized with 2D data.     
Since accretion flows are known to behave much differently in 3D than in 2D and the geometry of the perturber scenario is fundamentally three dimensional, it is therefore important to revisit this problem using full 3D simulations (and a fully general relativistic treatment of secondary black holes).
This is especially true for magnetically arrested accretion disks (MADs), where the axisymmetry causes the flow to be fully halted in 2D, while in 3D the gas can penetrate the magnetic barrier \citep{McKinney2012,White2019,Ripperda2020,Ripperda2022}.

Here we present our implementation of a  time-dependent binary black hole metric into {\tt Athena++} \citep{White2016,Athenapp} and present a series of appropriate test problems for time-dependent metrics. We use a superposed metric as in \cite{Combi2021}, generalized for arbitrary spins and eccentricities that result from solving the Post-Newtonian equations for the evolution of the black holes.
{\tt Athena++} is particularly suited for the binary black hole accretion problem with small mass-ratios due to its ability to use adaptive mesh refinement and excellent scalability.
As a conservative code using constrained transport for magnetic fields it conserves mass, energy, momentum, and magnetic flux to machine precision.   
It also now has full support for radiation, being the first GRMHD code to directly solve the GR Boltzmann transport equation \citep{White2023}.
Finally, it is widely used, public, and a ported version for use on GPUs will soon be publicly available ({\tt AthenaK}).

Although our implementation is general to any binary black hole configuration, in this work we focus on the study of small mass ratio (0.1) inspirals in galactic centers with thick disks.
In doing so we seek to improve on past work studying black hole-disk interactions with simulations.
We do this by evolving a 3D torus around a single supermassive black hole long enough to reach equilibrium out to $\sim$ 100 gravitational radii and then introducing a secondary black hole at some initial distance and inclination to the disk.  
The secondary then moves through the disk based on the solution to the post-Newtonian orbital equations and alters the accretion and outflow of the disk due to gravitational, electromagnetic, and plasma interactions.
We also study how the effects of spin-orbit coupling alter the accretion flow as the primary black hole spin axis changes direction as a function of time and study the accretion properties of the secondary.
We defer the study of radiatively cooled thin disks (more common for AGN) to later work.

This paper is organized as follows.  \S \ref{sec:methods} details the analytic and numerical methods we use to simulate binary black hole systems, \S \ref{sec:analytic} presents some useful analytic estimates, \S \ref{sec:results} describes our simulation results for small mass ratio inspirals, \S \ref{sec:comparison} compares our simulations to past work, \S \ref{sec:limitations} discusses the limitations of our work, and \S \ref{sec:conc} concludes.

\section{Methods}
\label{sec:methods}

\subsection{Time-Dependent Metrics in {\tt Athena++}}
We start with the publicly available multi-purpose fluid dynamics code {\tt Athena++} \citep{Athenapp}, particularly the extension for GRMHD \citep{White2016}, which solves the conservative equations in an arbitrary space-time set by the choice of coordinates and metric through the ``GRUser'' module.
In principle, this module can accommodate any choice of time-independent metric as long as the spatial derivatives (used to compute the connection coefficients) are supplied. 
We extend this framework to support time-dependent metrics (in a similar way to \citealt{Noble2012}).   
In the 3 + 1 conservative formulation of GRMHD (e.g., \citealt{Gammie2003}), the mass, energy, momentum, and magnetic field equations in a coordinate basis (and Lorentz-Heaviside units) are 
\begin{equation}
  \label{eq:cons_equations}
  \begin{aligned}
   \frac{\partial}{\partial t}\left(\gdet \rho u^t\right) &=  -\frac{\partial}{\partial x^i}\left(\gdet \rho u^i\right),  \\
       \frac{\partial}{\partial t}\left(\gdet T^t_\nu\right) &= - \frac{\partial}{\partial x^i}\left(\gdet T^i_\nu\right) + \gdet T^\alpha_\beta \Gamma^\beta_{\nu \alpha}, \\
       \frac{\partial}{\partial t}\left( \gdet B^i\right) &= - \frac{\partial}{\partial x^j}\left[\gdet \left(b^j u^i - b^i u^j\right)\right],
\end{aligned}
\end{equation}
where 
\begin{equation}
   \label{eq:Tmunu}
  T^{\mu \nu} = \left(\rho + \frac{\gamma}{\gamma-1} P + b^2 \right) u^\mu u^\nu + \left(P + \frac{b^2}{2}\right) g^{\mu\nu} - b^\mu b^\nu,
\end{equation}
is the stress-energy tensor, $t$ is the coordinate time, $x^i$ are the spatial coordinates ($x$, $y$, and $z$), $g$ is the determinant of the metric, $\rho$ is the rest-frame mass density, $u^\mu$ is the gas four-velocity, $\Gamma^\alpha_{\beta \kappa}$ is the connection coefficient, $\gamma$ is the adiabatic index, $P$ is the rest-frame gas pressure, $B^i$ is the magnetic field, $b^\mu$ is the four-magnetic field defined via
\begin{equation}
  \label{eq:four_b}
  \begin{aligned}
  b^t &= B^i u^\mu g_{i \mu} \\
  b^i &=  \frac{B^i + b^t u^i}{u^t},
\end{aligned}
\end{equation} 
and $g_{\mu \nu}$ is the metric. 
Here the indexes $i$ (not to be confused with the orbital inclination used later in this work) are limited to the spatial directions ($i \in \{1,2,3\}$), while Greek indexes include the time direction ($\mu \in \{0,1,2,3\})$.
The connection coefficients can be written in lowered form as 
\begin{equation}
  \label{eq:Gamma}
  \Gamma_{\alpha \beta \kappa} = \frac{1}{2} \left(\frac{\partial g_{\kappa \alpha}}{\partial x^\beta} +\frac{\partial g_{\kappa \beta} }{\partial x^\alpha} - \frac{\partial g_{\alpha \beta} }{\partial x^\kappa} \right),
\end{equation}

As formulated, equations (\ref{eq:cons_equations}--\ref{eq:Gamma}) are valid for both time-independent and time-dependent metrics.   
{\tt Athena++}, however, like all GRMHD codes designed for applications to stationary metrics, assumes that the terms proportional to $\partial g_{\mu \nu}/\partial_t$ in Equation \eqref{eq:Gamma} are 0.  
We are therefore required to generalize the calculation of $\Gamma_{\alpha \beta \kappa}$, or, more precisely, $T^\alpha_\beta \Gamma^\beta_{\nu \alpha}$.  
This is simplified by using the symmetry of the stress-energy tensor, $T_{\mu \nu} = T_{\nu \mu}$,  the symmetry of the metric, $g_{\mu \nu} = g_{\nu \mu}$, and the fact that the last two terms on the right hand side of Equation \eqref{eq:Gamma} are antisymmetric in the first and last index of $\Gamma_{\alpha \beta \kappa}$, so that
\begin{equation}
  T^\alpha_\beta \Gamma^\beta_{\nu \alpha} = T^{\alpha \beta} \Gamma_{\beta \nu \alpha} = \frac{1}{2} T^{\alpha \beta} \frac{\partial g_{\beta \alpha}}{\partial x^\nu}.
\end{equation}
This means that the nonzero $\partial g_{\beta \alpha}/\partial t$ only explicitly appears in the GRMHD equations as a source term in the conserved energy, $T^t_t$.
The other equations remain unchanged, although $\gdet$ can now depend on time.  

The set of equations \eqref{eq:cons_equations} are solved by {\tt Athena++} using a finite volume method, that is, integrated over space and time for each three-dimensional cell and time step.
For instance, integrated over volume, and assuming uniform spacing in discretization, the left-hand side of the hydrodynamic conservation equations in \eqref{eq:cons_equations} become (in one spatial dimension without loss of generality):
  \begin{multline}
\label{eq:time_derivative_discretized}
 \frac{ \left(\gdet \right)^{n+1,i}\left< U \right>_V^{n+1,i} -    \left(\gdet\right)^{n,i} \left< U \right>_V^{n,i}}{  \left(\gdet\right)^{n+1/2,i}  \Delta t}  \\
\approx  - \frac{1}{\left(\gdet\right)^{n+1/2,i}\Delta x^1} \left[  \left(\gdet\right)^{n+1/2,i+1/2} \left< F_{x^1}\right>_A^{n+1/2,i+1/2} \right. \\ 
  \shoveright{- \left.  \left(\gdet\right)^{n+1/2,i-1/2}\left< F_{x^1}\right>_A^{n+1/2,i-1/2}\right]}  \\  
+ \left<S\right>_V^{n+1/2,i} ,
\end{multline}
where $U$ is a conserved variable  (not including $B^i$), $F_{x^1}$ is the associated flux in the $x^1$ direction, $S$ is the source term,  $n$, $n+1/2$, and $n+1$ denote the time at the initial, half, and final stages of a given time step,  $i$ and $i \pm 1/2$ denote the cell center and cell faces of the ith grid cell,  $\Delta t$ is the time step, 
\begin{equation}
\left<F_{x^1}\right>_A = \frac{1}{\gdet \Delta x^2 \Delta x^3} \int \gdet F_{x^1} dx^2 dx^3
\end{equation}
denotes an average over the area of the face with normal in the $x^1$ direction at a fixed time, and 
\begin{equation}
  \left<U\right>_V = \frac{1}{\gdet \Delta x^1 \Delta x^2 \Delta x^3}\int \gdet U dx^1 dx^2 dx^3
  \end{equation} 
  denotes an average over the cell volume at fixed time. 
  Similarly, the magnetic field equation in the $x^1$ direction becomes (in 3D): 
   \begin{multline}
\label{eq:time_derivative_discretized_B}
 \frac{ \left(\gdet \right)^{n+1,i-1/2,j,k}\left< B_{x^1} \right>_A^{n+1,i-1/2,j,k} -    \left(\gdet\right)^{n,i-1/2,j,k} \left< B_{x^1} \right>_A^{n,i-1/2,j,k}}{  \left(\gdet\right)^{n+1/2,i-1/2,j,k}  \Delta t} \\
\shoveleft{\approx  - \frac{1}{ \left(\gdet\right)^{n+1/2,i-1/2,j,k}\Delta x^2}  \times} \\ 
\shoveright{\left[  \left(\gdet\right)^{n+1/2,i-1/2,j+1/2,k} \left< E_{x^3}\right>_L^{n+1/2,i-1/2,j+1/2,k} \right.} \\ 
\shoveright{-  \left. \left(\gdet\right)^{n+1/2,i-1/2,j-1/2,k}\left< E_{x^3}^i\right>_L^{n+1/2,i-1/2,j-1/2,k} \right]} \\
\shoveleft{ +\frac{1}{\left(\gdet\right)^{n+1/2,i-1/2,j,k}\Delta x^2}  \times} \\
\shoveright{\left[ \left(\gdet\right)^{n+1/2,i-1/2,j,k+1/2} \left< E_{x^2}\right>_L^{n+1/2,i-1/2,j,k+1/2} \right. } \\
 \shoveright{-\left. \left(\gdet\right)^{n+1/2,i-1/2,j,k-1/2}\left< E_{x^2}\right>_L^{n+1/2,i-1/2,j,k-1/2}\right],} \\
\end{multline}
where $B_{x^1}$ is shorthand notation for $B^i$, $E_{x^2} = b^3 u^1 - b^1 u^3$ and $E_{x^3} = b^1 u^2 - b^2 u^1$ are the electric fields (i.e., the covariant version of the standard $\mathbf{E} = - \mathbf{v} \times \mathbf{B}$), and 
\begin{equation}
  \langle E_{x^2}\rangle_L = \frac{1}{\gdet \Delta x^2 } \int \gdet E_{x^2}  dx^2, 
\end{equation} 
denotes an average over the length of an $x^2$ edge at a fixed time.

Note that for static metrics, the factors of $\gdet$ on the left-hand side of Equations \eqref{eq:time_derivative_discretized} and \eqref{eq:time_derivative_discretized_B} cancel, but for general time-dependent metrics they need to be accounted for.

In Appendix \ref{app:1d_test}, we present a 1D test of a particular time-dependent metric and demonstrate the code's second-order convergence when the metric is updated every sub-timestep.
When updated only every timestep or any multiple of a timestep the code converges at first order.

\subsection{Approximate Binary Metric}
\label{sec:approx_metric}
To establish notation and for explicitness, in this subsection, we review the approximate binary black hole metric presented in \citet{Combi2021} in which the authors superimpose two boosted individual Kerr metrics\footnote{The metric could in principle be made even simpler by neglecting the boost terms (e.g., \citealt{Davelaar2022}).  We note, however, that the lowest order boost terms corresponding to a Galilean transformation need to be included for the metric to be an accurate representation of a moving black hole.}.

As in \cite{Lopez2021}, we will use Kerr-Schild gauge of the superposition but following the construction in \citet{Combi2021}. In the rest frame of an isolated Kerr black hole, the metric in Cartesian Kerr-Schild coordinates (CKS,  \citealt{Kerr1963}) is 
\begin{equation} 
  g_{\mu \nu} = \eta_{\mu \nu} + f l_\mu l_\nu \\
\end{equation}
where $\eta_{\mu \nu}$ is the Minkowski metric,
\begin{equation}
  f = M_{\rm BH} \frac{2r^3}{r^4 + a_iX^i},
\end{equation}
\begin{equation}
  \begin{aligned}
l_\mu = \left[\rule{0cm}{0.6cm} 1 \right.,  &\frac{r X - a_Y Z +a_Z Y + \left(a_iX^i\right)\frac{a_X}{r}}{r^2+a^2},  \\
 & \frac{r Y - a_Z X +a_X Z + \left(a_iX^i\right)\frac{a_Y}{r}}{r^2+a^2},  \\ 
 & \left. \frac{r Z - a_X Y +a_Y X + \left(a_iX^i\right)\frac{a_Z}{r}}{r^2+a^2} \right], 
 \label{eq:lmu}
\end{aligned}
\end{equation}   
$M_{\rm BH}$ is the black hole mass,  
\begin{equation}
  \begin{aligned}
  R^2 &= X^2 + Y^2 + Z^2 \\
  r^2 &= \frac{R^2 - a^2 + \sqrt{\left(R^2-a^2\right)^2 + 4a_iX^i} }{2},
  \label{eq:radius}
\end{aligned}
\end{equation}
$a_i = (a_X,a_Y, a_Z)$ is the dimensionless black hole spin in the $X$, $Y$, and $Z$ directions, and $a = \sqrt{a_X^2 + a_Y^2 + a_Z^2}$. {This useful form of the KS metric for arbitrary spin direction was presented in the Appendix of \citet{ma2021extending} for the first time (as far as we know)}.
Here, $X^i  = (X,Y,Z)$ are the black hole rest frame spatial coordinates.

Now consider a Kerr black hole moving on a trajectory given by $s^\mu(\tau) = [t_{\tau}(\tau),s^x(t_\tau),s^y(t_\tau),s^z(t_\tau)]$, where $\tau$ is the proper time of the black hole. 
This is related to the time in the inertial lab frame, 
\begin{equation}
  \tau = \int \limits_{0}^{t_{\tau}} \frac{dt}{\Gamma(t)} ,
\end{equation}
where $\Gamma(t)^2 = [1-\beta(t)^2]^{-1}$, and $\beta^i(t)  = ds^i(t)/dt$.
Following \citeauthor{Combi2021} (\citeyear{Combi2021}; see also \citealt{Mashhoon2002}), for every space time event $x^\mu = (t,x,y,z)$ in the lab frame we construct a coordinate system centered on the black hole with a proper time $\tau$ such that the given event and the black hole trajectory are simultaneous.  
The lab frame time corresponding to this $\tau$, $t_{\tau}(\tau)$,  is not the same as the lab frame event time $t$ unless the black hole is not moving. 
Mathematically, this corresponds to the the relation:
\begin{equation}
  x^\mu = s^\mu\left(\tau\right) + \Lambda^\mu_i\left(t_\tau\right)X^i,
  \label{eq:coordinate_definition}
\end{equation}
where $\Lambda^\mu_\nu$ is the instantaneous Lorentz transformation:
\begin{equation}
  \Lambda^\mu_\nu = 
  \begin{pmatrix}
\Gamma &  \Gamma \beta^x &  \Gamma \beta^y & \Gamma \beta^z \\
 \Gamma \beta^x & 1 +\left( \Gamma-1\right) n_x^2  & \left( \Gamma-1\right) n_xn_y & \left( \Gamma-1\right) n_x n_z \\
 \Gamma \beta^y  & \left( \Gamma-1\right) n_x n_y &  1 +\left( \Gamma-1\right) n_y^2  & \left( \Gamma-1\right) n_y n_z   \\
 \Gamma \beta^z & \left( \Gamma-1\right) n_x n_z & \left( \Gamma-1\right) n_y n_z &  1 +\left( \Gamma-1\right) n_z^2 
\end{pmatrix},
\label{eq:Lambda}
\end{equation}
 where $n_i=n^i = \beta^i/\beta$ and the time dependence of $\beta$ and $\Gamma$ is assumed implicitly.  
 Equation \eqref{eq:coordinate_definition} defines the relationship between $X^\mu = (\tau,X,Y,Z)$ and $x^\mu = (t,x,y,z)$ at every point in time and space given the black hole trajectory.   
 Note that we have chosen coordinates where the lab frame $x,y,$ and $z$ axes are aligned with the black hole frame $X,Y,$ and $Z$ axes (since the spin axis of the black hole can be in an arbitrary direction this choice comes without loss of generality).
 
 Equation \eqref{eq:coordinate_definition} results in a nonlinear set of equations (see Equations \ref{eq:coordinates} and \ref{eq:coordinates_inverse} in Appendix \ref{app:nonlinear_expression}), which, given a black hole trajectory can be solved numerically at each lab frame location and time to obtain $t_\tau$ and thus determine $X^\mu$ in terms of $x^\mu$.  
Doing so, however, would introduce a significant extra computational cost for computing the metric and potentially reducing the overall speed of the simulations (though it is not obvious by how much).
Not only that, but the coordinates become ill-defined for larger distances from the black hole if the trajectory is accelerating.   
Primarily motivated by this coordinate issue, instead of solving the nonlinear coordinate transformation we make the approximation that $s^i(t_\tau) \approx s^i(t) + \beta^i (t) (t_\tau - t) $ and $\beta^i(t_\tau) \approx \beta^i(t)$.  
In general, this is a good approximation close to the black hole where $|t_\tau - t|$ is small but it has the potential to be highly inaccurate at large distances from the black hole when $|t_\tau - t|$ is large.
For the specific case of an orbiting black hole, however, 
there is an upper limit to the error incurred due to this approximation on the rest frame distance from the black hole.
This error is roughly equal to the typical $\beta^2$ of the orbit (e.g., $\lesssim 5 \%$ for a black hole on a circular orbit around a much larger black hole at a separation of $\gtrsim 20$ light crossing times of the larger black hole).  
This is because the relative difference in black hole position for any two given points along the orbit become negligible at larger distances from the system.  
The potential $\lesssim 5 \%$ error in distance could cause small differences in the gas distribution at large distances from the black hole, however, these differences would be smaller than the uncertainties in, e.g., the initial conditions of black hole accretion flows.

Using this approximation, Equations (\ref{eq:coordinates}--\ref{eq:coordinates_inverse}) become:
     \begin{equation}
  \begin{aligned}
    t_\tau\left(\tau\right)  ={} & t - \Gamma^2 \beta^x\left[x-s^x(t)\right] -  \Gamma^2 \beta^y\left[y-s^y( t)\right] - \Gamma^2\beta^z\left[z-s^z(t)\right] \\
    X =&  \left[1 + \left( \Gamma-1\right) \left(\frac{\beta^x}{\beta}\right)^2\right] \left[x-s^x(t)\right] + \left( \Gamma-1\right) \frac{\beta^x\beta^y}{\beta^2} \left[y-s^y(t)\right] \\ 
    &+\left(  \Gamma-1\right) \frac{\beta^x\beta^z}{\beta^2}\left[z-s^z(t)\right]\\
        Y =& \left(  \Gamma-1\right) \frac{\beta^x\beta^y}{\beta^2} \left[x-s^x(t)\right] +  \left[1 + \left( \Gamma-1\right) \left(\frac{\beta^y}{\beta}\right)^2\right] \left[y-s^y(t)\right]  \\ 
        &+ \left(  \Gamma-1\right) \frac{\beta^y\beta^z}{\beta^2}\left[z-s^z(t)\right] \\
        Z =&   \left(  \Gamma-1\right) \frac{\beta^x\beta^z}{\beta^2} \left[x-s^x(t)\right]  +  \left(  \Gamma-1\right) \frac{\beta^y\beta^z}{\beta^2} \left[y-s^y(t)\right] \\ 
        &+ \left[1 + \left( \Gamma-1\right) \left(\frac{\beta^z}{\beta}\right)^2\right]\left[z-s^z(t)\right],
  \end{aligned}
  \label{eq:coordinates_linear}
\end{equation}
and 
\begin{equation}
  \begin{aligned}
    t  ={} & t_\tau(\tau) + \Gamma \beta^x  X + \Gamma \beta^y  Y + \Gamma \beta^z  Z \\
    x =&  s^x\left(t\right) + \left[1 + \left( \frac{1}{\Gamma}-1\right) \left(\frac{\beta^x}{\beta}\right)^2\right] X + \left(  \frac{1}{\Gamma}-1\right) \frac{\beta^x\beta^y}{\beta^2} Y \\ 
    &+\left( \frac{1}{\Gamma}-1\right) \frac{\beta^x\beta^z}{\beta^2}Z\\
        y =& s^y\left(t\right) + \left(  \frac{1}{\Gamma}-1\right) \frac{\beta^x\beta^y}{\beta^2} X+  \left[1 + \left( \frac{1}{\Gamma}-1\right) \left(\frac{\beta^y}{\beta}\right)^2\right] Y  \\ 
        &+ \left( \frac{1}{\Gamma}-1\right) \frac{\beta^y\beta^z}{\beta^2}Z\\
        z =&  s^z\left(t\right)+  \left( \frac{1}{\Gamma}-1\right) \frac{\beta^x\beta^z}{\beta^2} X  +  \left(  \frac{1}{\Gamma}-1\right) \frac{\beta^y\beta^z}{\beta^2} Y \\ 
        &+ \left[1 + \left(  \frac{1}{\Gamma}-1\right) \left(\frac{\beta^z}{\beta}\right)^2\right]Z,
  \end{aligned}
  \label{eq:coordinates_linear_inverse}
\end{equation}   
where $\beta^i$ and $\Gamma$ are now evaluated at $t$ instead of $t_{\tau}$.  
 Taking the derivatives of Equations (\ref{eq:coordinates_linear}--\ref{eq:coordinates_linear_inverse}), one can show that $dx^\mu = \Lambda^\mu_\alpha(t) dX^\alpha$, where $\Lambda^\mu_\nu$ is the standard Lorentz transformation given by Equation \eqref{eq:Lambda}.
 
Using this boost and the coordinate transformation we obtain our approximate binary metric:
\begin{equation}
  \label{eq:boosted_metric}
  \begin{aligned}
  g_{\mu \nu} =  \eta_{\mu \nu} & + \left[ f \left(X^\mu\right) \left(\Lambda^{-1}\right)^\alpha_\mu l_\alpha \left(X^\mu\right) \left(\Lambda^{-1}\right)^\kappa_\nu l_\kappa \left(X^\mu\right) \right]_1 \\ & +\left[ f \left(X^\mu\right) \left(\Lambda^{-1}\right)^\alpha_\mu l_\alpha \left(X^\mu\right) \left(\Lambda^{-1}\right)^\kappa_\nu l_\kappa \left(X^\mu\right)\right]_2,
  \end{aligned}
\end{equation} 
where the subscripts $1$ and $2$ indicate the primary and secondary black holes.    
We obtain the inverse metric, $g^{\mu \nu}$, as well as the spatial and temporal derivatives numerically. 

In Appendices \ref{app:bondi_test} and \ref{app:bhl_test} we present two different tests involving moving black holes using this metric, demonstrating our code's convergence properties and the accuracy of our implementation.

\subsection{Post-Newtonian Orbits of the Black Holes}
\label{sec:PN_orbits}
The above metric has as an input the black hole trajectories, which have to be solved for independently.
To do this, we use the public code {\tt CBwaves} \citep{cbwaves} to evolve the trajectories of the two black holes starting from a given eccentricity, separation distance, and initial spin directions and magnitude with respect to the orbital plane.  
 {\tt CBwaves} is a fast C code that solves the post-Newtonian equations of motion \citep{blanchet2014gravitational} up to the 4th PN order and includes all the relevant acceleration terms, radiation reaction, spin-spin coupling, and spin-orbit coupling. It has been tested in \citet{cbwaves} and is one of the few public codes to solve the PN equations (a coupled set of ordinary differential equations with lengthy terms) in full generality.  The outputs of the code are the two black hole 3D positions, 3D velocities, and 3D spins as a function of time up to merger. 
For simplicity, in this work, we always use zero initial eccentricity but allow for the black hole spin and angular momentum vectors to be misaligned.
This can lead to significant precession due to spin-orbit coupling, as we shall see in some of our simulations.  
We define the initial inclination angle, $i_0$, such that for $i_0=0$, the orbit is initially prograde with the accretion disk in the midplane, and for $i_0=90^\circ$ the orbit is initially perpendicular to the midplane and moves clockwise in the $y$-$z$ plane.
We distinguish between $i$ as the current inclination of the orbit (always defined with respect to the fixed $x$-$y$ plane) and $i_0$, the initial inclination of the orbit.

\subsection{Further Approximations Specific to This Work}

In this work, we apply the boosted binary metric to a system in which the mass ratio, $M_2/M_1 \equiv q \ll 1$, where $M_{1,2}$ are the masses of the primary and secondary black hole, respectively. 
Therefore, we approximate the primary black hole as stationary and located at the origin of the lab frame. 
This approximation would break down for higher mass ratios, $q\lesssim 1$, at which point the qualitative picture of a perturber black hole impacting an established AGN disk becomes inaccurate.
We also neglect self-gravity of the accretion flow and radiative effects.
This formally limits the applicability of our study to lower mass accretion rates where these effects are negligible.  
At higher accretion rates self-gravity could introduce enough dynamical friction and drag to alter the black hole orbits, while radiation would significantly cool the disk and reduce its scale height, altering the flow dynamics substantially.  
The latter regime we plan on studying in future work.

\subsection{Algorithmic Details}

For stability purposes, the time step used by {\tt Athena++} is set by the light crossing time across the shortest $x$, $y$, or $z$ length of a cell. 
In particular, we choose a Courant-Friedrichs-Lewy (CFL) number of 0.3.  
Because of this, the timestep can be significantly shorter than the characteristic time for the metric to change (e.g., an orbital time for a black hole binary metric, which for the orbits we consider in this work is proportional to $v\lesssim 0.1c$).
Taking advantage of this fact, we only update the metric every 10 timesteps.  
This means that the MHD equations are first solved including the time-dependent source terms described at the beginning of \S \ref{sec:methods} but otherwise as if the metric were time-independent.  
That is, the conversion between conserved and primitive variables (and vice-versa) is done using the metric at the time of the last update, \emph{not} the metric at the current time of a given step or substep of the algorithm.  

As a result, the code is formally only first-order accurate in time.
However, when the metric changes at a rate much slower than that of the fluid (i.e., when the black holes move at velocities much slower than the characteristic GRMHD velocities), the errors incurred by the first-order metric update can still be much less than the errors incurred by the second-order GRMHD evolution.
Quantifying this more precisely is difficult and is likely different for every simulation and choice of parameters.
That said, for two 3D GRMHD accretion problems with moving  black holes that are similar in several ways to our target problem (see Appendix \ref{app:bondi_test}, Appendix \ref{app:bhl_test}, and Figures \ref{fig:bondi_contour}--\ref{fig:bhl_contour}) we have found that updating the metric once every 10 timesteps still results in satisfactory agreement with the expected solutions.
This is true even though the black hole velocities in those test problems are much higher ($0.9c$) than those we expect in our simulations ($\lesssim 0.25c$ for orbits with separations $\gtrsim 20\rg$).  
Moreover, when discontinuities or shocks are present in the flow (as they are in turbulent accretion simulations in general but especially for the Bondi-Hoyle-Lyttleton-type flows we expect near each black hole) all methods reduce to first-order anyway.  

 \subsection{Floors and Patches to the Metric}
 

To avoid coordinate singularities within the event horizon of the black holes 
we modify each black hole's coordinates when $r\le 0.8 r_{\rm H}$ (defined in each black hole's rest frame), where $r_{\rm H} = M_{\rm BH} + \sqrt{M_{\rm BH}^2 - a^2}$ is the event horizon radius for an isolated black hole.  For $r\le 0.8 r_{\rm H}$ we first calculate
\begin{equation}
  \begin{aligned}
  \theta &= \arccos(l_z) \\
  \varphi &= \mathrm{arctan2}(l_y,l_x),
  \end{aligned}
\end{equation}
 where $l_\mu$ is defined in Equation \eqref{eq:lmu}. Then we set $r=0.8 r_{\rm H}$ and recalculate 
 \begin{equation}
   \begin{aligned}
     X = r \sin(\theta) \cos (\varphi)  + a_Y \cos(\theta) - a_Z \sin(\theta) \sin(\varphi)\\
     Y = r \sin(\theta) \sin (\varphi) + a_Z \sin(\theta) \cos(\varphi) - a_X \cos(\theta)   \\
     Z = r \cos(\theta) + a_X \sin(\theta) \sin(\varphi) - a_Y \sin(\theta) \cos(\varphi)      
      \end{aligned}
 \end{equation}
 from the new $r$ and old $\theta$ and $\varphi$.
 Since this coordinate modification is only applied well within the event horizon it should have no effect on the simulation outside the horizon and it prevents occasional $\mathrm{NaN}$s from crashing the simulation.
 
 Within the horizon of each black hole, we also set the gas to be moving along with the black hole by setting $u^\mu = \Lambda^\mu_\nu (u_{\rm rest}^\prime)^\nu$, where $(u_{\rm rest}^\prime)^\mu$ is the four velocity of a stationary observer in the instantaneous black hole rest frame:
\begin{equation}
  \begin{aligned}
    (u_{\rm rest}^\prime)^t &= \frac{1}{\alpha} \\
    (u_{\rm rest}^\prime)^i &= - \alpha g^{t i},
  \end{aligned}
\end{equation}
where $\alpha \equiv 1/\sqrt{-g^{tt}}$ and $\Lambda^\mu_\nu$ is the Lorentz transformation defined in Equation \eqref{eq:Lambda}.
This helps prevent gas and magnetic fields from within the event horizon from `leaking' out into the rest of the computational domain as the black hole moves across the grid. 
In particular, without enforcing this velocity condition we have found that `magnetic explosions' caused by unphysically large magnetic fields leaking out of the horizon can ruin the simulation.

For the MHD quantities, the density floor is $10^{-6} (r/r_{\rm g})^{-3/2}$ and the pressure floor is $3.33 \times 10^{-9} (r/r_{\rm g})^{-5/2}$, with $\sigma \equiv b^2/\rho \le 100$ and $\beta \ge 0.001$ enforced via additional density and pressure floors, respectively.  Here $\beta$ is the ratio between the thermal and magnetic pressures while $b^2$ is twice the magnetic pressure in Lorentz-Heaviside units.  Additionally, the velocity of the gas is limited such that the maximum Lorentz factor is 50.  
The radial power law indices of the pressure and density floors are chosen to be consistent with spherical Bondi-type accretion flows appropriate for a non-rotating, low density atmosphere surrounding the accretion flow.
The precise magnitudes of these floors have negligible effects on the accretion flow \citep{Porth2019} and are chosen to be several orders of magnitude less than the initial density maximum ($=$1 in code units).
The $\sigma $ and plasma $\beta $ limits help prevent primitive inversion failures in strongly magnetized regions while the limit on Lorentz factor helps localize failures; the values we use are based on those found to be fairly robust in GRMHD torus simulations \citep{Porth2019}.
The resulting Lorentz factor in the very low density/highly magnetized jet of the simulations can directly depend on these limits and thus should not be over-interpreted. 

  \begin{figure}
\includegraphics[width=0.49\textwidth]{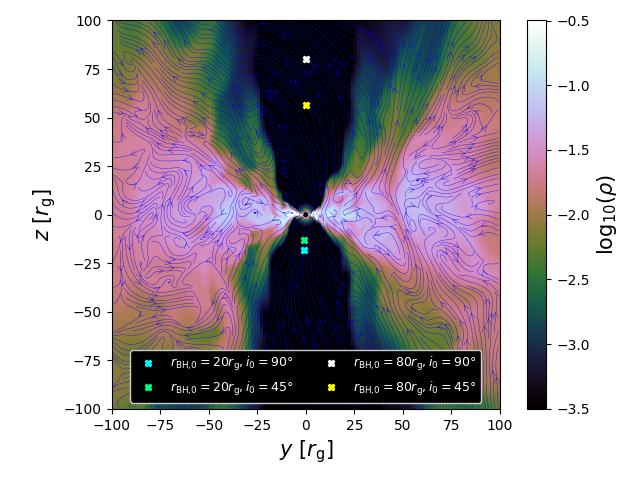}
\caption{2D slice of density overplotted with magnetic field lines at $t=100{, }000 M$ for the single  $a=0.9375$ black hole torus simulation used as the `initial' state of our small mass ratio binary simulations. 
The black circle at the center represents the event horizon.  
The flow is thick, turbulent, magnetically arrested, and is associated with a powerful jet in the direction of the spin axis ($+z$).
To this simulation, we introduce perturber black holes on various orbits to study the effect they have on outflows and accretion.
The initial projected locations of the secondaries are marked by thick `X's.
} 
\label{fig:init_torus}
\end{figure}
 
\subsection{Initial Conditions}
\label{sec:inits}
Before adding the secondary black hole into the system, we run a single black hole simulation with a stationary Kerr metric initialized with a \citet{Fishbone1976} torus with inner radius $20 r_{\rm g}$ and pressure maximum at $41 r_{\rm g}$.  
Note that we define $r_{\rm g} = G M_1/c^2$ in terms of the mass of the primary, which also sets the timescale $\rg/c = M_1$.
A large, single loop of magnetic field is seeded in the torus with $\max{P}/\max{P_{B}} = 100$, where $P_{B} = b^\mu b_\mu /2$ is the magnetic pressure.
The black hole has dimensionless spin $a=0.937$ in the $+z$ direction.

The grid encompasses $(1000 r_{\rm g})^3$ and includes a $128^3$ base resolution with 8 additional levels of static mesh refinement (SMR), increasing the resolution by a factor of 2 every factor of $\approx$ 2 in radius. 
The finest resolution is concentrated within $-6.25 r_{\rm g} \le x,y,z \le 6.25 r_{\rm g}$ with cell size $\approx (0.1 r_{\rm g})^3$.
This resolution is comparable or better than the highest resolution (192$^3$) spherical modified Kerr-Schild simulations in the Event Horizon Code Comparison Project \citep{Porth2019} that were found to be converged for most fluid quantities. 
Specifically, at ($r=12\rg, \theta = {\rm \pi}/2$), our simulations are better resolved in most of the domain by a factor of $\sim$ 1.5 in the radial direction and $\sim$ 1.875 in the azimuthal direction but less resolved by a factor of $\sim$ 0.75 in the $\theta$ direction at the midplane where the modified Kerr-Schild coordinates focus the highest resolution.  
Within $-3.125 r_{\rm g} \le x,y,z \le 3.125 r_{\rm g}$ our grid becomes comparably less resolved by a factor of $\sim$ 2 than the rest of our simulations because we do not place a 9th level of mesh refinement in this region (as would be required for effectively logarithmic radial spacing).
We do this to save computational cost since our focus in this paper is predominantly on the flow at larger radii and there are still many cells contained within the event horizon.

We use piecewise-linear reconstruction and the HLLE Riemann solver.

The simulation is run for 100{,}000 $M_1$ to obtain the initial conditions for our binary simulations and then an additional 50{,}000 $M_1$ for comparison purposes. 

\begin{figure}
\includegraphics[width=0.47\textwidth]{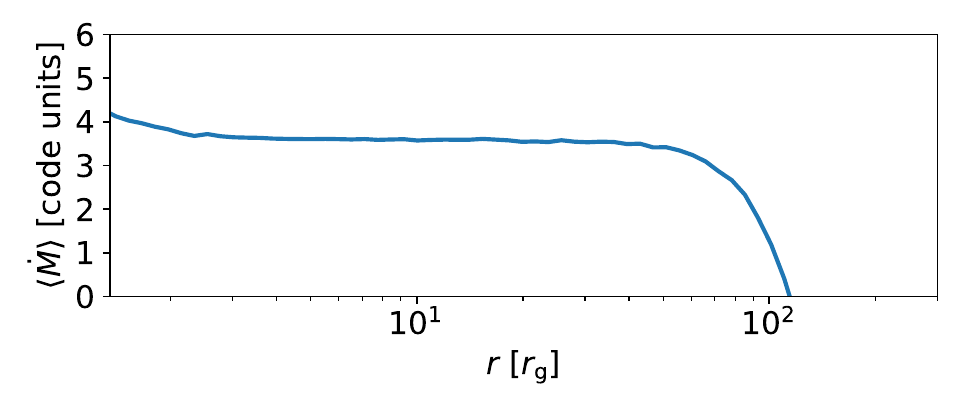}
\includegraphics[width=0.49\textwidth]{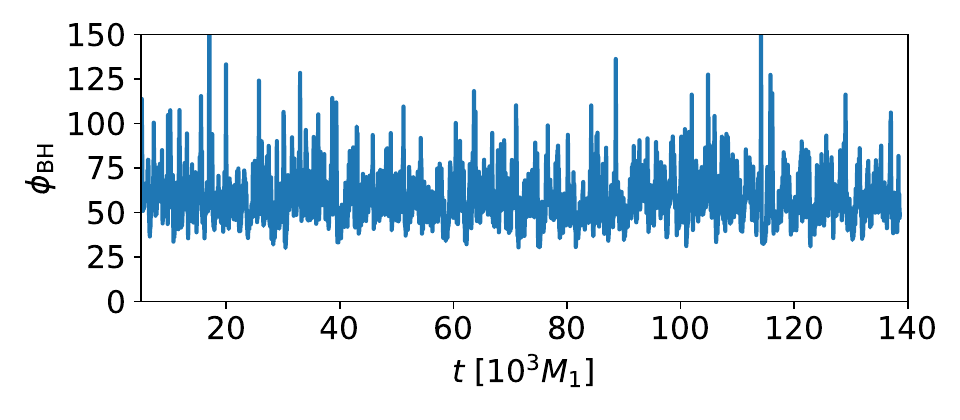}
\caption{
Properties of the single black hole accretion flow used as initial conditions for our binary simulations.  
Top: time-averaged accretion rate, $\langle \dot M \rangle$ vs.\ radius (averaged over 50{,}000--100{,}000 $M_1$).  Bottom: dimensionless black hole flux threading the black hole, $\phi_{\rm BH}$, vs.\ time.  
$\langle \dot M 
\rangle$ is roughly constant out to $\approx$ 70--100 $r_{\rm g}$, indicating that the flow has reached inflow equilibrium out to this distance. 
The accretion flow is firmly in the MAD state, with saturated magnetic flux around $\phi_{\rm BH} \approx$ 60 \citep{Sasha2011} and going through cycles of build-up followed by dissipation every $\sim$ 3{,}000 M.  
} 
\label{fig:1d_torus_plots}
\end{figure}

Figure \ref{fig:init_torus} plots a 2D poloidal density slice at this time, showing a thick, turbulent accretion flow with a narrow jet in the $z$ direction.  
This flow is magnetically arrested and in equilibrium out to $70$--$100$ $r_{\rm g}$, as shown in the dimensionless black hole flux, $\phi_{\rm BH}$ vs.\ time and net mass accretion rate vs.\ radius in Figure \ref{fig:1d_torus_plots}.
These are defined as 
\begin{equation}
\label{eq:mdot}
  \dot M \equiv  \iint \rho u^r d\Omega,
\end{equation}
\begin{equation}
\label{eq:phibh}
  \phi_{\rm BH} \equiv \frac{\sqrt{4{\rm \pi}} \iint |B^r| d\Omega}{2 \sqrt{|\dot M|}},
\end{equation}
where $u^r$ and $B^r$ are the radial component of the four-velocity and magnetic three-vector (converted from Cartesian to spherical CKS coordinates), $d\Omega = \gdet d\theta d\varphi$, and the expressions are evaluated at $r= 5\rg$\footnote{$\dot M$ and $\phi_{\rm BH} $ at $r=5\rg$ 
 are very similar to $\dot M$ and $\phi_{\rm BH} $ at the event horizon but less noisy.}.

\subsection{Introducing The Secondary Black Hole}

At $t= 100{, }000 M_1$, we instantaneously change the metric in the simulation from that of a stationary, single black hole to the binary metric described in \S \ref{sec:approx_metric}, starting with the initial conditions of the Post-Newtonian orbit of the secondary (\S \ref{sec:PN_orbits}).
These initial conditions are chosen such that the perturber initially is located at $x=y=0$ and $z = \rbhi$.
This location is chosen to be within the jet so that any artifact of the sudden addition of the secondary black hole has a negligible effect on the accretion flow.

To conserve fluid quantities, after the instantaneous change in the metric we rescale each conserved variable $U \in [\rho u^t,T^t_t,T^t_i]$ 
and the magnetic field $B^i$ via
\begin{equation}
  \begin{aligned}
  U_{\rm new} &= U_{\rm old} \frac{\sqrt{-g_{\rm old}}}{\sqrt{-g_{\rm new}}} \\
  B^i_{\rm new} &= B^i_{\rm old} \frac{\sqrt{-g_{\rm old}}}{\sqrt{-g_{\rm new}}}.
  \end{aligned}
\end{equation} 
This ensures that the conserved energy, momentum, mass, and the divergence of the magnetic field are the same before and after the introduction of the secondary.
Though we see no obvious artifact of the instantaneous introduction of the secondary in simulation quantities, we argue that even if such artifacts are present they would have a negligible effect on our results.  
Firstly, since we study small mass ratios, $q\ll 1$, the metric only significantly changes very close to the initial location of the secondary black hole (within a few $\rg$), where the matter related quantities are predominantly set by the floors anyway. 
Secondly, the flow in this region is rapidly outflowing and so any potential artificial feature would be quickly swept away to larger radii and out of the domain of interest.

  \begin{figure*}
\includegraphics[width=0.99\textwidth]{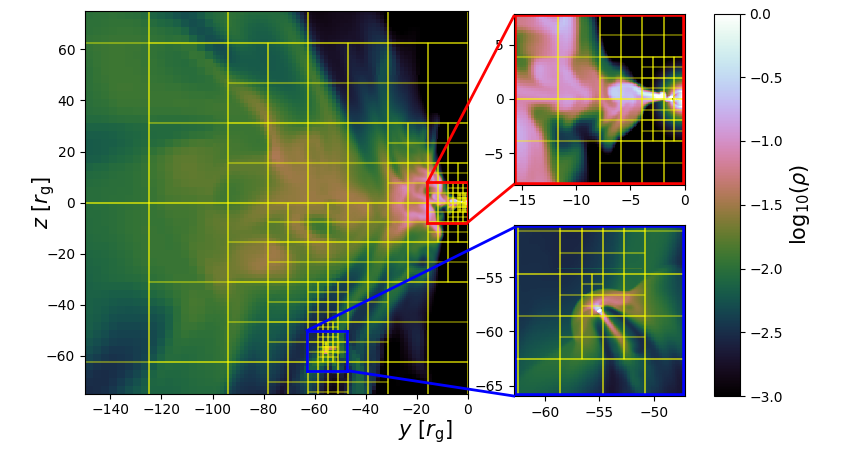}
\caption{Representative example of our static and adaptive mesh-refined grid.
A mass density contour is plotted in the $y$-$z$ plane for our $\rbh=80^\circ$, $i_0=90^\circ$ simulation, with yellow lines demarcating blocks of $16^3$ cells.  
The red and blue outlined subplots zoom in on the primary and secondary black holes, respectively.  
The mesh refinement effectively allows us to resolve multiple scales, particularly surrounding the two black holes.
} 
\label{fig:mesh_plot}
\end{figure*}

In addition to the 8 levels of SMR centered on the primary, we add another 8 levels of adaptive mesh refinement (AMR) centered on the secondary.  
The highest level of refinement is contained within $|X_2|, |Y_2|, |Z_2|  \le 3.125 q \rg$, where $X_2$, $Y_2$, $Z_2$ are the secondary's rest frame coordinates (with the secondary as the origin), the second highest level of refinement is contained within $|X_2|, |Y_2|, |Z_2|  \le 6.25 q M_1$, and so on.
More precisely, the $n$th level of AMR is contained within $|X_2|, |Y_2|, |Z_2|  \le 3.125 (2^{n_{\rm max}-n}) q M_1$, where $n_{\rm max}$ is the maximum AMR level.
This results in a cell size of $(0.1 \rg)^3$ at the maximum refinement level (which means that the secondary horizon radius is $\sim$ 2 cells in length for $q=0.1$ and a non-spinning secondary).
As an example of how this works in our simulations, Figure \ref{fig:mesh_plot} shows the grid structure at a representative time in our $\rbhi=80 \rg$, $i_0=90^\circ$ simulation, plotted over a 2D contour of density in the $y$-$z$ plane.  
Each $16^3$ block of cells is outlined by a yellow square.  
This demonstrates how our grid effectively focuses resolution on the two black holes and resolves multiple scales. 

Once the secondary is introduced, we run the simulations an additional 50{,}000 $M_1$ for $\rbhi=80\rg$ and 40{,}000 $M_1$ for $\rbhi=20\rg$.
This time is sufficient for $\gtrsim 10$ orbits for secondary black holes located at $\rbhi \lesssim 80 \rg$ and long enough to see spin-orbit effects for orbits around $\rbhi \sim 20 \rg$.

\subsection{Suite of Runs}
The primary goals of this work are to 1) demonstrate the basic properties of a thick accretion disk around a supermassive black hole perturbed by a smaller black hole on an inclined orbit and 2) to describe, test, and demonstrate capabilities of the new time-dependent metric version of {\tt Athena++}.  
We do not therefore seek to either simulate an exhaustive sweep of parameter space nor do we specifically focus on a target astrophysical system.  
Instead, we choose a select few simulations to run that we expect can represent some of the more general possibilities in such a system.  
Namely, we fix $q=0.1$, and use two different initial black hole separations, $\rbhi = 20 r_{\rm g}$ and $\rbhi=80 r_{\rm g}$.  
We also use two different initial orbital inclinations, $i_0=90^\circ$ (i.e., an orbit initially passing perpendicularly through the accretion disk), and $i_0=45^\circ$.  
These orbits are initiated as quasi-circular by using an eccentricity reduction procedure; in particular, we change the initial velocities iteratively until the quantity $e = (r_{\rm max} - r_{\min})/(r_{\rm max} + r_{\min})$ is below $0.01$, where $e$ is the eccentricity, and $r_{\rm min/max}$ is the minimum/maximum distances between the two black holes.
All simulations use $a_2 = 0$, that is, the secondary black hole is non-spinning. 

We choose to focus on $q=0.1$ because it is the highest mass ratio at which we feel our approximation of a stationary primary black hole is justified, while for much smaller mass ratios we have found the effects of the secondary on the primary accretion flow to be almost undetectable.
We choose $\rbhi=20\rg$ and $\rbhi=80 \rg$ because these represent the minimum and maximum initial separation distances at which we can reasonably trust our results.  
For $\rbhi < 20\rg$ our metric approximation of a linear superposition of two boosted Kerr metrics becomes poor, while for $\rbhi > 80 \rg$ the secondary would be traveling through regions of the primary accretion flow that have not yet reached a steady state.
We choose $i_0=90^\circ$ and $i_0=45^\circ$ to bracket the two extremes of orbits still in the ``collision regime'' for our thick primary accretion disk. 
$i_0=90^\circ$ orbits are completely perpendicular to the disk and thus the impact velocity of the secondary is maximized.
$i_0=45^\circ$ orbits on the other hand only graze the edge of the disk with much smaller tangential velocity (as mentioned in \S \ref{sec:analytic_general}).

\begin{figure}
\includegraphics[width=0.47\textwidth]{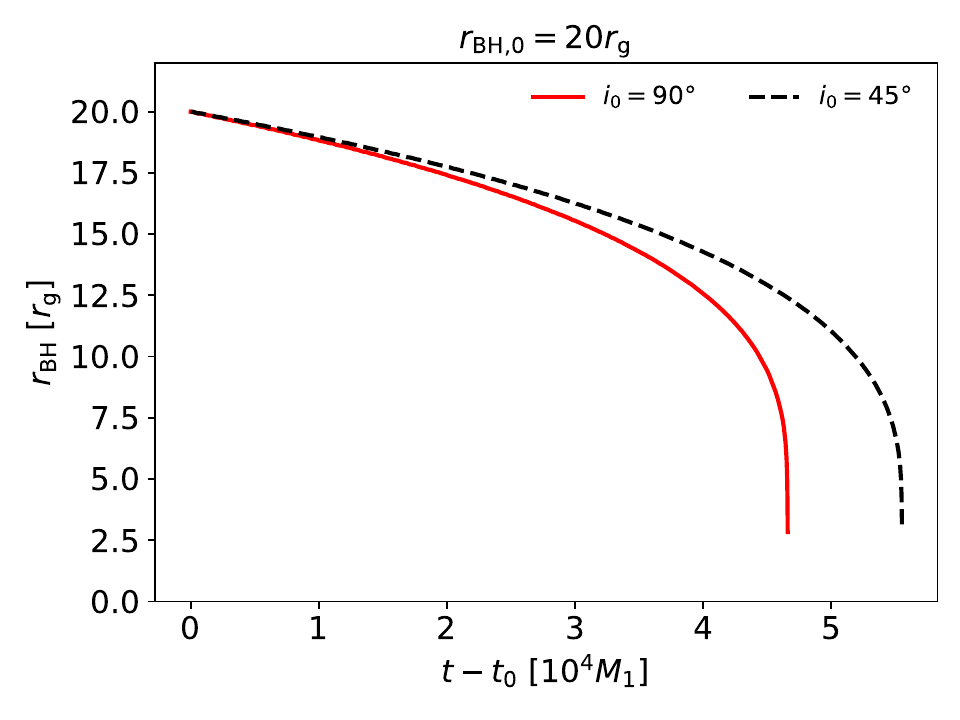}
\includegraphics[width=0.47\textwidth]{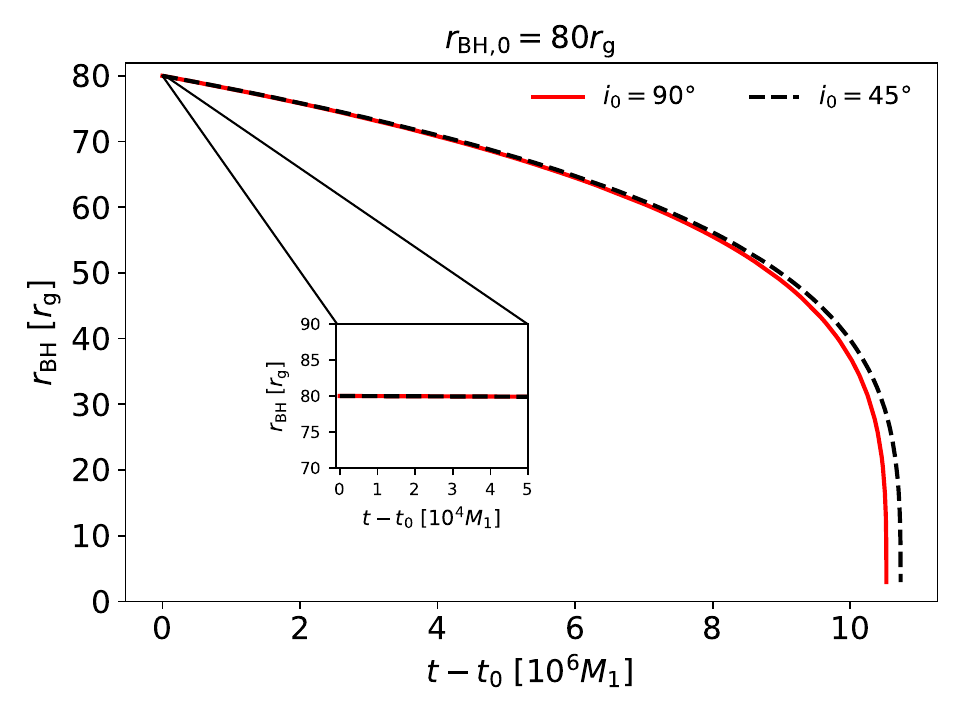}
\caption{
Binary separation distance vs.\ time for the 4 different orbits used in this work.  
Top: $\rbhi = 20 r_{\rm g}$ for $i_0=90^\circ$ (red, solid) and $i_0 = 45^\circ$ (black, dashed).
Bottom: $\rbhi = 80 r_{\rm g}$ for $i_0=90^\circ$ (red, solid) and $i_0 = 45^\circ$ (black, dashed), where the inset zooms in on the time covered by our simulations. 
For orbits with initial separations of $\rbhi = 20 r_{\rm g}$, the timescale for the orbital radius to significantly change is $\sim$ $10^4 M_1$, much shorter than the $\sim$ $10^6 M_1$ timescale for the $\rbhi=80 r_{\rm g}$ orbits.
Since $10^6 M_1$ is much longer than the length of our simulations, for $\rbhi=80 r_{\rm g}$ we safely assume Keplerian circular orbits that do not change in time.  
For $\rbhi=20 r_{\rm g}$ simulations, however, we include the full orbital evolution and can expect to see significant changes on the timescales we can simulate.
The differences between $i_0=90^\circ$ and $i_0=45^\circ$ binary separation distances are relatively small until close to merger ($\sim 4\times 10^4$ for $\rbhi = 20\rg$ and $\sim 9\times 10^6$ for $\rbhi = 80\rg$).
} 
\label{fig:r_orbits}
\end{figure}

\begin{figure}
\includegraphics[width=0.49\textwidth]{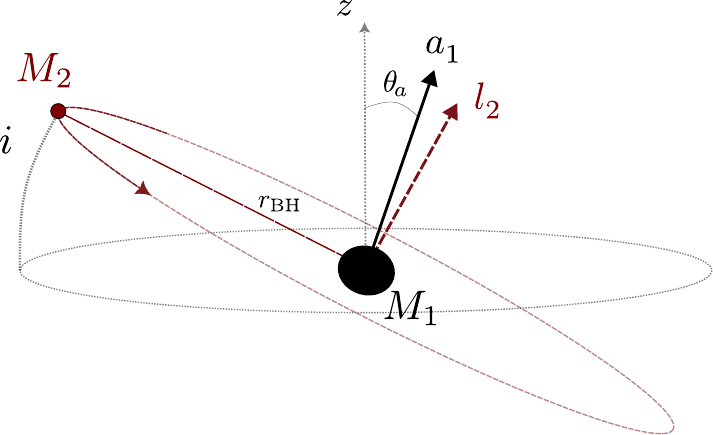}
\caption{ Angles and distances describing the orbits in our simulations.  $r_{\rm BH}$ is the separation between the primary and secondary black holes, $a_1$ is the primary black hole spin, $l_2$ is the secondary's orbital angular momentum, $i$ is the angle between the $x$-$y$ plane and the orbital plane (where $i=0$ corresponds to clockwise motion in the $x$-$y$ plane), and $\theta_a$ is the angle between the primary black hole spin direction and the $z$-axis (the initial primary black hole spin direction).
}
\label{fig:binary_diagram}
\end{figure}

The resulting binary separation distances as a function of time given by {solving the PN equations using} {\tt CBwaves} for these 4 different orbits are plotted in Figure \ref{fig:r_orbits}.  
For $\rbhi = 20 \rg$, merger would happen after $\approx$ 4.5--5.5 $\times 10^4 M_1$  (earlier for $i_0=90^\circ$, later for $i_0=45^\circ$), with significant changes to the binary separation happening on timescales of $\sim$ $10^4 M_1$ (note  that these are comparable to the $4 \times 10^4 M_1$ runtime of our simulations). 
For $\rbhi = 80 \rg$, merger would happen after $\approx$ 10.3--10.7 $\times 10^6 M_1$  (earlier for $i_0=90^\circ$, later for $i_0=45^\circ$), with significant changes to the binary separation happening on timescales of $\sim$ $10^6 M_1$ (note that these are much longer than the $4 \times 10^4 M_1$ runtime of our simulations).
The same timescales are seen in the evolution of the orbital angular momentum of the secondary and primary black hole spin, which we quantify using the inclination of the orbit, $i$, as well as the angle between the primary black hole spin and its initial direction {along the z-axis}, $\theta_a$.
These are defined as 
\begin{equation}
\label{eq:inc}
i \equiv \arccos\left(\frac{l_{2,z}}{l_2}\right)
\end{equation}
and
\begin{equation}
\label{eq:theta_a}
\theta_a \equiv \arccos\left(\frac{a_{1,z}}{a_1}\right),
\end{equation}
where $a_{1,i}$ is the spin vector of the primary, $a_1 = (a_{1,i}a_1^i)^{1/2}$, $l_{2,i}$ is the specific angular momentum of the secondary (e.g., $l_{2,x} = y_2  v_{2,z} -  z_2 v_{2,y}$), $l_2 = (l_{2,i}l_2^i)^{1/2}$, $x_2$, $y_2$, and $z_2$ are the $x$, $y$, and $z$ positions of the secondary, and $v_{2,i}$ are the velocities of the secondary.
These angles are diagrammed schematically in Figure \ref{fig:binary_diagram}.

\begin{figure*}
\gridline{\includegraphics[width=0.49\textwidth]{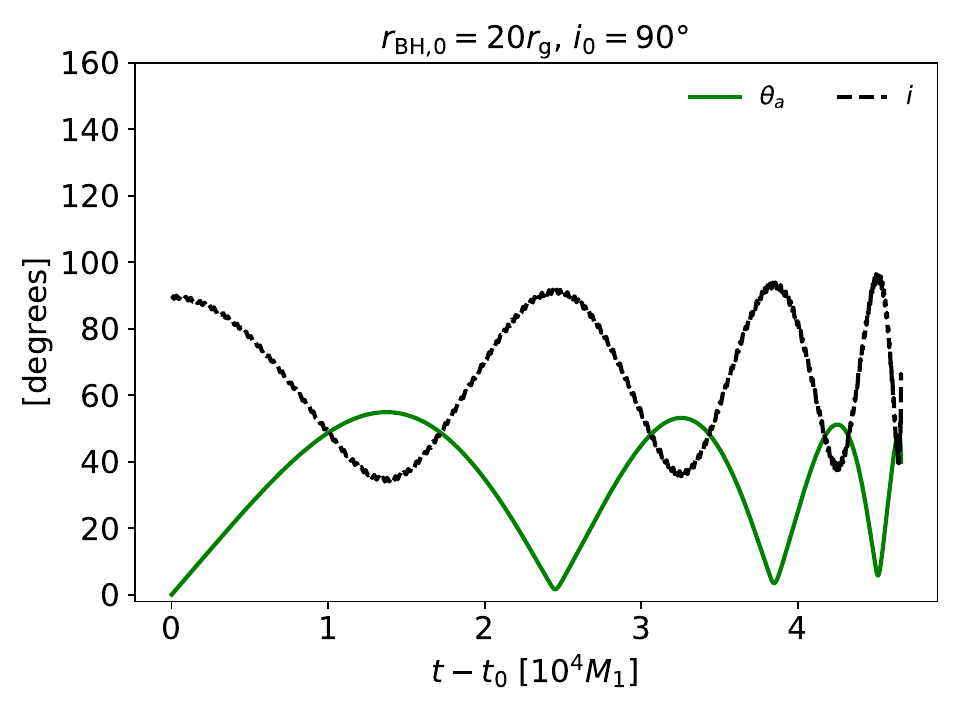}
\includegraphics[width=0.49\textwidth]{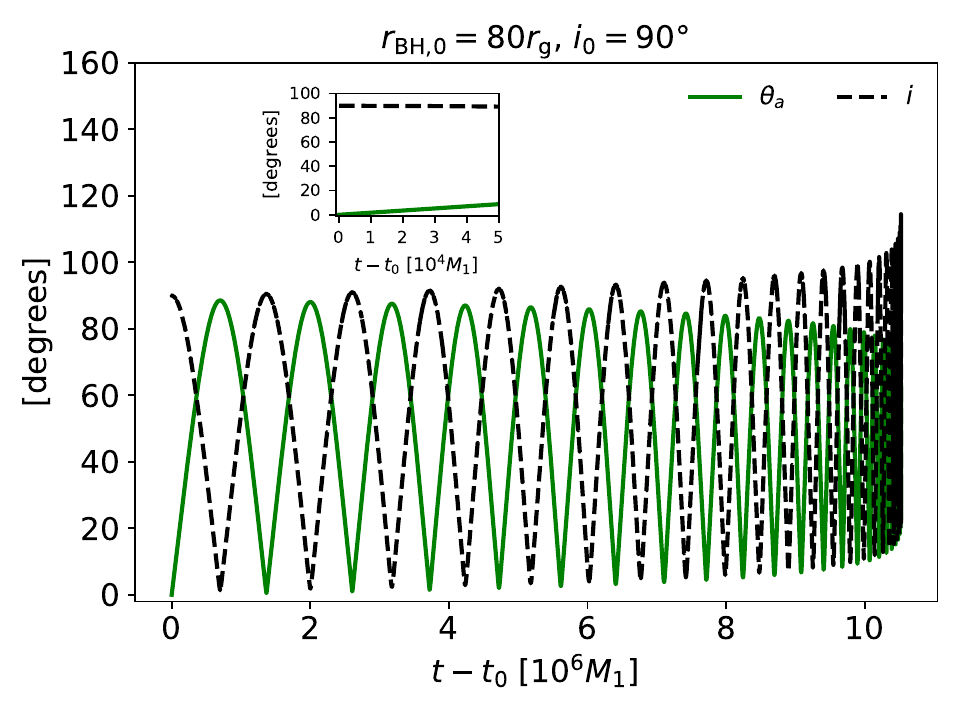}}
\gridline{\includegraphics[width=0.49\textwidth]{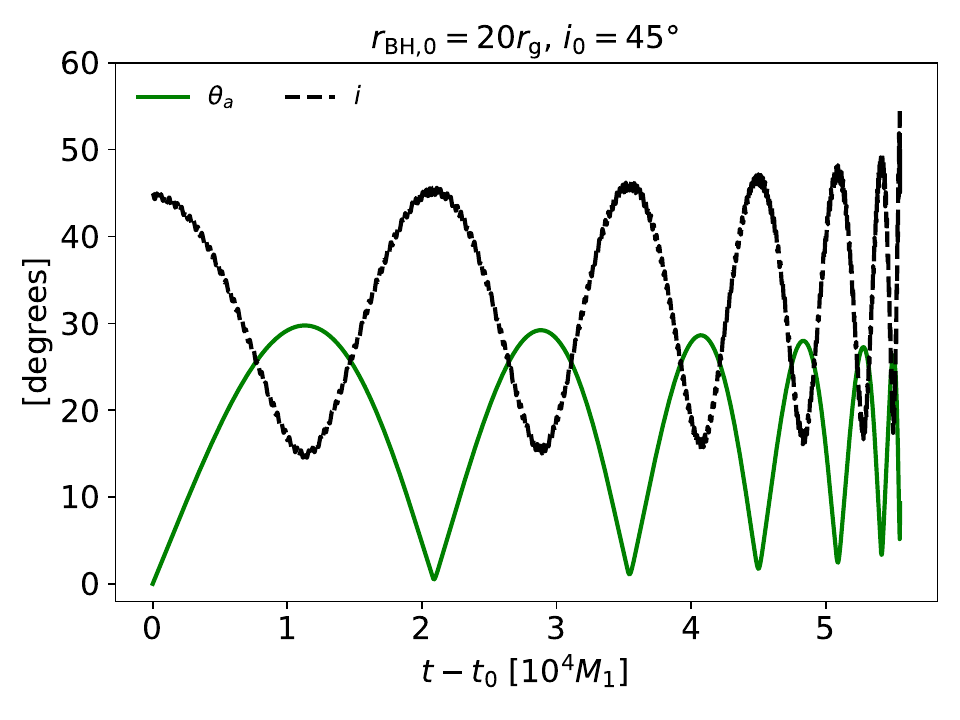}
\includegraphics[width=0.49\textwidth]{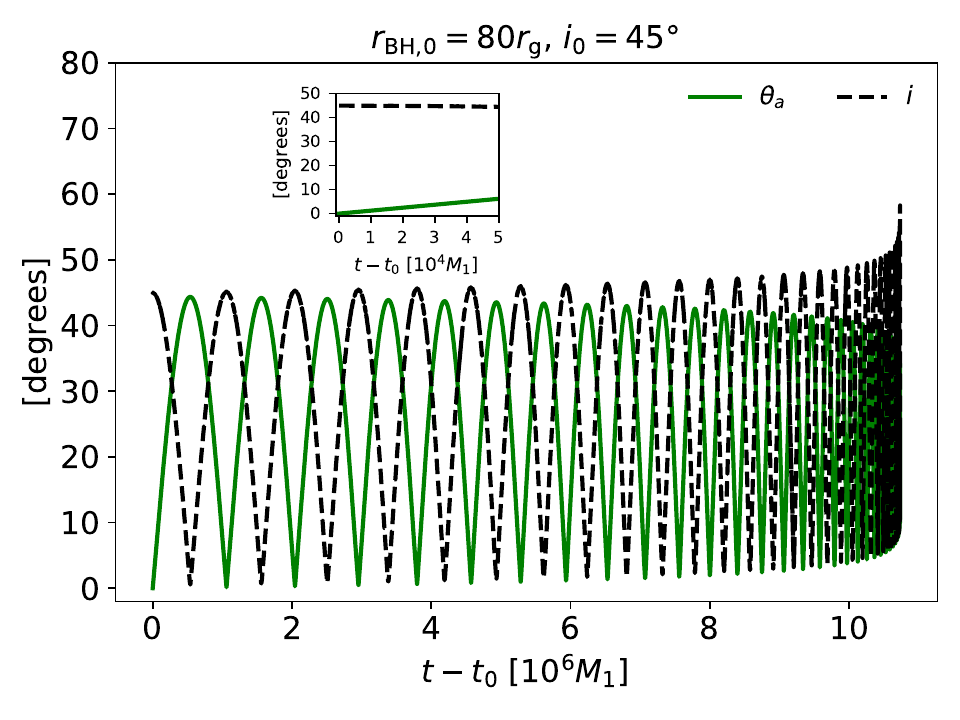}}
\caption{
Angle between the primary black hole spin vector and its initial direction, $\theta_{a}$ (green solid), and the angle between the orbital angular momentum vector and the $x$-$y$ plane, $i$ (black dashed), vs.\ time for the orbital choices in our simulations (all with $q=0.1$).  
Top left: $\rbhi=20\rg, i_0 = 90^\circ$. 
Top right: $\rbhi=80\rg, i_0 = 90^\circ$.
Bottom left: $\rbhi=20\rg, i_0 = 45^\circ$. 
Bottom right: $\rbhi=80\rg, i_0 = 45^\circ$.
In all simulations, both the primary black hole spin and orbital angular momentum change directions significantly over time. 
$i$ changes by $45^\circ$--$130^\circ$ while $\theta_a$ peaks at around $30^\circ$--$90^\circ$.
Generally, the $\rbhi=20\rg$ orbits have stronger amplitude variation in $i$, while the $\rbhi=80\rg$ orbits have stronger amplitude variation in $\theta_a$.  
The timescales for these changes are the same as for the binary separation distances shown in Figure \ref{fig:r_orbits} so that we expect to see significant change in primary spin and orbital angular momentum in $\rbhi=20\rg$ simulations but not in $\rbhi=80\rg$ simulations due to the run times.
For comparison, the insets in the $\rbh=80 \rg$ plots show $\theta_a$ and $i$ zoomed in on the timescale of our simulations.}
\label{fig:spin_orbit}
\end{figure*}

 The angles $i$ and $\theta_a$ are plotted in Figure \ref{fig:spin_orbit} for all four orbital configurations.
For $\rbhi=80 \rg$, the direction of the primary spin of the black hole changes by $\approx$ 45$^\circ$ for $i_0=45^\circ$ in the first $\sim$ $10^6 M_1$ and $\approx$ 90$^\circ$ for $i_0=90^\circ$ in the first $\sim$ $1.5\times 10^6 M_1$, while the direction of the orbital angular momentum changes similar amounts during the same times.
For $\rbhi=20 \rg$, the direction of the primary spin of the black hole changes by $\approx$ 30$^\circ$ for $i_0=45^\circ$ in the first $\sim$ $10^4 M_1$ and $\approx$ 55$^\circ$ for $i_0=90^\circ$ in the first $\sim$ $1.5\times 10^4 M_1$, while the direction of the orbital angular momentum changes by $\approx$ 60$^\circ$ for $i_0=45^\circ$ and $\approx$ 125$^\circ$ for $i_0=90^\circ$ during the same times.
The orbital and spin directions then continue to oscillate back and forth from the initial values on shorter and shorter timescales until merger (which happens after 3--6 oscillations for $\rbhi=20 \rg$ and $\sim$ 20 oscillations for $\rbhi=80\rg$).

Since we can only reasonably simulate timescales $\lesssim 10^5 M_1$, for the $\rbhi=80\rg$ simulations we neglect orbital changes and assume circular Keplerian orbits, evolving for $\approx$ 9 orbits.
For $\rbhi=20 \rg$, however, we could in principle simulate all the way to merger, although at that point the approximation used in superimposing the two black hole metrics without any interaction terms would break down.
Instead, for $\rbhi=20\rg$ we simulate up to separations of $\sim$ $
14 \rg$ ($\sim$ 85 orbits) and $\sim$ $12 \rg$ ($\sim$ 79 orbits) for $i_0=45^\circ$ and $i_0=90^\circ$, respectively, so we see significant changes in both the primary black hole spin and orbital directions throughout our simulations.

We emphasize that the orbital and spin evolutions used in our simulations that we have just described do not include any fluid effects like drag or dynamical friction.  
Instead, they represent the solution to the Post-Newtonian orbital equations for two black holes in a vacuum.
This is a good approximation for moderate and low accretion rates, but fluid effects could become important for the highest accretion rates (close to and exceeding Eddington for either the primary or the secondary), at least for larger separations (e.g., our $80 \rg$ case) where there is significant time for accumulated drag and friction to affect the secondary before merger.
Incorporating such effects in the orbital evolution of the binary would require coupling the PN orbital equations to the GRMHD simulation via additional source terms that account for the accretion of linear/angular momentum and the gravitational effects of the stress-energy tensor of the surrounding plasma.  
Naively one might expect drag and friction to reduce the orbital velocity of the secondary (and thus increasing the accretion rate onto the secondary) and reduce the time it takes for the binary to merge. 
On the other hand, previous studies have found that flows around compact objects with significant outflows can have \emph{negative} dynamical friction \citep{Gruzinov2020,Li2020,Kaaz2023}.
In that case the orbital velocity of the secondary and the time it takes for the binary to merge could increase.  
The cumulative long term effect that drag and friction have on binary evolution is still controversial and requires further numerical study.
We also note the additional challenge that this fluid back-reaction is likely only significant on timescales much longer than the orbital time during which the secondary can sample several different regions and realizations of the flow.


\section{Black-hole disk interaction: Analytical Considerations}
\label{sec:analytic}

The passage of a smaller, secondary black hole through a thick accretion disk surrounding a more massive primary black hole can be compared to a Bondi-Hoyle-Lyttleton-type flow where a uniform wind impacts a massive object (\citealt{Hoyle1939,BHoyle1944,Edgar2004}; quite different from a BH-disk collision in thin cooled disks cf.\ \citealt{Ivanov1998}). This is true specifically for inclined orbits on small spatial scales close to the secondary and short time-scales where the orbital motion of the wind can be approximated as linear. 

It is thus instructive to consider that solution in the context of our simulations.  
Doing so allows us to get rough estimates of what we expect to find in the simulations and gives us a conceptual framework to interpret our results.
As the simulations confirm the basic paradigm described by the model, it also allows us to make predictions about regions of parameter space that we have not simulated.

For the purposes of this section, we use the variable $r$ for the radial distance away from the primary and $R$ for the radial distance from the secondary.

\subsection{90 Degree Inclined Orbits}

Assuming that the secondary is on a fixed circular orbit with inclination $i=90^\circ$ around the primary at $\rbh$, then the asymptotic impact velocity {with respect to the secondary black hole} is $v_\infty^2 \approx (GM_1 / \rbh) (1 + \alpha_{\rm kep}^2)$, where $\alpha_{\rm kep}$ is the rotational velocity of the accretion disk divided by the Keplerian velocity ($\alpha_{\rm kep} \approx 0.5$ for radiatively inefficient MADs, see \citealt{Narayan2012,Ressler2020}), and the asymptotic impact density is $\rho (r= \rbh) \approx \rho_{\rm H} (r_{\rm H}/r)^{0.8}$, where $r_{\rm H}$ is the event horizon radius of the primary, $\rho_{\rm H} \equiv \rho(r=r_{\rm H})$, and we have assumed that the density scales as $r^{-0.8}$ in the radial range of interest (\citealt{SCAF}; we will show later that this is a good assumption in our simulations).
For simplicity we also have taken the density to be independent of angle.
The asymptotic sound speed is expected to be some fixed fraction of the Keplerian velocity, which we measure to be $\approx 0.3$ for $r\gtrsim 10 \rg$ in our simulations ($c_{\rm s,\infty}^2 \approx 0.3 GM_1 / \rbh$).
The accretion radius is then 
\begin{equation}
  \label{eq:BHL_radius}
R_{\rm BHL} \approx \frac{2 G M_{2} }{ v_\infty^2 + c_{\rm s,\infty}^2 }\approx 1.3 q  \rbh.
\end{equation}
This can be compared with the Hill radius (inside of which the gravity of the secondary dominates over the gravity of the primary): $R_{\rm Hill} \approx \{q/[3(1+q)]\}^{1/3}\rbh$.
For these parameters, when $q\lesssim 0.34$, $R_{\rm BHL} < R_{\rm Hill}$ and thus the effective influence radius of the secondary is determined by $R_{\rm BHL}$.
Gas outside this radius will be relatively unaffected by the secondary black hole while gas inside this radius will be accreted.
The approximate accretion rate onto the secondary is 
\begin{equation}
\dot M _{\rm BHL} \approx \frac{4 {\rm \pi}G^2 M_{2}^2  \rho_{\infty}}{\left(v_\infty^2 + c_{\rm s,\infty}^2 \right)^{1.5} } \approx 2.1 {\rm \pi} q^2 \rho_{\rm H} c \rg^2 \left(\frac{r_{\rm H}}{\rg}\right)^{0.8} \left(\frac{\rbh}{\rg}\right)^{0.7} \propto q^2\rbh^{0.7},  
\end{equation}
so we expect the secondary's size of influence and accretion rate to increase with orbital radius.  

The accretion disk is not spherically symmetric, however, but has magnetically dominated, matter-deficient polar regions.   
For a rapidly spinning black hole in a MAD state as we study here, there will also be a powerful, electromagnetically dominated jet pushing outwards.   
As the secondary passes through the disk into the pole, we might expect it to bring with it the amount of mass contained within $R \le R_{\rm BHL}$.
This will be 
\begin{equation}
\label{eq:Mblob}
  M_{\rm blob} \approx \frac{4 {\rm \pi}}{3} \rho_{\infty} R_{\rm BHL}^3  \approx 2.9 {\rm \pi} q^3 \rho_{\rm H} \rg^3 \left(\frac{r_{\rm H}}{\rg}\right)^{0.8}  \left( \frac{\rbh}{\rg} \right)^{2.2} \propto q^3 \rbh^{2.2}.
  \end{equation}  
Now this mass will be deposited into the jet region in a time 
\begin{equation}
  T_{\rm deposit} \approx \frac{2 R_{\rm BHL}}{v_{\rm orbit} }\approx 2.6 q \frac{\rg}{c} \left(\frac{\rbh}{\rg}\right)^{1.5},
\end{equation}
so that the passage of the secondary from the disk into the polar region should increase the unbound outflow rate (assuming it is outside of the stagnation surface) by 
\begin{equation}
\label{eq:delta_mdot_out_approx}
\begin{aligned}
  \Delta \dot M_{\rm unbound} &\approx \frac{M_{\rm blob}}{T_{\rm deposit}} \\ &\approx \frac{2 {\rm \pi}}{3} \rho_\infty v_{\rm orbit} R_{\rm BHL}^2 \approx 1.1 {\rm \pi} q^2  \rho_{\rm H} c \rg^2 \left(\frac{r_{\rm H}}{\rg}\right)^{0.8} \left(\frac{\rbh}{\rg}\right)^{0.7}.
  \end{aligned}
\end{equation}
Now, we can compare this with the expected scaling of the unbound outflow rate for the accretion disk itself (i.e., material blown off the disk in the process of accretion, not the highly relativistic jet material).
In radiatively inefficient flows with significant outflows, the mass inflow ($\dot M_{\rm in}$) and outflow rates will be roughly equal in magnitude with the inflow speeds being some fraction of the Keplerian speed $v_{\rm kep}$ \citep{ADIOS}.
Then $\dot M _{\rm unbound } \approx  |\dot M_{\rm in}| \propto \rho v_{\rm kep} r^2  \propto r^{0.7}$, which scales the same way with $r$ as $\Delta \dot M_{\rm unbound}$ scales with $\rbh$.  
We have confirmed that this scaling holds in the single black hole simulation described in \S \ref{sec:inits} for $r\gtrsim 10\rg$ (not plotted). 
Thus we expect that the ratio between $\Delta \dot M_{\rm unbound}$ and $\dot M_{\rm unbound}$ will be similar if measured at $r=\rbh$ for all $\rbh$.  

To estimate the impact the secondary might have on the primary accretion flow,
as an upper limit we can think of the secondary as effectively screening a fraction of the inflowing material, determined by the area that it sweeps out in the disk over the course of its orbit on a spherical shell located at $r=\rbh$. 
This area is 
\begin{equation}
    A_{\rm orbit} \approx 2 \left(\frac{R_{\rm BHL}}{\rbh}\right) \left(\frac{H}{\sqrt{H^2+\rbh^2}}\right) \rbh^2,
\end{equation}
where $H$ is the scale height of the disk and we have used $R_{\rm BHL} \ll \rbh$.
The effective area of the inflowing accretion disk is similarly 
\begin{equation}
A_{\rm disk} \approx 4 {\rm \pi} \left(\frac{H}{\sqrt{H^2+\rbh^2}}\right)\rbh^2.
\end{equation}
A rough estimate of the amount by which the net accretion rate could change is then
\begin{equation}
\label{eq:delta_mdot}
    \left |\frac{\Delta \dot M }{\dot M} \right| \approx \frac{A_{\rm orbit}}{A_{\rm disk}} \approx \frac{1}{2 {\rm \pi}} \left(\frac{R_{\rm BHL}}{\rbh}\right) \approx 0.2 q \ll 1.
\end{equation}
Therefore, for $i_0=90^\circ$ orbits we expect the secondary to have a minimal effect on the net accretion flow of the primary (as we show later).  

In this brief analysis we have neglected many considerations that might be important in the simulations, including magnetic fields (which can change the structure of the Bondi-Hoyle-Lyttleton accretion flow, \citealt{Kaaz2023,Gracia-Linares2023}), the velocity gradient in the wind provided by the accretion disk (which can induce turbulence and also change the structure of the flow, \citealt{Xu2019}), turbulence (which could introduce stochastic variability to the predicted quantities), the time-dependent nature of accretion (which could introduce secular variability to the predicted quantities), and the variation of density with angle (which could lead to smaller-than-predicted mass outbursts since the density on the surface of the disk is smaller than the midplane).  
These approximate values, however, give us a good set of comparisons for our numerical results.

\subsection{More General Expressions}
\label{sec:analytic_general}
The above analysis can also be done for orbital planes closer to the midplane of the disk.  
This will have the effect of either increasing or decreasing $v_\infty$ in the frame of the secondary depending on whether the orbit is prograde or retrograde to the accretion disk.  
It will also increase the time it takes to deposit matter outside the disk (for orbits sufficiently inclined that the secondary still passes out of the disk) because the component of the velocity perpendicular to the disk, $v_{\rm perp}$, will be reduced.  
Both $v_\infty$ and $v_{\rm perp}$ depend on the particular location of the secondary in its orbit when it crosses the surface of the disk.
However, we can approximately evaluate them when the secondary crosses the midplane as $v_{\infty}^2 =v_{\rm orbit}^2 \{\sin(i)^2 + [\alpha_{\rm kep}-\cos(i)]^2\}$ and $v_{\rm perp} = \sin(i) v_{\rm orbit}$.
These expressions are approximately valid if the accretion disk is not too thick ($\lesssim 30^\circ$ above and below the midplane).  
We can also parameterize the sound speed of the disk as $c_{\rm s,\infty}^2 = \alpha_{\rm s}^2 G M_1/\rbh$ and the rotational velocity of the disk as $v_{\rm rot}^2 = \alpha_{\rm kep}^2 G M_1/\rbh$.
Repeating the same calculation as in the previous subsection, this results in 
\begin{equation}
\label{eq:BHL_radius_gen}
  R_{\rm BHL} \approx \frac{2}{\sin(i)^2 + [\alpha_{\rm kep}-\cos(i)]^2 + \alpha_{\rm s}^2} q \rbh,
\end{equation}
\begin{equation}
\begin{aligned}
\dot M _{\rm BHL} \approx &\frac{4 {\rm \pi}}{\left\{\sin(i)^2 + [\alpha_{\rm kep}-\cos(i)]^2 + \alpha_{\rm s}^2\right\}^{3/2}} \\
&\times q^2 \rho_{\rm H} c \rg^2 \left(\frac{r_{\rm H}}{\rg}\right)^{0.8} \left(\frac{\rbh}{\rg}\right)^{0.7}, 
\end{aligned}
\end{equation}
and 
\begin{equation}
\label{eq:delta_mdot_out_approx_general}
\begin{aligned}
  \Delta \dot M_{\rm unbound}  \approx  &\frac{2 {\rm \pi}}{3} \frac{\sin(i)}{\left\{\sin(i)^2 + [\alpha_{\rm kep}-\cos(i)]^2+ \alpha_{\rm s}^2\right\}^{2} }  \\
  & \times q^2 \rho_{\rm H} c \rg^2 \left(\frac{r_{\rm H}}{\rg}\right)^{0.8} \left(\frac{\rbh}{\rg}\right)^{0.7}.
  \end{aligned}
\end{equation}

Substituting in $i_0=45^\circ$, $\alpha_{\rm kep} = 0.5 $, and $\alpha_{\rm s}^2 = 0.3$, we find that $R_{\rm BHL}$, $\dot M_{\rm BHL}$, and $\Delta \dot M_{\rm unbound}$ are larger than the $i_0=90^\circ$ expressions by factors of $\approx$ 1.8, 2.5, and 2.4, respectively.  
Note, however, that Equation \eqref{eq:delta_mdot_out_approx_general} for $\Delta \dot M_{\rm unbound}$ crucially depends on the assumption that the orbit of the secondary brings it out of the disk into the polar region.  If the disk is too thick or the orbit not inclined enough the actual value of $\Delta \dot M_{\rm unbound}$ will be much less.  
This is true in our simulations for $i_0=45^\circ$ (note the thickness of the disk in Figure \ref{fig:init_torus}), where the orbit only grazes the edge of the disk instead of plunging out into the polar region.  
Thus we might expect the relative $\Delta \dot M_{\rm unbound}$ to be smaller, though it is not obvious by how much.
Note additionally that for low inclinations $R_{\rm Hill}$ can become less than $R_{\rm BHL}$.  
For these parameters, at $q=0.1$ this happens for $-24^\circ \lesssim i\lesssim 24^\circ$ (meaning that for all the simulations in this work, the Bondi-Hoyle-Lyttleton radius determines the influence radius). 

The general expression for the area swept out in the disk by the orbit of the secondary on a spherical shell located at $\rbh$ is 
\begin{equation}
A_{\rm orbit} \approx 
    \begin{cases}
        \frac{2}{|\sin(i)|} \left(\frac{R_{\rm BHL}}{\rbh}\right) \left(\frac{H}{\sqrt{H^2+\rbh^2}}\right) \rbh^2 &  |\sin(i)|> \frac{H}{\sqrt{H^2+\rbh^2}} \\
         2 \left(\frac{R_{\rm BHL}}{\rbh}\right)  \rbh^2 &  |\sin(i)|\le \frac{H}{\sqrt{H^2+\rbh^2}}
    \end{cases}
\end{equation}
so that 
\begin{equation}
\label{eq:delta_mdot_gen}
\left|\frac{\Delta \dot M }{\dot M}\right| \approx 
    \begin{cases}
        \frac{1}{2 {\rm \pi}}\frac{1}{|\sin(i)|} \left(\frac{R_{\rm BHL}}{\rbh}\right)  &  |\sin(i)|> \frac{H}{\sqrt{H^2+\rbh^2}} \\
         \frac{1}{2 {\rm \pi}} \left(\frac{R_{\rm BHL}}{\rbh}\right) \left(\frac{H}{\sqrt{H^2+\rbh^2}}\right)^{-1}   &  |\sin(i)|\le \frac{H}{\sqrt{H^2+\rbh^2}},
    \end{cases}
\end{equation}
where the top expression is used when the orbit of the secondary passes out of the disk at some point in its orbit and the second expression is used when the orbit is entirely contained within the disk.  
For the thick disks we study in this work, the term related to the scale height is relatively close to $1$, so for all orbits the maximum $|\Delta \dot  M / \dot M|$ predicted by Equation \eqref{eq:delta_mdot_gen} is $\approx 0.6 q \ll 1$ .

It is interesting to note that Equation \eqref{eq:delta_mdot_gen} predicts that for a thinner disk with $H/\rbh \ll 1$, secondaries with significantly inclined  orbits would still have a relatively small effect on the primary accretion disk.
On the other hand, for orbits with low inclination, $ 2 {\rm \pi} |\sin(i)| \lesssim R_{\rm BHL}/\rbh $, Equation \eqref{eq:delta_mdot_gen} predicts $|\Delta \dot  M / \dot M| \sim 1$, at which point the disk structure would likely significantly change and this approximation would break down.
If we assume $\alpha_{\rm kep}\approx 1$ and $\alpha_{\rm s} \approx 0$ for a thin disk, then this would happen for $-18^\circ \lesssim i \lesssim 18^\circ$ when $q=0.1$.
However, we again emphasize that this estimate is simplistic and neglects thermal effects that could be significant, especially for thin disks.  
The problem of a secondary black hole impacting a thin accretion disk has also been studied analytically with significantly more detail in other works (e.g., \citealt{Ivanov1998,Ivanov1999,Pihajoki2016}).
Our numerical method will be able to study such systems when combined with optically thin radiative cooling or a full treatment of radiation (e.g., \citealt{White2023}).

  \begin{figure*}
  \begin{centering}
\includegraphics[width=0.99\textwidth]{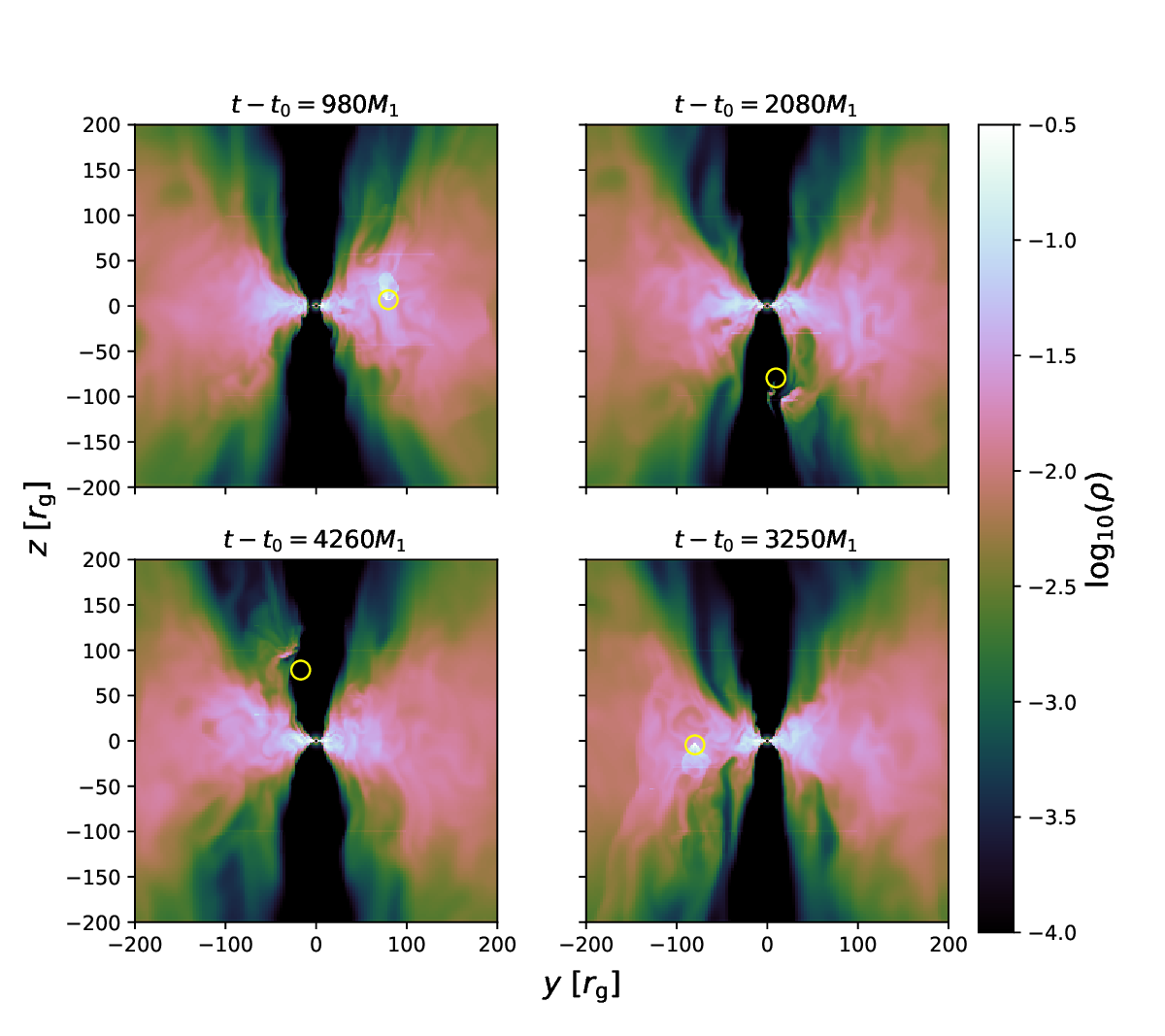}
\end{centering}
\caption{2D slice of density for our $q=0.1$, $\rbhi=80 r_{\rm g}$, and $i_0=90^\circ$ binary black hole perturber simulation at four different times. 
The yellow circles represent the influence radii predicted from Equation \eqref{eq:BHL_radius_gen}.
Starting from the upper left panel, and proceeding clockwise, the panels represent $980 M_1$, $2080 M_1$, $3250 M_1$, and $4260 M_1$ since the secondary was introduced.
As it passes through the accretion flow, the secondary creates a Bondi-Hoyle-Lyttleton-esque shock front (upper left and lower right panels; see also Appendix \ref{app:bhl_test}) caused by its supersonic trajectory.
Each time black hole passes through the corona of the disk into the jet, it carries with it some a small amount of matter that is then quickly swept away by the jet (e.g. upper right and lower left panel). 
For an animated version of this figure, see \url{
https://youtu.be/lU82eT18HLU}.}
\label{fig:rbh80_contour}
\end{figure*}

\section{GRMHD Results}
\label{sec:results}
To highlight the basic features of our binary simulations, Figure \ref{fig:rbh80_contour} shows the time evolution of mass density in our $\rbhi = 80 r_{\rm g}$, $i_0 = 90^\circ$ simulation.  
The secondary black hole travels supersonically through the accretion disk, forming a bow shock that propagates through the flow.
As the secondary continues along its orbit and crosses out of the disk into the jet region, it carries with it a small amount of matter that gets deposited into the funnel region and subsequently blown away by the jet.  
Evidence for the fact that the gas is being accelerated by the jet is in the fact that the time-averaged electromagnetic outflow energy (e.g., the Poynting flux, not shown) is reduced in the binary simulations when compared with the single black hole simulation\footnote{The electromagnetic outflow energy is highest in the single black hole simulation, lower for the $\rbhi=80\rg$ simulations, and lowest for the $\rbhi=20\rg$ simulations.  
We interpret this as the gas ejected by the black holes on $\rbhi=20\rg$ orbits requiring more energy to unbind.}.
This process continues in a periodic or quasi-periodic way for the duration of the simulation.  

\begin{figure*}
\gridline{\includegraphics[width=0.49\textwidth]{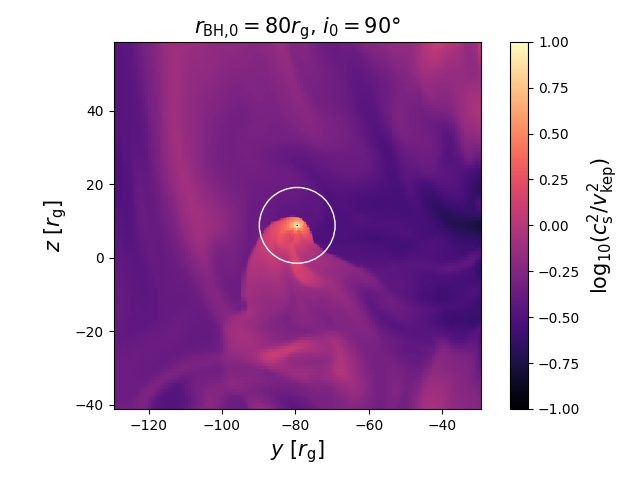}
\includegraphics[width=0.49\textwidth]{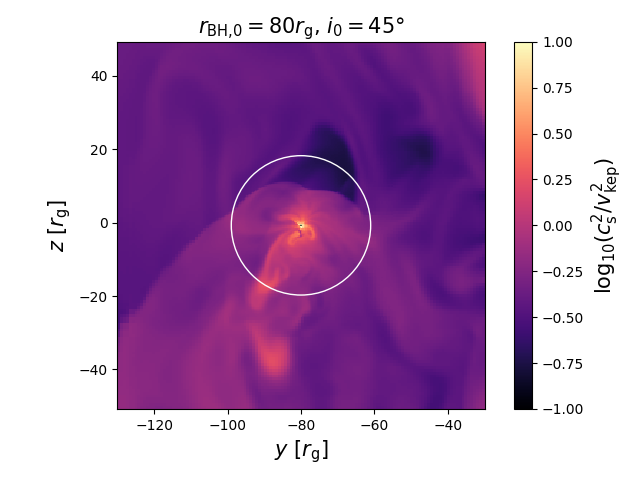}}
\gridline{\includegraphics[width=0.49\textwidth]{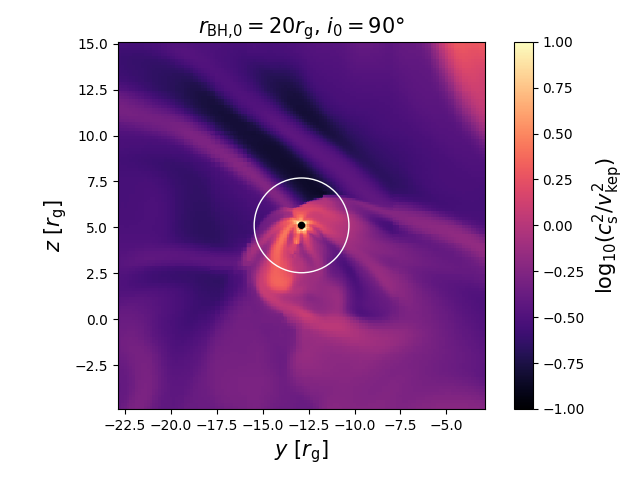}
\includegraphics[width=0.49\textwidth]{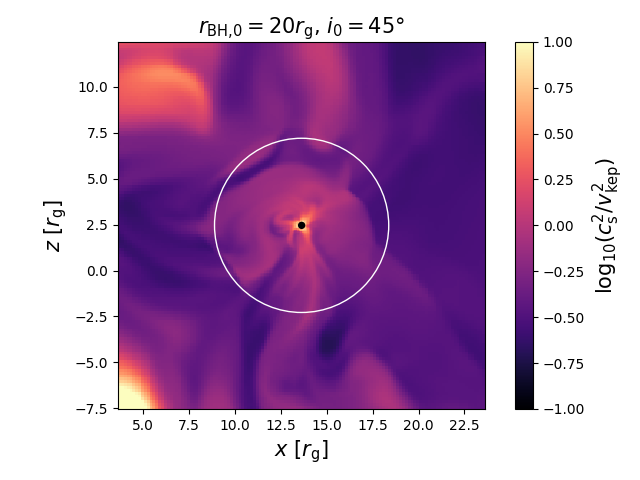}}
\caption{2D slices of the square sound speed divided by the square Keplerian velocity, $c_{\rm s}^2/v_{\rm kep}^2$, where $c_{\rm s}^2 = \gamma P/\rho  $ and $v_{\rm kep}^2 = 1/\sqrt{r}$ in these units. The contours are centered on the secondary black holes as they pass through the midplane of the accretion disk during representative times in our simulations.
The white circles represent the influence radii predicted from Equation \eqref{eq:BHL_radius_gen}.
The moving black hole creates a bow shock in the gas of varying shapes and sizes.  
All simulations generally have moderate shocks with temperatures (note that $c_{\rm s}^2$ is directly proportional to temperature) increasing by factors of $\sim $ a few from the average gas temperature at the orbital radius. 
The $i_0=45^\circ$ (right) simulations generally display a larger secondary influence radii but slightly weaker shocks than the $i_0=90^\circ$ simulations (left) due to the reduced relative gas motion.   
Similarly, the $\rbhi=80 \rg$ simulations (top) display a larger secondary influence radius than the $\rbhi=20 \rg$ simulations (bottom).
All simulations show good agreement with the predicted influence radii from Equation \eqref{eq:BHL_radius_gen}. }
\label{fig:thetap_shock}
\end{figure*}

Figure \ref{fig:thetap_shock} zooms in on the secondary black holes in the lab frame as they pass through the midplane for our four simulations, displaying 2D contours of the square sound speed normalized to the square Keplerian velocity, $c_{\rm s}^2/v_{\rm kep}^2$ ($\propto T_{\rm g} r$, where $T_{\rm g}$ is the gas temperature), in the $x=x_{\rm BH}$ plane (parallel to $y$-$z$) at representative times for our four binary simulations (except for the $\rbhi=20\rg$, $i_0=45^\circ$ simulation which is plotted in the $y=y_{\rm BH}$ plane parallel to $x$-$z$ to better display the motion of the secondary through the gas).
Around the secondary, the flow resembles a turbulent Bondi-Hoyle-Lyttleton-type flow with a shock front and wake. 
The shocks are stronger for orbits with $i_0 = 90^\circ$ than $i_0=45^\circ$ due to the reduced relative motion of the gas in the latter, but all simulations have moderate shocks with temperature increasing by factors of $\sim$ a few from the average temperature at the orbital radius.
These temperatures are consistent with Bondi-Hoyle-Lyttleton simulations for flows with Mach numbers moderately greater than 1 and $\lesssim$ 2.
Another consequence of the reduced relative velocities of the gas is that the $i_0=45^\circ$ simulations have larger secondary influence radii than their $i_0=90^\circ$ counterparts.
The sizes of the influence radii of the secondaries in the $\rbhi=80\rg$ simulations are also about 4 times larger than those in the $\rbhi=20\rg$ simulations due to the slower orbital velocity.

\subsection{Effects on the Primary Accretion Flow}
 
\begin{figure}
\includegraphics[width=0.48\textwidth]{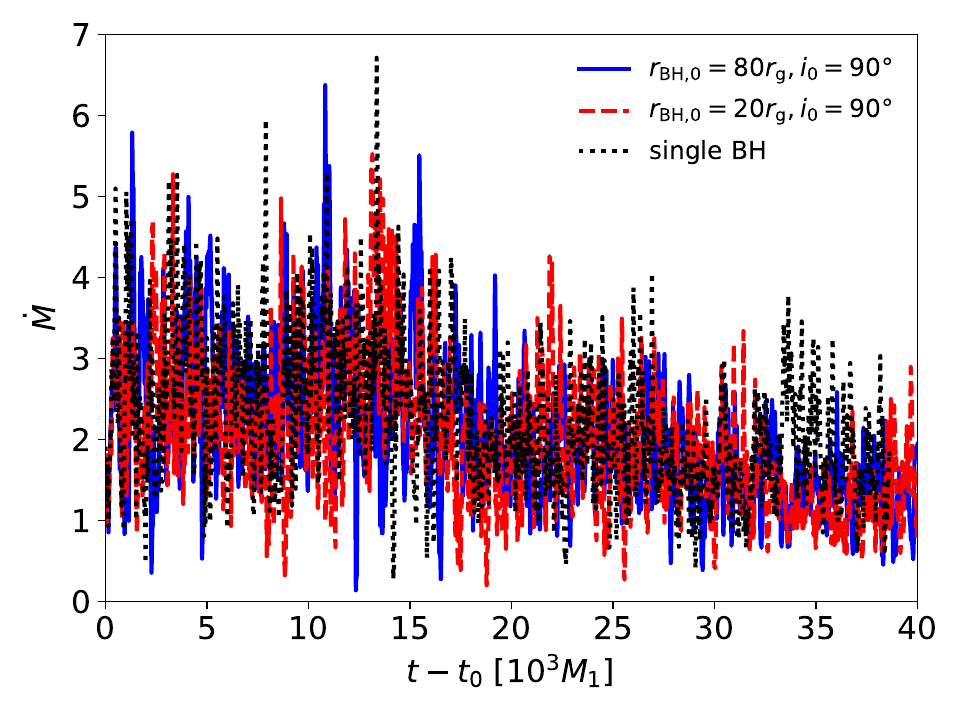}
\includegraphics[width=0.48\textwidth]{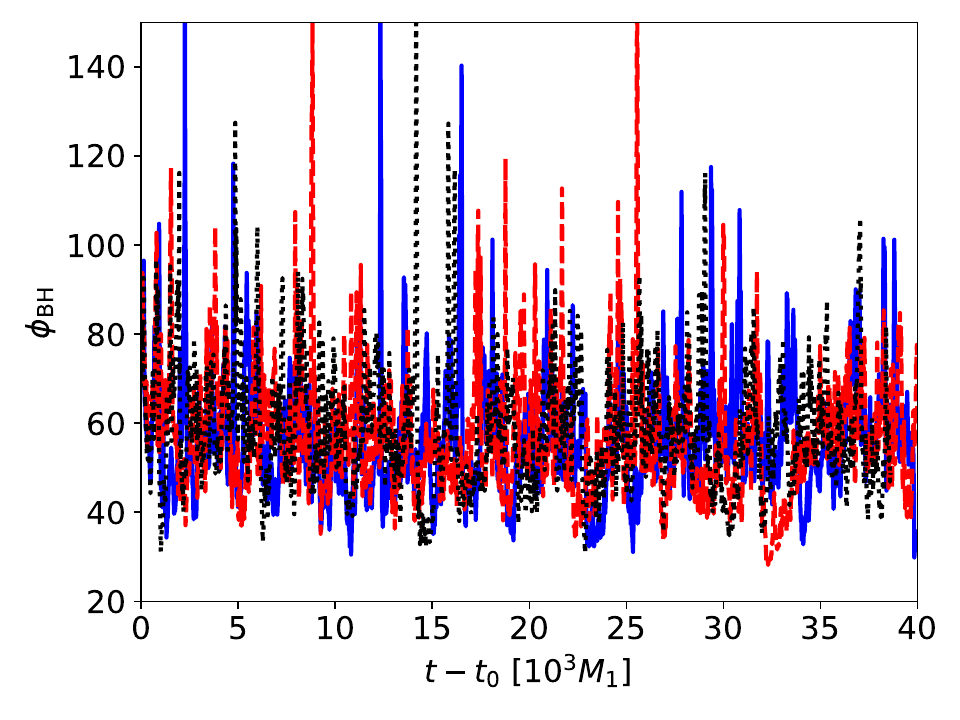}
\caption{
Accretion rate, $\dot M$, and dimensionless magnetic flux threading the primary black hole, $\phi_{\rm BH}$, as a function of time for the single black hole simulation (black dotted line) compared with the $q=0.1$, $\rbhi=80 r_{\rm g}$, $i_0=90^\circ$ (solid blue line) and $q=0.1$, $\rbhi=20 r_{\rm g}$, $i_0=90^\circ$ (dashed red line) binary simulations.
Neither the accretion rate nor the magnetic flux show any clear signature of the secondary orbit's periodicity.  
In fact, both quantities are very similar in the binary and single black hole simulations, with mainly stochastic differences.  
This shows that the effect of the secondary is small on the near-horizon dynamics of the flow.
We argue that this is because the area swept out in the disk by the orbit of the secondary on a spherical shell is small compared with the area of the disk itself at the orbital radius.
Note that $\dot M$ and $\phi_{\rm BH}$ for the $i_0=45^\circ$ simulations are similar and are not plotted to avoid clutter.
} 
\label{fig:mdot_phibh_rbh_80}
\end{figure}

In Figure \ref{fig:rbh80_contour} neither the shock from the secondary nor the ejection of matter from the disk seem to dramatically affect the accretion flow dynamics {onto the primary}. 
To quantify this more directly, we plot the accretion rate onto the primary black hole and dimensionless magnetic flux threading the primary black hole vs.\ time in Figure \ref{fig:mdot_phibh_rbh_80} for the binary simulations compared with the single black hole simulation.
These quantities are calculated in the same way as in Equations \eqref{eq:mdot} and \eqref{eq:phibh}, that is, using a single black hole Kerr metric (which is a good approximation near the primary event horizon).   
The quantities for the binary simulation show statistically almost identical behavior to those in the single black hole simulation.
The five simulations have average $\dot M =[2.2, 2.1, 2.1, 2.1, 2.3]$, average $\varphi_{\rm BH} = [56, 59, 56, 56, 58]$, standard deviation of $\dot M$ $= [0.85, 0.93, 0.85, 0.87, 0.86]$, and standard deviation of $\varphi_{\rm BH}$ $= [13, 13, 14, 14, 13]$ for the [($\rbhi=80\rg$ $i_0=90^\circ$), ($\rbhi=80\rg$ $i_0=45^\circ$), ($\rbhi=20\rg$ $i_0=90^\circ$), ($\rbhi=20\rg$ $i_0=45^\circ$), single black hole] simulations, which display only slight differences. 
In Figure \ref{fig:mdot_phibh_rbh_80} there are also no clear signatures of the periodicity of the secondary orbit.

\begin{figure}
\includegraphics[width=0.49\textwidth]{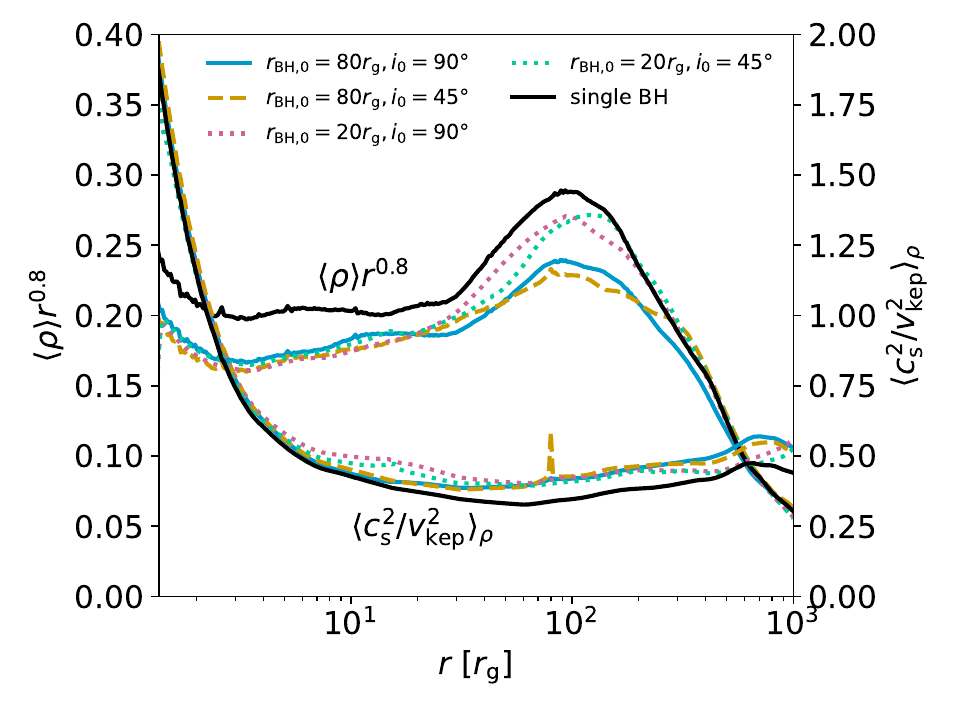}
\caption{
Time and angle-averaged radial profiles of density multiplied by $r^{0.8}$ (top set of lines), $\langle \rho \rangle r^{0.8}$, and mass-weighted square sound speed divided by the square Keplerian speed (bottom set of lines), $\langle c_{\rm s}^2/v_{\rm kep}^2\rangle_\rho $ for our four different binary simulations compared with the single black hole simulation (black solid line).  
The secondary black hole has only a small effect on the average density and temperature, providing a small increase in temperature around the orbital radius (caused by the bow shock following the secondary) and a small decrease in density near and at smaller radii than orbital radius (caused by the expulsion of mass from the disk).  
} 
\label{fig:radial_profiles}
\end{figure}

The time and angle-averaged radial profiles in the binary simulations are also quite similar to those in the single black hole simulation as seen in Figure \ref{fig:radial_profiles} for the density and square sound speed, where we perform time averages over the range $120{, }000$--$140{, }000$ $M_{1}$.
For radii near and within the orbital radius of the secondary, the density is decreased relative to the single black hole simulation by $\sim$ 20\%.
This is caused by a combination of matter being expelled from the disk by the secondary and by matter being accreted onto the secondary.
The density in all simulations agrees reasonably well with an $r$ $\tilde \propto$ $r^{-0.8}$ dependence between $3 \rg \lesssim r \lesssim 80 \rg$ as used for our analytic estimates in \S \ref{sec:analytic} \citep{SCAF}.
The temperatures (directly proportional to $c_{\rm s}^2$) of the binary simulations are slightly hotter (by $\sim$ 20\%) in the bulk of the disk for $r \gtrsim 10 \rg$ caused by the bow shock propagating through the flow.
This is particularly evident near the orbital radii where there are small peaks in temperature. 
The $\rbhi=80\rg$, $i_0=45^\circ$ simulation has an especially prominent peak at the orbital radius because it spends a large fraction of time within the disk and so the shocked temperature contributes more to the time-averaged temperature at that radius. 

These findings agree with our analytical estimates in \S \ref{sec:analytic_general}, particularly Equations \eqref{eq:delta_mdot} and \eqref{eq:delta_mdot_gen}, where we argued that the effect of the secondary on the primary accretion flow would be quite small for all orbital inclinations.  

\subsection{Quasi-Periodic Outbursts/Eruptions}

  \begin{figure*}
\includegraphics[width=0.93\textwidth]{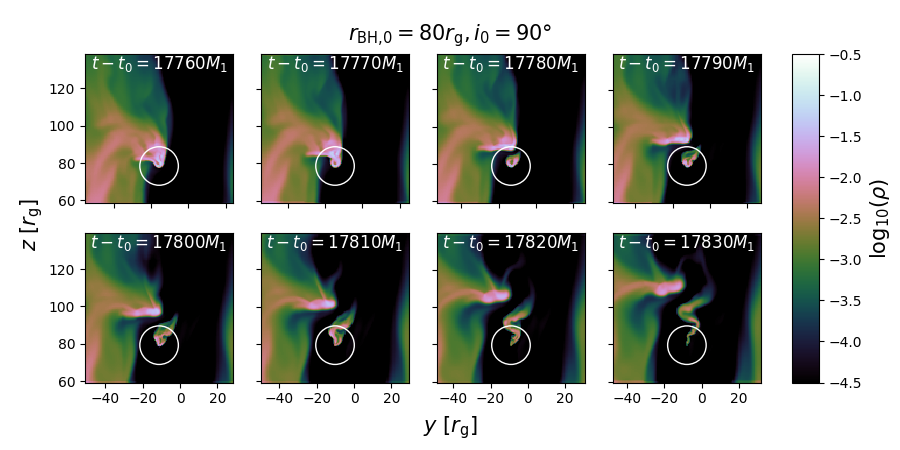}
\includegraphics[width=0.93\textwidth]{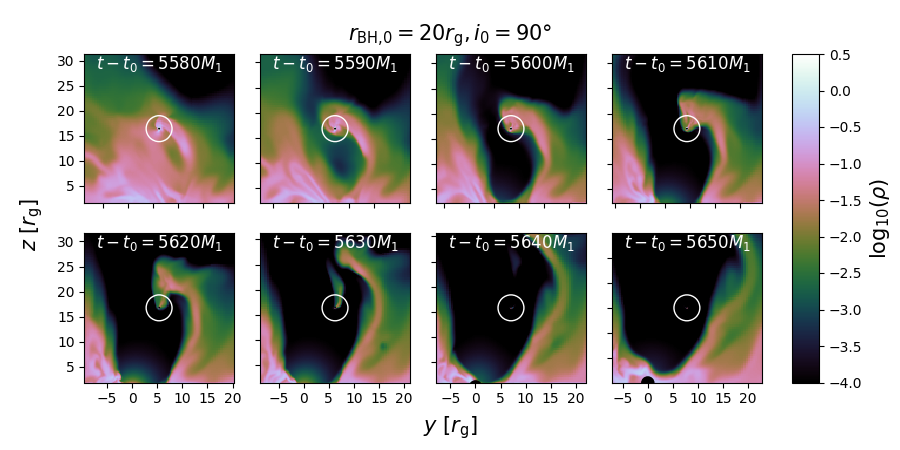}
\caption{Evolution of an ejected blob in our $\rbhi=80\rg$, $i_0=90^\circ$ (top) and $\rbhi=20\rg$, $i_0=90^\circ$ (bottom) simulations.  Plotted is a 2D slice of mass density in the $x=x_{\rm BH}$ plane (parallel to $y$-$z$) at 8 different times for each simulation (time proceeds from top left to top right and then bottom left to bottom right).
The white circles represent the influence radii predicted from Equation \eqref{eq:BHL_radius_gen}.
Note that the two sets of contours are on different spatial scales.
The secondary black hole pulls gas from the accretion disk into the polar region and then the electromagnetically dominated jet blows this gas away. 
The sizes of these blobs are consistent with the Bondi-Hoyle-Lyttleton radius predicted by Equation \eqref{eq:BHL_radius_gen},  $\sim$ $10\rg$ in radius for $\rbhi=80^\rg$ and $\sim$ $2.5 \rg$ in radius for $\rbhi=20 \rg$.
} 
\label{fig:qpe_contour}
\end{figure*}

Even though the primary accretion flow dynamics are not significantly affected by the secondary black hole, there are, however, clear signatures of the secondary in the outflow. 
Zooming in on times when the secondary passes out of the disk into the polar regions, 
we plot 2D contours of mass density in the $x=x_{\rm BH}$ plane (parallel to $y$-$z$) centered on the secondary black hole as it crosses into the jet in Figure \ref{fig:qpe_contour} for two particular representative time series in the $\rbhi=80 \rg$, $i_0=90^\circ$ and $\rbhi=20 \rg$, $i_0=90^\circ$ simulations.
In both simulations, as the secondary passes from the disk/jet boundary region to the jet itself, it brings with it a blob that subsequently expands, shears out, and gets blown away by the jet.  
The sizes of the initial blobs are consistent with the Bondi-Hoyle-Lyttleton radius (Equation \ref{eq:BHL_radius}), roughly $10 \rg$ and $2.5 \rg$ in radius for $\rbhi=80\rg$ and $\rbhi=20 \rg$ when mapped to a sphere.
The temperatures of the blobs are approximately virial, with $c_{\rm s} \approx v_{\rm kep}$.

  \begin{figure}
\includegraphics[width=0.49\textwidth]{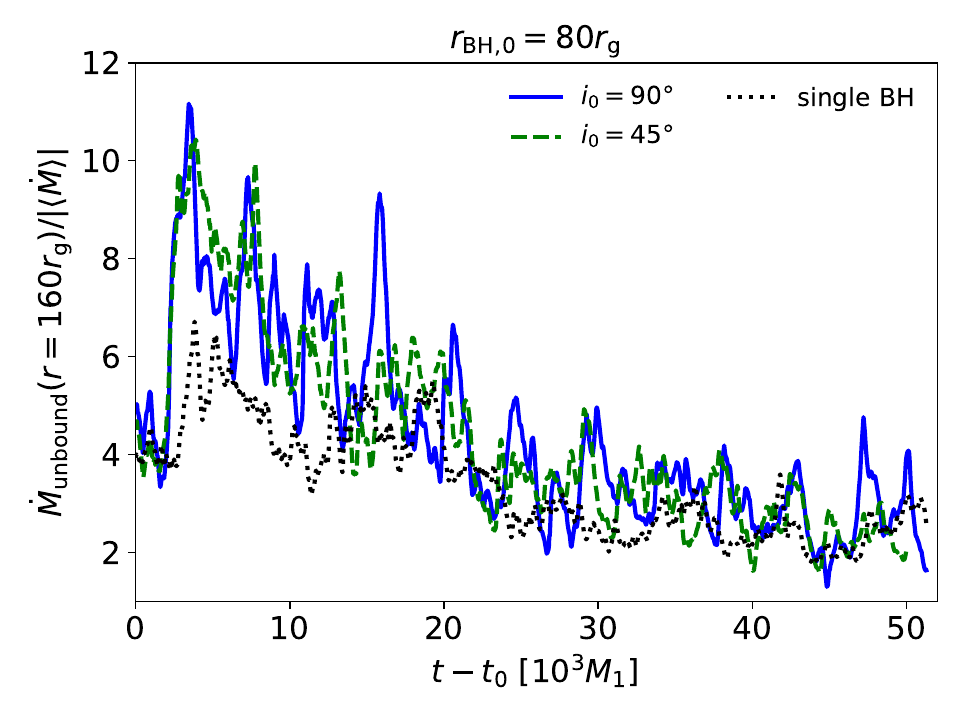}
\includegraphics[width=0.49\textwidth]{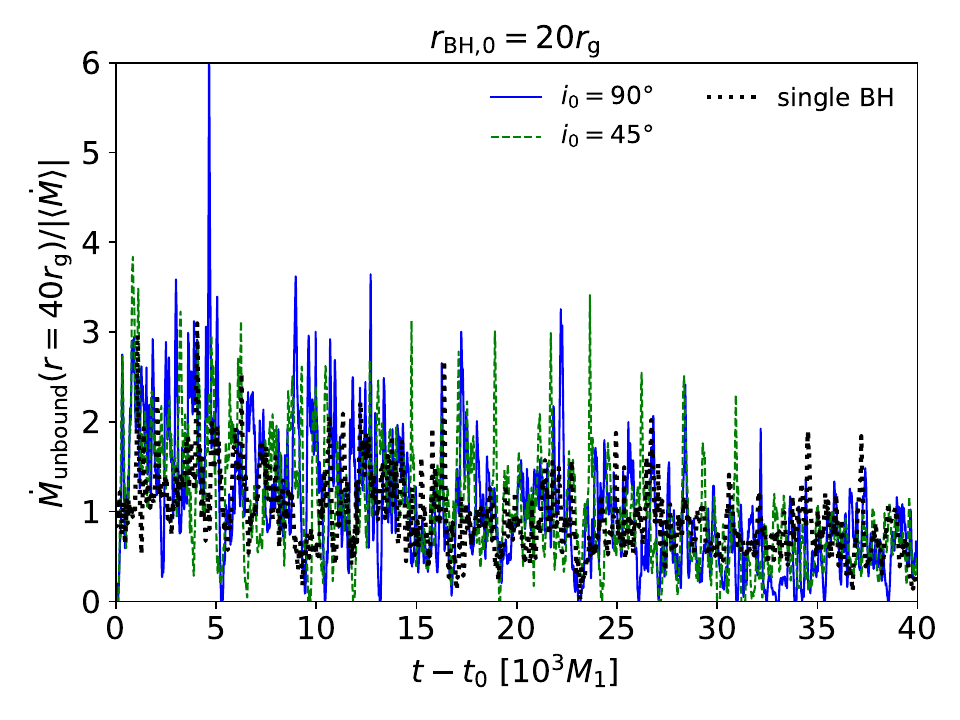}
\caption{
Unbound mass outflow rate, $\dot M_{\rm unbound}$, plotted vs.\ time elapsed since the secondary was introduced, $t-t_0$, in our $\rbhi= 80 \rg$ (top) and $\rbhi=20 \rg$ (bottom) simulations for $i_0=90^\circ$ (blue solid line) and $i_0=45^\circ$ (green dashed line), compared to the single black hole simulation (black dotted line).  
$\dot M_{\rm unbound}$ is measured at $r=160 \rg$ for $\rbhi=80 \rg$ and $r=40 \rg$ for $\rbhi=20\rg$. 
The curves have been normalized by the time-averaged accretion rate over the interval ($0\le t-t_0 \le 40{, }000 M_1$).
The secondary increases the unbound mass outflow rate at quasi-period intervals of order the orbital period ($\approx$ 4500 $M_1$ for $\rbhi=80\rg$ and $\approx$ 560 $M_1$ for $\rbhi=20\rg$) by factors of $\sim$ 2--4 compared to the same quantity in the single black hole simulation.
Note that as in all GRMHD torus simulations there is a secular decline in all outflow rates due to the mass in the torus depleting over time.
} 
\label{fig:mdot_out_all}
\end{figure}

  \begin{figure}
\includegraphics[width=0.49\textwidth]{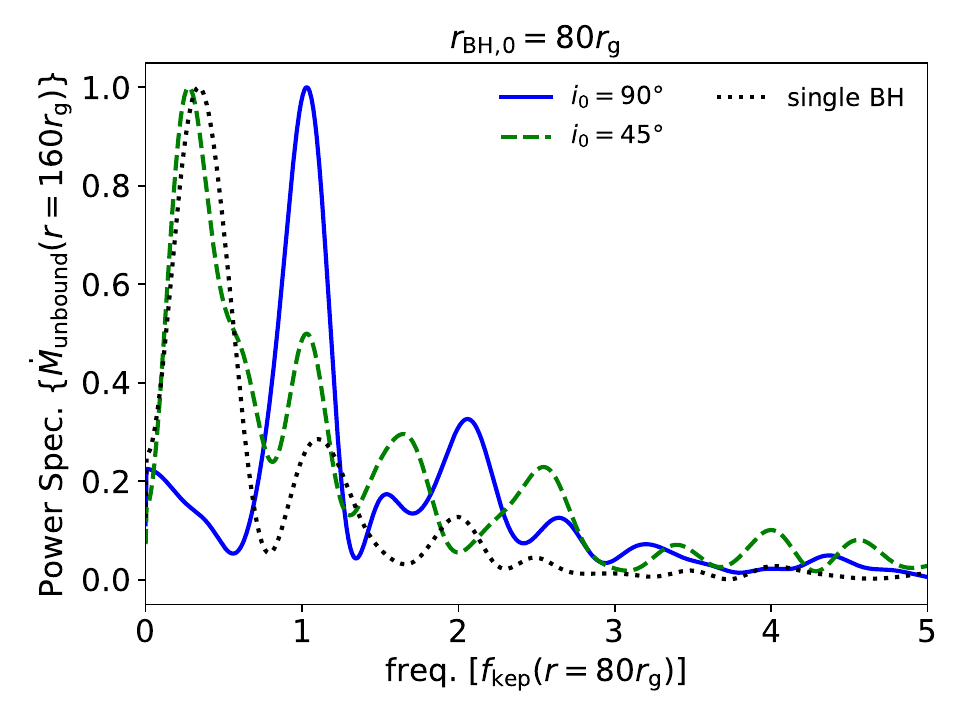}
\includegraphics[width=0.49\textwidth]{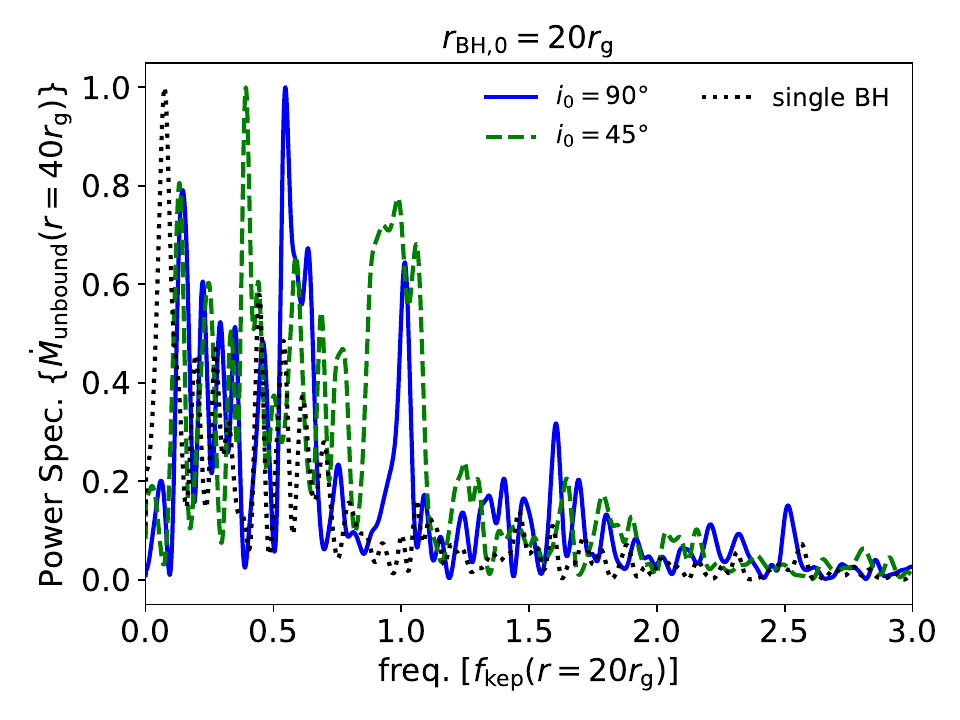}
\caption{
Power spectra of unbound mass outflow rates, $\dot M_{\rm unbound}$, for orbits with $i_0=90^\circ$ (blue solid) and $i_0=45^\circ$ (green dashed) compared to the single black hole simulation (black dotted).
Each spectrum is normalized such that the highest peak is 1.
Top: $\rbhi=80 \rg$, with $\dot M_{\rm unbound}$ measured at $r=160\rg$ and the $x$-axis measuring frequency in units of the orbital frequency at $r=80\rg$.
Bottom: $\rbhi=20\rg$, with $\dot M_{\rm unbound}$ measured at $r=40\rg$ and the $x$-axis measuring frequency in units of the orbital frequency at $r=20\rg$.
For $\rbhi=80\rg$, the $i_0=90^\circ$ PSD shows clear peaks at the orbital frequency and twice the orbital frequency, with other smaller peaks.
The $i=45^\circ$ does have a peak at the orbital frequency but it is subdominant when compared with the lower frequency peak also seen in the single black hole simulation.
For $\rbhi=20\rg$, there are significant peaks at the orbital frequency for both $i_0=90^\circ$ and $i_0=45^\circ$, but these are lower than some peaks at lower frequencies.  
This is likely due to propagation effects that spread out mass ejected by the secondaries in radius (and correspondingly, time in $\dot M_{\rm unbound}$).
} 
\label{fig:psd_all}
\end{figure}

To measure the effect of these ejected blobs on the outflow, we particularly focus on the unbound mass outflow rate, defined as
\begin{equation}
  \dot M_{\rm unbound} (r) \equiv \int \rho u^r (\rm{Be}>0) d \Omega,
  \label{eq:mdot_out}
\end{equation}
where $u^r$ is the radial component of the four velocity, converted from CKS coordinates in the rest frame of the primary, 
\begin{equation}
\rm{Be} = -\left(1 + \frac{\gamma}{\gamma-1} \frac{P}{\rho} + \frac{b^2}{\rho} \right) u_t -1
\end{equation}
is the relativistic Bernoulli parameter (where $\rm{Be}>0$ implies unbound material, \citealt{Penna2013}),
$d \Omega = \gdet \sin(\theta) d\theta d\varphi$, and $\theta$ and $\varphi$ are the polar and azimuthal angles converted from CKS coordinates in the rest frame of the primary.  
$\dot M_{\rm unbound}$ is shown as a function of time in Figure \ref{fig:mdot_out_all} for our binary simulations measured at $r = 2\rbhi$, compared to the same quantities in the single black hole simulation measured at the same radii. 
The unbound outflow rates display clear quasi-periodic peaks on timescales comparable to an orbital period ($\approx$ 4500 $M_1$ for $\rbhi=80\rg$ and $\approx$ 560 $M_1$ for $\rbhi=20\rg$), with durations $\sim$ 2000 $M_1$  for $\rbhi=80 \rg$ and $\sim$ 200 $M_1$ (or $\sim$ 1/2 an orbital period for each).
Each ``outburst'' is of varying intensity when compared to the single black hole simulation.
For some of the peaks, $\dot M_{\rm unbound}$ is only a few percent higher than the corresponding $\dot M_{\rm unbound}$ in the single black hole simulation, while at others it reaches $\sim$ 2--4 times $\dot M_{\rm unbound}$ in the single black hole simulation.
The fact that the peaks for both $\rbhi$ values are about the same magnitude relative to the single black hole simulation is expected based on our analysis in \S \ref{sec:analytic}. 
Also consistent with our analysis is that the absolute magnitude of the peaks in $\dot M_{\rm unbound}$ are larger for $\rbhi=80 \rg$ than for $\rbhi=20\rg$ (when measured at the same multiple of $\rbhi$).  
The former peaks are larger by a factor of $\sim$ 2--3, which is in good agreement with our estimate (Equation \ref{eq:delta_mdot_out_approx_general}) of $\sim 4^{0.7} \approx 2.6$.
The peaks in the $i_0=45^\circ$ simulations are generally about the same magnitude or slightly lower on average than those in the $i_0=90^\circ$ simulations.
This is because the orbits for $i_0=45^\circ$ do not bring the secondary fully out of the disk into the polar regions and are thus less efficient at depositing matter there (even though their influence radii are larger due to the reduced relative gas speed, see \S \ref{sec:analytic_general}).

To more quantitatively measure the frequencies of the outbursts in unbound outflow rates we compute power spectra of the $\dot M_{\rm unbound}$ curves by first linearly de-trending the data and then using Welch's method, which averages periodograms for overlapping windows of the data.  
We plot the resulting power spectra in Figure \ref{fig:psd_all} using a window size of $20{,000} M_1$ and scale the frequency to the orbital frequency, $f_{\rm kep}(r=\rbhi)$.
Due to the limited sample size of the data and the complex variability of the primary MAD accretion disk, the power spectra for the five simulations display a number of peaks at different frequencies, many of which are sensitive to the precise method/averaging used to compute the periodogram. 
Because of this, specific features in Figure 
\ref{fig:psd_all} should not be over-interpreted, rather, our focus is on the general behavior.  
$\dot M_{\rm unbound}$ for the single black hole simulation generally shows the most power at lower frequencies (e.g., periods $\sim$ 17{,}000 $M_1$ at $r=160\rg$ and $\sim$ 8000 $M_1$ at $r=40\rg$). 
At $2\rbhi$, The binary simulations generally show a peak at the orbital frequency of the binary and sometimes twice that frequency (i.e., every half orbit), with the cleanest example being $\rbhi=80\rg$, $i_0=90^\circ$, where the two highest peaks are located at $f_{\rm kep}(r=80 \rg)$ and $2f_{\rm kep}(r=80 \rg)$.  
The $\rbhi=80\rg$, $i_0=45^\circ$ power spectrum also shows a peak at $f_{\rm kep}(r=80 \rg)$, but it is subdominant compared to the lower frequency peak also seen in the single black hole simulation, likely due to the fact that the orbit does not bring the secondary fully out of the disk to create as distinctive outbursts as $i_0=90^\circ$.
The $\rbhi=20\rg$ simulations show a diverse set of frequencies that stand out in the $\dot M_{\rm unbound}$ power spectrum.  
For $i_0=90^\circ$ the orbital frequency is the third highest peak, with the highest located at a little over half the orbital frequency (or a period of $\sim$ 1000 $M_1$), while for $i_0=45^\circ$ the orbital frequency is the second highest peak, with the highest located at $\approx$ 0.4 times the orbital frequency (or a period of $\sim$ 1400 $M_1$).
Both simulations have several other prominent peaks located at lower frequencies.  
This is at least partly due to propagation effects.
As the secondary brings matter into the polar regions, the jet accelerates and unbinds this matter and it is propelled to larger radii. 
The gas then can spread out in radius (resulting in peaks in $\dot M_{\rm unbound}$ with a broader spread in time) and even catch up with previously ejected matter (resulting in merged peaks in $\dot M_{\rm unbound}$ showing up in lower frequencies in the power spectrum).

  \begin{figure}
\includegraphics[width=0.49\textwidth]{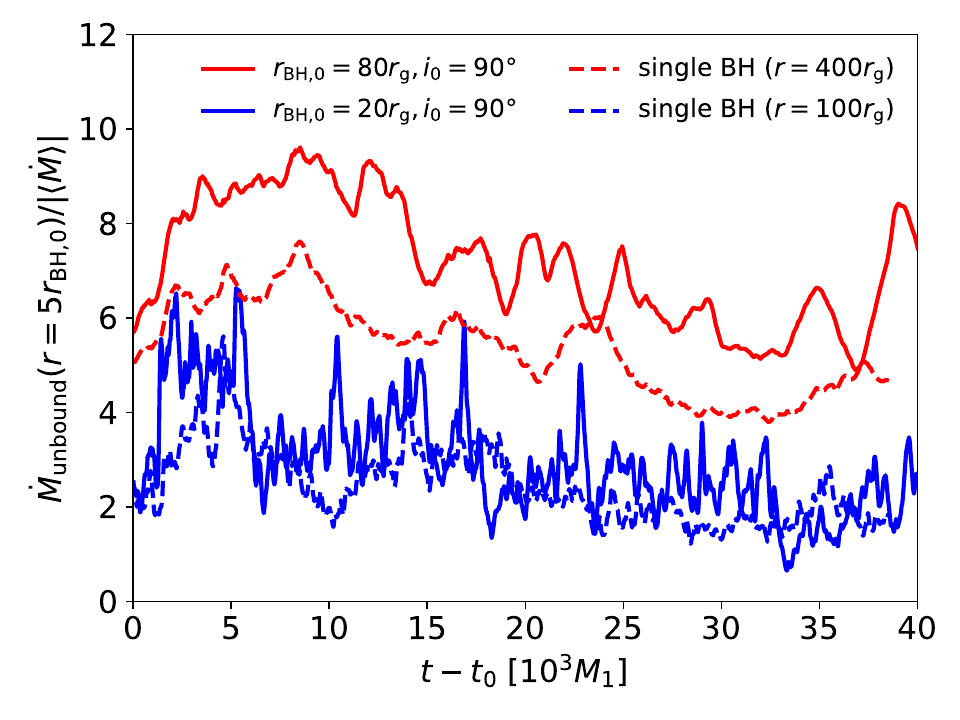}
\includegraphics[width=0.49\textwidth]{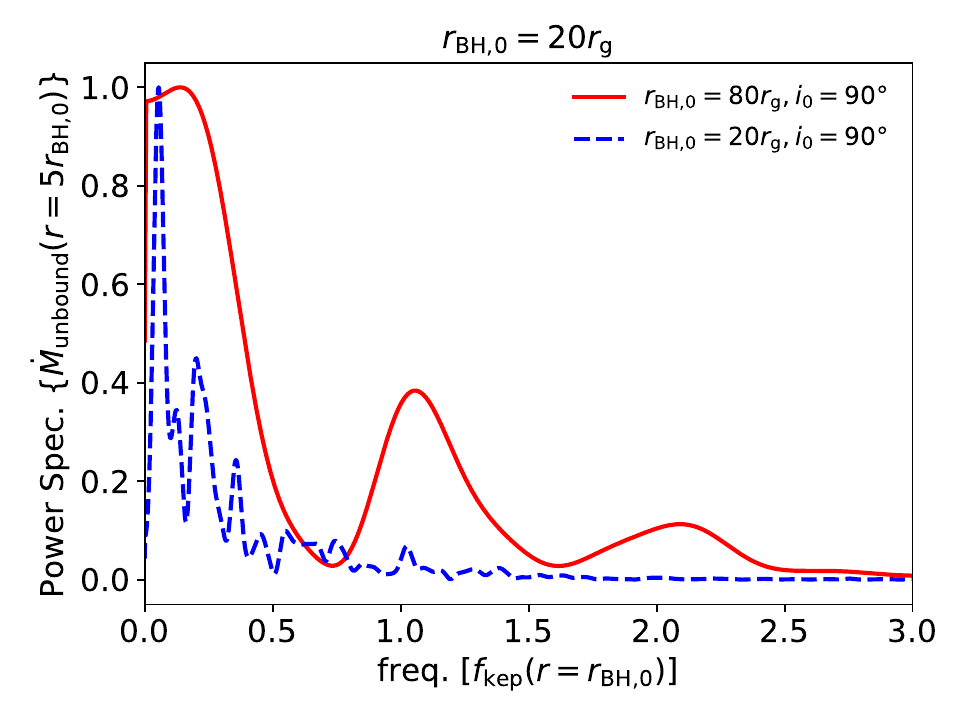}
\caption{
Top: Unbound mass outflow rate, $\dot M_{\rm unbound}$, plotted vs.\ time elapsed since the secondary was introduced, $t-t_0$, in our $\rbhi= 80 \rg,i_0=90^\circ$ (red solid line) and $\rbhi=20 \rg,i_0=90^\circ$ (blue dashed line) simulations measured at $r=5\rbhi$ compared to the single black hole simulation measured at the same radii (red and blue dotted lines).  
$\dot M_{\rm unbound}$ normalized by the time-averaged accretion rate over the interval ($0\le t-t_0 \le 40{, }000 M_1$). 
Bottom: Power spectrum for $\dot M_{\rm unbound}$ in the $\rbhi= 80 \rg,i_0=90^\circ$ and $\rbhi=20 \rg,i_0=90^\circ$ simulations, each normalized such that the highest peak is 1.
Note that the frequency resolution compared with the orbital frequency at $\rbhi$, $f_{\rm kep}(r=\rbhi)$, is much higher for $\rbhi=20\rg$ than $\rbhi=80\rg$, resulting in smoother and broader peaks for the $\rbhi=80\rg$ power spectrum.
Compared with Figures \ref{fig:mdot_out_all} and \ref{fig:psd_all}, power has shifted to lower frequencies than the orbital frequencies at $\rbhi$.
 Peaks in $\dot M_{\rm unbound}$ have become more broad in time.
 We propose that this is due to propagation effects as the mass outflow provided by the secondary spreads out in radius as it is accelerated outwards by the jet.  
 For $\rbhi=80\rg$ this results in a more continuous increase in unbound outflow rates compared with the single black hole simulation, while for $\rbhi=20\rg$ there are still distinct, relatively high-contrast peaks.
As in all GRMHD torus simulations, there is a secular decline in all outflow rates due to the mass in the torus depleting over time.
 } 
\label{fig:mdot_out_all_further_out}
\end{figure}

The unbound outflow rates we have studied thus far have been measured at $r=2\rbhi$.  
Perhaps a more observationally meaningful measurement of outflow is to measure $\dot M_{\rm unbound}$ at infinity or very large distances from the primary black hole. 
Practically, we have only a limited range of radial distances included in our simulation ($r \lesssim 1600 \rg $), and so use $r=5\rbhi$ as a proxy for larger radii.   
$\dot M_{\rm unbound}$ at $r=5\rbhi$ and the associated power spectra are shown for the $i_0=90^\circ$ simulations in Figure \ref{fig:mdot_out_all_further_out} (again compared to the single black hole simulation values at each radius).  
When averaged over time, the unbound outflow rates in the binary simulations are now 20--30\% larger than the single black hole unbound outflow rates at the same radii. 
For $\rbhi=80\rg$ the peaks seen in Figure \ref{fig:mdot_out_all} have become less distinct and spread out of over time (with the $\dot M_{\rm unbound}$ curve now clearly rising above the single black hole $\dot M_{\rm unbound}$ curve at almost all times).
For $\rbhi=20 \rg$, on the other hand, there are still peaks with relatively high contrast (factors of 2--3), with much of the $\dot M_{\rm unbound}$ curve lying close to the single black hole $\dot M_{\rm unbound}$ curve.
In terms of the power spectra, this behavior corresponds to a shift in power to lower frequencies.
The $\rbhi=80 \rg$ power spectrum still shows noticeable peaks at $f_{\rm kep}(r = \rbhi)$ and  $2f_{\rm kep}(r = \rbhi)$, but they are shorter than the peak at low frequencies.  
The $\rbhi=20 \rg$ power spectrum has almost no peak at $f_{\rm kep}(r = \rbhi)$ but instead has several peaks $\lesssim 0.5 f_{\rm kep}(r = \rbhi)$ (or periods $\gtrsim 1100 M_1$).
Again, we interpret this behavior as the dispersion of the unbound matter provided by the secondary as it travels outwards in radius.  
Indeed, starting from $\rbhi$ and plotting $\dot M_{\rm unbound}$ at progressively larger radii (not shown) we see that the peaks spread out in time and even merge together.
Eventually the quasi-periodicity in the unbound outflow rate disappears entirely for $r\gtrsim 10 \rbhi$ (also not shown), with the power spectra shifting to the lowest frequencies.
The resulting $\dot M_{\rm uunbound}$ is then a more continuous 20--30\% increase above the ``background'' unbound outflow rate. 
This means that we expect quasi-periodic signatures in the outflow to be only significant for a limited range of radii that depend on the orbital period: $\rbhi \lesssim r\lesssim 10 \rbhi$.

We note that the fact that the peaks in the power spectra for $\rbhi=80\rg$ are broader and smoother in Figure \ref{fig:mdot_out_all_further_out} than those for $\rbhi=20 \rg$ is due to the fact that $f_{\rm kep}(r=80\rg)$ is only $\sim$ 4 times the frequency resolution (i.e., the orbital period, $\sim$ $4500 M_1$, is $\sim$ 1/4 the window size of $20{,}000 M_1$).

\subsection{Spin-Orbit Coupling}

\begin{figure}
\includegraphics[width=0.47\textwidth]{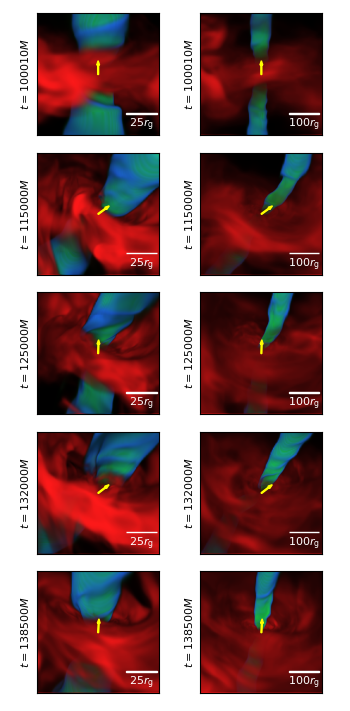}
\caption{Volume renderings of our $\rbhi=20\rg$, $i_0=90^\circ$ simulation at 5 different times on a $(100 \rg)^3$ (left) and $(400 \rg)^3$ (right) scale, with the primary black hole spin axis direction indicated by a yellow arrow.
In this figure, $x$ is the horizontal direction (positive to the left) and $z$ is the vertical direction (positive up), while $y$ is into and out of the page (positive out of the page).  
The volume rendering highlights the highest $\sigma=b^2/\rho$ (green/blue) and density (red), showing how the disk and jet tilt and precess as a function of time due to binary spin-orbit coupling.  
Both disk and jet align tend to align with the changing black hole spin axis, particularly at smaller scales.
For an animated version of this figure, see \url{https://youtu.be/GdgAINSolOY}.
}
\label{fig:jet_disk_3D}
\end{figure}

In this subsection, we highlight the features of the $\rbhi=20\rg$ simulations where the spin-orbit coupling between the binary orbit and the primary black hole spin is evident on simulated timescales. For instance, Figure \ref{fig:jet_disk_3D} shows a volume rendering of the accretion flow at five different times and two different spatial scales for the $\rbhi=20\rg$, $i_0=90^\circ$ simulation.
Regions with high $\sigma = b^2/\rho$ and regions with high $\rho$ are highlighted with green/blue and red, respectively.
Initially, when the perturber is first introduced, the accretion flow, electromagnetically dominated jet, and the primary black hole spin axis are all aligned.  
As the primary black hole spin tilts at later times due to relativistic spin-orbit coupling with the perturber,  both the accretion flow and the jet adjust so that near the peak tilt (see Figure \ref{fig:spin_orbit}) the jet and disk angular momentum are mostly aligned with the new spin axis.
As the spin returns to its original orientation and this process repeats, the disk and jet orientation on these scales for the most part trace the black hole spin axis (especially at small radii).  
This is because the primary black hole spin axis evolves on relatively long timescales (see Figure \ref{fig:spin_orbit}) so that at these radii the flow can adjust approximately adiabatically.

\begin{figure}
\includegraphics[width=0.47\textwidth]{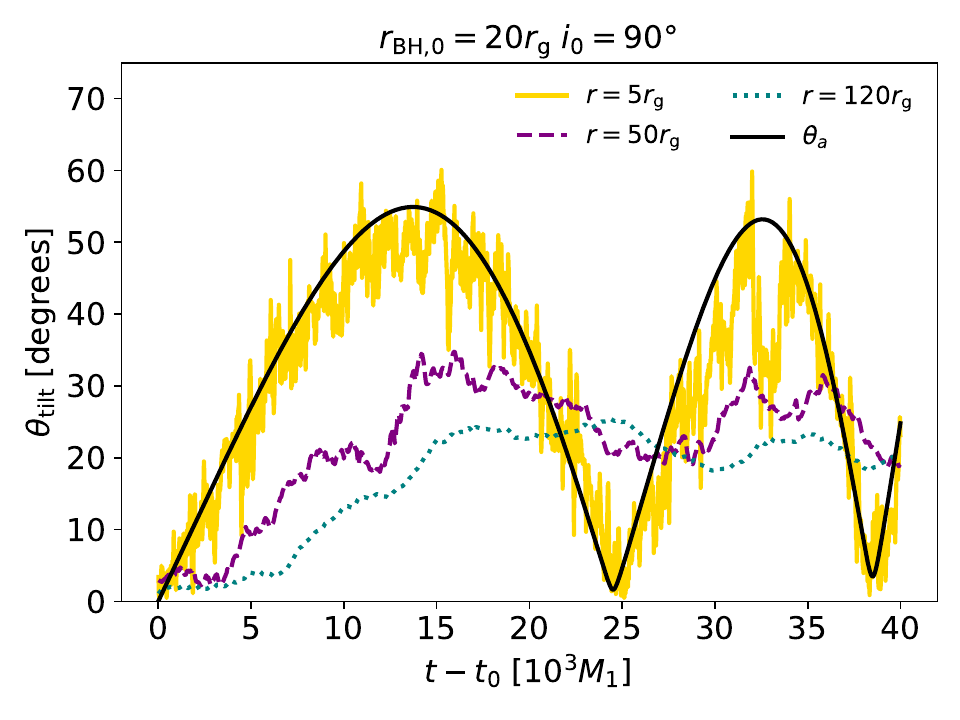}
\includegraphics[width=0.47\textwidth]{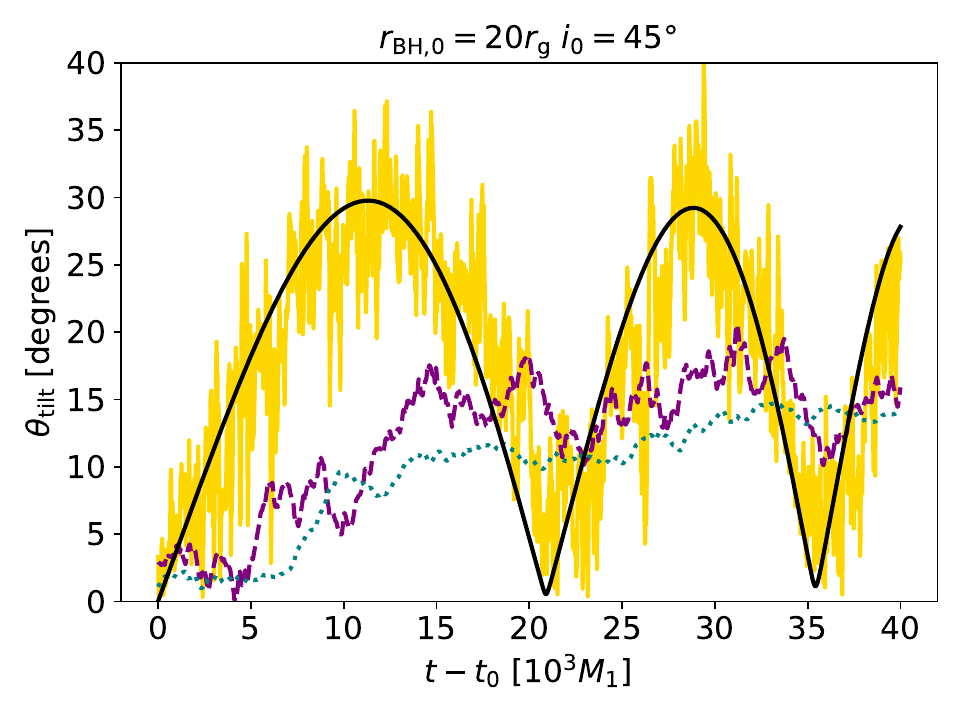}
\caption{ Angle between the angle-averaged angular momentum of the gas
and its initial direction, $\theta_{\rm tilt}$, vs.\ time measured at $r = 5 \rg$ (solid gold), $r = 50 \rg$ (dashed purple), and $r = 120 \rg$ (dotted teal), compared to the tilt angle of the primary black hole spin axis, $\theta_a$ (solid black), for the $\rbhi=20 \rg$ simulations with $i_0=90^\circ$ (top) and $i_0=45^\circ$ (bottom).  
As the black hole spin axis changes due to spin-orbit coupling (see Figure \ref{fig:spin_orbit}), frame-dragging and torques from magnetic fields cause the flow quickly to align with the gas at small radii $r \approx 5\rg$ so that $\theta_{\rm tilt}(r=5\rg)$ mostly traces $\theta_a$ with added stochasticity. 
At larger radii ($r\approx 50\rg$ and above), however, the gas tilt slowly rises to a saturation value of $\sim 30^\circ$ and $\sim 15^\circ$ for $i_0=90^\circ$ and $i =45^\circ$, respectively, and never returns back to $\theta_{\rm tilt}=0$.
This is likely because the flow at these radii effectively sees the time-averaged central black hole spin.
    }
\label{fig:th_tilt}
\end{figure}

To see this more quantitatively, Figure \ref{fig:th_tilt} compares the tilt of the primary black hole spin axis, $\theta_a$, to the tilt angle of the accretion flow as a function of time for the two $\rbhi=20\rg$ simulations at $r = 5\rg$, $r=50\rg$, and $r=120 \rg$. 
We define this tilt angle with respect to the $z$-direction (i.e., the original spin axis before the perturber is introduced) as (e.g., \citealt{White2019b})
\begin{equation}
  \theta_{\rm tilt} = \arctan\left(\frac{\langle L_z\rangle}{\sqrt{\langle L_x\rangle^2 + \langle L_y\rangle^2 + \langle L_z\rangle^2}}\right),
  \end{equation}
where $L_x = \rho (y u^z- z u^y)$, $L_y = \rho (z u^x- x u^z)$, $L_z = \rho (x u^y- y u^x)$, and $\langle\rangle$ denotes an angle average.  
In Figure \ref{fig:th_tilt}, $\theta_{\rm tilt}$ at $r=5\rg$ for both simulations displays significant stochastic temporal variability but on average follows the $\theta_a$ curve, i.e., the disk angular momentum axis quickly aligns with the primary black hole spin axis.
This corresponds to a peak average tilt of $\sim 60^\circ$ for $i_0=90^\circ$ and $\sim 30^\circ$ for $i_0=45^\circ$, though the $i_0=45^\circ$ simulation has times with larger $\theta_{\rm tilt}$ (e.g., $\sim 40^\circ$).
At this radius the angular momentum of the gas varies by as much as $\approx$ $20^\circ$ on short ($\lesssim 1000 M_1$) timescales.
At larger radii the gas angular momentum  is not only less variable but slower to respond to the change in the primary black hole axis.
In both simulations the angular momentum direction at $r=50 \rg$ and $r=120 \rg$ first tilts along with the primary black hole spin axis but at roughly half the rate of the disk at smaller radii.  
When the primary black hole spin axis starts returning to its original value, however, the gas does not similarly return to its original orientation.  
Instead, it remains tilted for the rest of the simulations, effectively saturating at $\theta_{\rm tilt} \sim 30^\circ$ for $i_0=90^\circ$ and $\theta_{\rm tilt} \sim 15^\circ$ for $i_0=45^\circ$.  
This is likely because the timescale for the primary black hole spin to change is smaller than the timescale for the larger radii flow to align and so the larger radii flow effectively sees a time-averaged black hole spin direction ($\sim$ $30^\circ$ from the $z$-axis for $i_0=90^\circ$ and $\sim$ $15^\circ$ from the $z$-axis for $i_0=45^\circ$).

\begin{figure}
\includegraphics[width=0.47\textwidth]{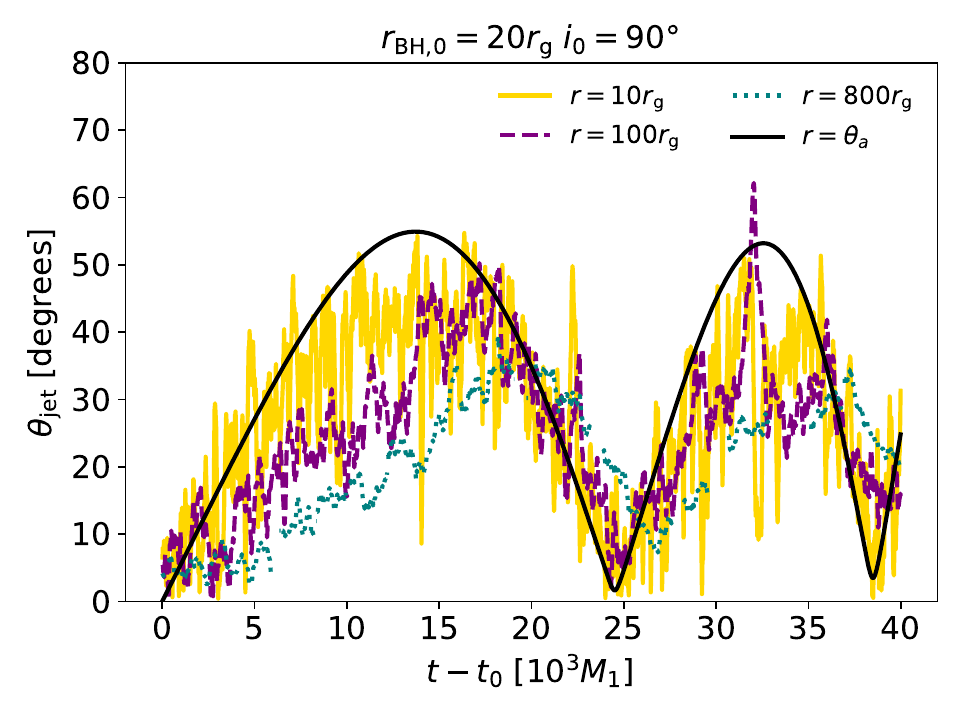}
\includegraphics[width=0.47\textwidth]{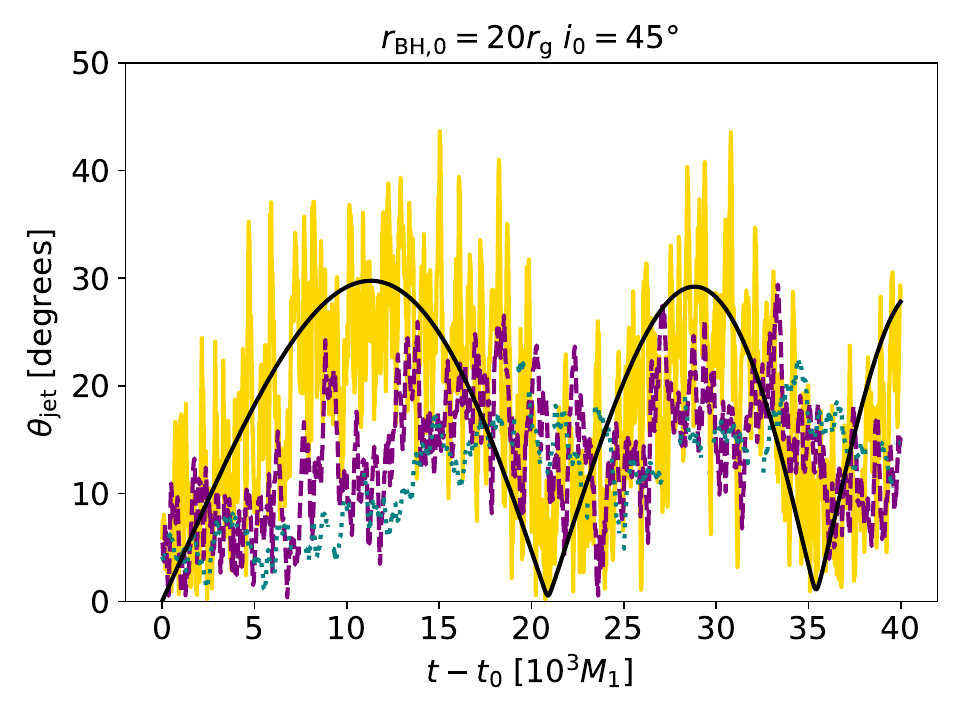}
\caption{Angle between the electromagnetically dominated upper jet (that is, the jet with polar angle $\theta< {\rm \pi}/2$) and its initial direction, $\theta_{\rm tilt}$, vs.\ time measured at $r = 10 \rg$ (solid gold), $r = 100 \rg$ (dashed purple), and $r = 800 \rg$ (dotted teal), compared with the primary black hole spin tilt angle, $\theta_a$ (solid black), for the $\rbhi=20 \rg$ simulations with $i_0=90^\circ$ (top) and $i_0=45^\circ$ (bottom).   The jets at all radii tend to align with the black hole spin but at a rate that decreases with increasing distance from the central black hole.
Unlike the disk (Figure \ref{fig:th_tilt}), the jet does tend to return to $\theta_{\rm jet}=0$ with the black hole spin due to the shorter timescales associated with the relativistic outflow speeds.
    }
\label{fig:th_jet}
\end{figure}

We perform a similar analysis with the jet in Figure \ref{fig:th_jet}, which shows how the tilt angle of the jet, $\theta_{\rm jet}$, changes as a function of time for $r=10\rg$, $r=100 \rg$, and $r=800\rg$ compared to $\theta_a$.  
Here we define $\theta_{\rm jet}$ by measuring the angle between the $z$-axis and the $\sigma$-weighted position vector of either the upper or lower jet.  
For simplicity, we only plot $\theta_{\rm jet}$ for the upper jet in Figure \ref{fig:th_jet}, but the quantity looks similar for the lower jet.
Generally, for both simulations the behavior of $\theta_{\rm jet}$ is similar to $\theta_{\rm tilt}$ at $r=5\rg$, that is, the jet direction mostly follows the black hole spin axis with added stochastic variability.
The jet at larger radii ($r=100 \rg$ and $r=800 \rg$) like the disk also similarly lags behind the black hole spin axis initially, with a slower change in direction than at smaller radii (though to a lesser extent than the disk, with peak tilt angles around $\sim 40$--$50^\circ$ and $\sim 15$--20$^\circ$ for $i_0=90^\circ$ and $i_0=45^\circ$, respectively).  
Unlike the disk, however, the jet at large radii does return to the initial orientation and follow the black hole spin axis, at least in part.  
This is likely because changes in the jet propagate at roughly the speed of light which is $\gg$ than the average radial velocity in the bulk of the disk

The fact that the secondary causes such a dramatic change in the orientation of the accretion flow and jet in these two $\rbhi=20\rg$ simulations makes it all the more remarkable that their accretion rates and dimensionless black hole flux values were so similar to the single black hole simulation in Figure \ref{fig:mdot_phibh_rbh_80}.
In fact, the time and angle-averaged gas quantities are also almost unchanged as we showed in Figure \ref{fig:radial_profiles}.  
This is likely because the timescales for the central black hole to tilt are so much longer than the dynamical times of the accretion flow at near horizon scales that the flow can adiabatically adjust as it tilts.  

Observationally speaking, the jet precession we see in our binary simulations could generally be detected (or at least inferred) from radio observations of jet morphologies.  
The most direct signature of precession would be quasi-periodic variations in the radio position angles and fluxes of observed jets (e.g., \citealt{Britzen2018,britzen2023precession,Cui2023,vonFellenberg2023}). 
The latter effect is caused by variations in the relativistic Doppler beaming as the jet points more or less toward the Earth.
Other evidence for precession could be more subtle, such as the appearance of what looks like multiple jet components at different locations and directions (caused by the jet changing direction over time; \citealt{Lister2013,Nandi2021}) or variations in uncollimated outflow features \citep{Falceta2010,Britzen2019,Krause2019}.

\subsection{Accretion and Magnetic Flux Accumulation on the Secondary}

\begin{figure}
\includegraphics[width=0.47\textwidth]{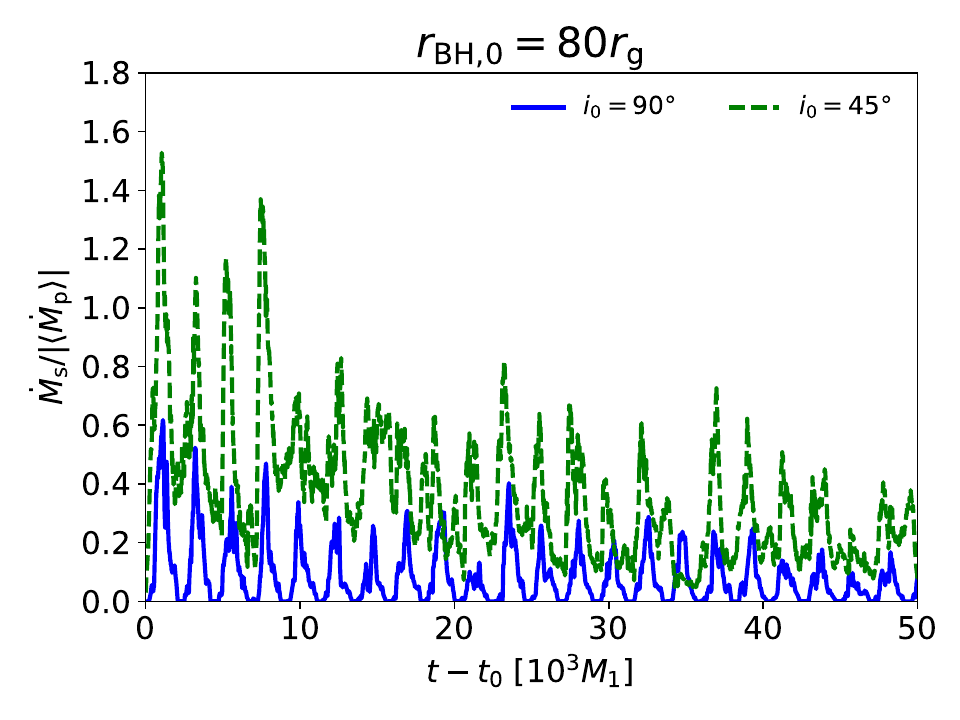}
\includegraphics[width=0.47\textwidth]{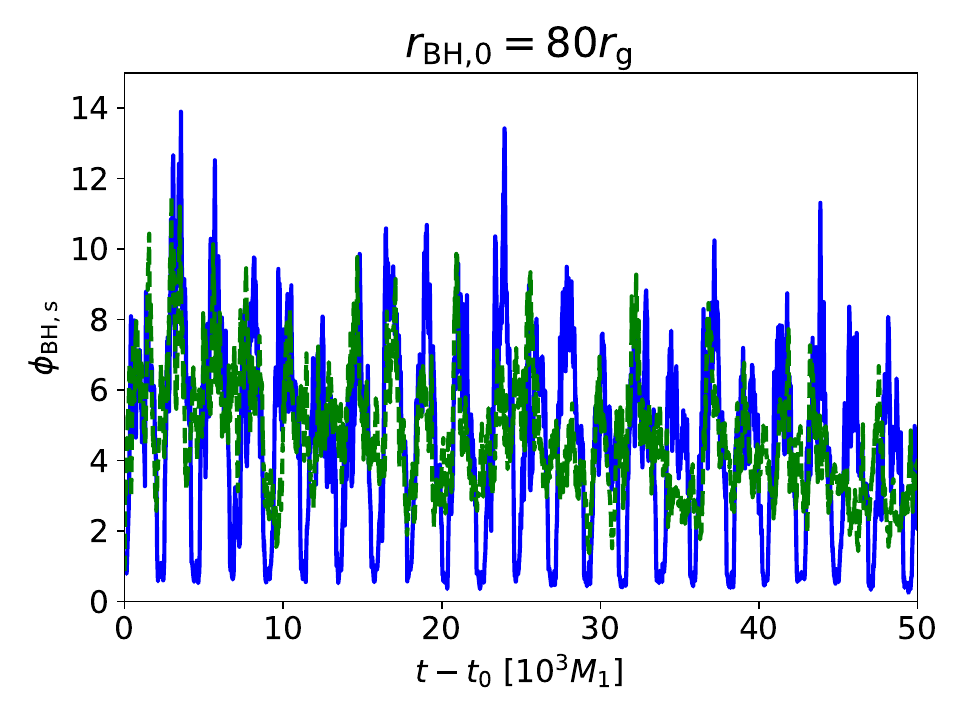}
\caption{Accretion properties of the secondary black hole for $\rbhi=80\rg$ for $i_0=90^\circ$ (solid blue ) and $i_0=45^\circ$ (dashed green).  
Top: Accretion rate normalized to the time-averaged accretion rate onto the primary, $\dot M_{\rm s}/|\langle \dot M_{\rm p}\rangle|$.
Bottom: Dimensionless black hole flux, $\phi_{\rm BH}$, calculated using the time-averaged accretion rate onto the secondary.
The secondary accretes a significant amount of gas while passing through the disk ($\sim$ 0.2 $\dot M_{\rm p}$ for $i_0=90^\circ$ and $\sim$ 0.6 $\dot M_{\rm p}$ for $i_0=45^\circ$).
Relative flux accumulation is also significant ($\phi_{\rm bH} \sim$ 6--10 for both inclinations), but does not attain to the MAD level.  
Note that the total magnetic flux, $\Phi_{\rm BH}$, 
 is larger in the $i_0=45^\circ$ simulation than the $i_0=90^\circ$ simulation by roughly the same factor as $\sqrt{|\dot M_{\rm s}|}$, resulting in similar values for $\phi_{\rm BH}$. 
} 
\label{fig:secondary_accretion_rbh_80}
\end{figure}

\begin{figure}
\includegraphics[width=0.47\textwidth]{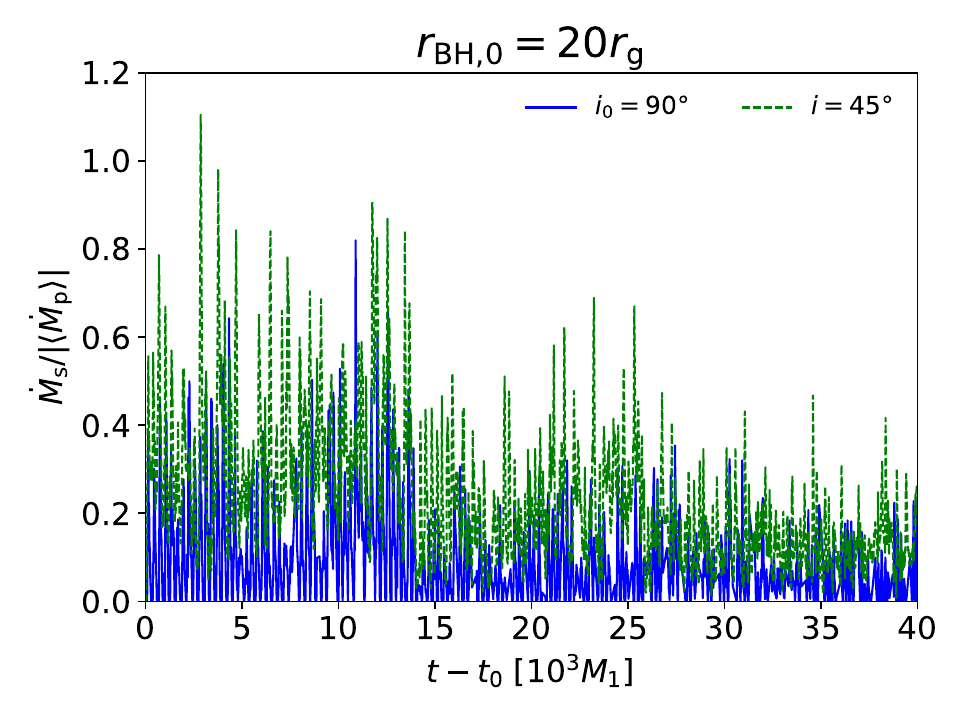}
\includegraphics[width=0.47\textwidth]{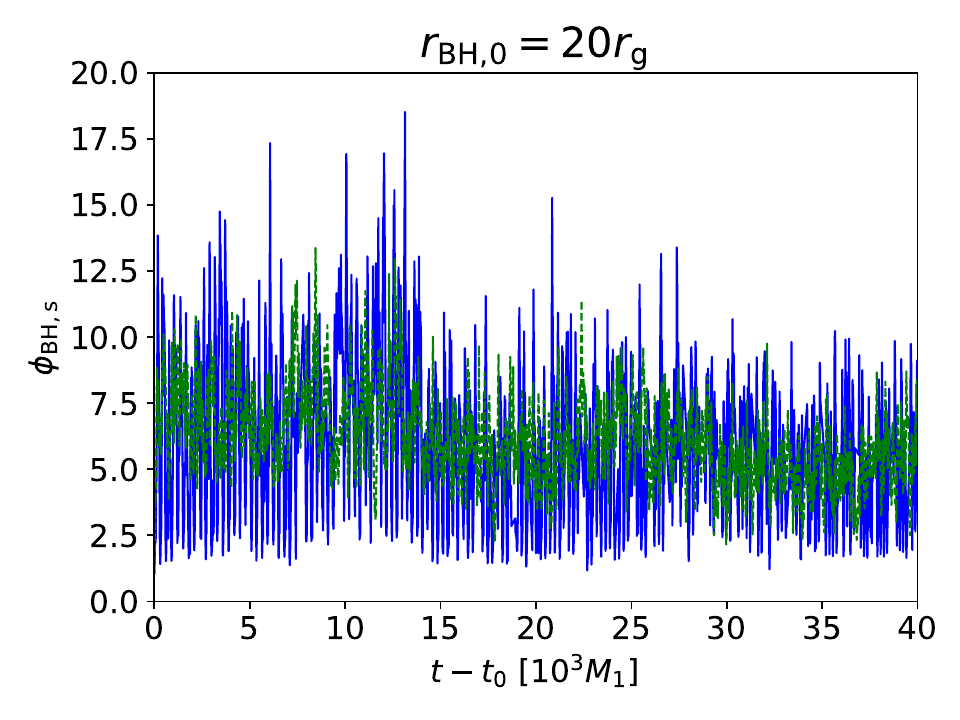}
\caption{Accretion properties of the secondary black hole for $\rbhi=20\rg$ for $i_0=90^\circ$ (solid blue ) and $i_0=45^\circ$ (dashed green).  
Top: Accretion rate normalized to the time-averaged accretion rate onto the primary, $\dot M_{\rm s}/|\langle \dot M_{\rm p}\rangle|$.
Bottom: Dimensionless black hole flux, $\phi_{\rm BH}$, calculated using the time-averaged accretion rate onto the secondary.
The secondary accretes less than the $\rbhi=80\rg$ simulations, but still a significant amount ($\sim$ 0.1--0.2 $\dot M_{\rm p}$ for $i_0=90^\circ$ and $\sim$ 0.3--0.5 $\dot M_{\rm p}$ for $i_0=45^\circ$).
The accumulated flux is roughly comparable to the $\rbhi=80\rg$ simulations, $\phi_{\rm bH} \sim$ 5--8.
Note that the total magnetic flux, $\Phi_{\rm BH}$, 
 is larger in the $i_0=45^\circ$ simulation than the $i_0=90^\circ$ simulation by roughly the same factor as $\sqrt{|\dot M_{\rm s}|}$, resulting in similar values for $\phi_{\rm BH}$.
} 
\label{fig:secondary_accretion_rbh_20}
\end{figure}

In this section, we describe the accretion properties onto the secondary black hole.  
To do this, we boost the simulation data into the rest frame of the secondary using the coordinate transformations described in \S \ref{sec:approx_metric} and then convert and interpolate the data onto a spherical grid in local Kerr-Schild coordinates. 
The transformation into spherical Kerr-Schild coordinates utilizes the single-black hole expressions (neglecting the effects of the primary on the metric), which is only appropriate for small distances from the secondary. 
We then calculate the accretion rate (relative to the time-averaged single black hole accretion rate) and dimensionless flux threading the secondary's horizon in the usual way (Equations \ref{eq:mdot} and \ref{eq:phibh}).
Figures \ref{fig:secondary_accretion_rbh_80} and \ref{fig:secondary_accretion_rbh_20} show these two quantities for our $\rbhi=80 \rg$ and $\rbhi = 20 \rg$ simulations, respectively.   
For both $i_0=90^\circ$ simulations, the accretion rate and black hole flux vary from 0 (when the secondary passes through the jet) to some peak value (when the secondary passes through the midplane of the disk) and then back to 0 every half orbit.  
The peak values of $|\dot M_{\rm s}|/| \langle \dot M_{\rm p}\rangle|$ are generally larger for $\rbhi=80\rg, i_0=90^\circ$ ($\sim$ 0.3--0.6) than $\rbhi=20\rg,i_0=90^\circ$ ($\sim$ 0.1--0.6).   
This is because the $\rbhi=80\rg$ black hole is moving slower than the $\rbhi =20\rg$ black hole and so it can accrete more gas even though it is surrounded by lower densities.  
We showed in \S \ref{sec:analytic} that these two effects compete to result in a scaling of $\dot M_{\rm s} \propto \rbhi^{0.7}$, which is $(80/20)^{0.7} \approx$ 2.6 for $\rbhi =20\rg$ and $\rbhi=80 \rg$, consistent with our findings.
The peak values of $|\dot M_{\rm s}|/| \langle \dot M_{\rm p}\rangle|$ are significantly higher for $i_0=45^\circ$ than $i_0=90^\circ$ for both sets of simulations by factors of 2--3.  
This is because the black holes with $i_0=45^\circ$ orbits are travelling with prograde motion through the accretion disk and so the net velocity of the gas in the frame of the secondary is reduced, leading to a larger accretion radius and accretion rate.  
The magnitude of this increase is consistent with our analysis in \S \ref{sec:analytic_general}, where we predicted that the accretion rates for $i_0=45^\circ$ would be $\approx$ 2.5 times the $i_0=90^\circ$ accretion rates.

Note that since the secondary black hole is 10 times less massive than the primary, for all simulations the Eddington ratio for the secondary is larger than the primary (at times by as much as a factor $\gtrsim 10$).

In terms of magnetic flux, all simulations show a significant amount of flux accumulation but not enough to reach the MAD state. 
All simulations show similar time-averaged values of $\phi_{\rm BH} \sim$ 5--8, with quasi-periodic variation.    
There is no indication that the secondaries in any simulation will eventually accumulate enough magnetic flux to become MAD.
This is partially because the accretion rate is not steady; as the secondary passes away from the midplane, the reduced accretion rate allows magnetic flux to be expelled.

\section{Comparison to Previous Work}
\label{sec:comparison}
\subsection{ Accretion Disk Perturber Simulations}
There have not been many studies of impacts of smaller objects with supermassive black hole accretion disks in GRMHD. 
The most relevant to the current work is \citeauthor{Sukova2021} (\citeyear{Sukova2021}; hereafter S21), in which the authors simulated the passage of objects (including stars and black holes) of various sizes through a MAD accretion disk (see also \citealt{Pasham2024} where similar simulations are used to make a case for the existence of a secondary black hole in the source ASASSN-20qc).  
This was done by enforcing that all the gas within a specified radial distance from the center of the objects has the same velocity as the objects, which are on geodesic orbits calculated alongside the simulations.  
The authors investigated a range of orbital distances (10--50 $\rg$) and influence radii (0.1--10 $\rg$). 
Note that the latter quantity does not necessarily correspond to the radius of the object itself but the radial range where the secondary has a significant effect on the accretion flow.
Even for an influence radius of $1 \rg$, in 2D S21 found that the secondary significantly altered the accretion flow, resulting in quasi-periodic oscillations of the accretion rate onto the primary, effectively shutting off accretion with every passage of the object through the midplane.
This same motion produced quasi-periodic relativistic mass outflow rates with peaks that were 1--2 orders of magnitude higher than the ``background'' mass outflow rates.  
In the one 3D simulation\footnote{This simulation ran for a shorter time ($5000 M$) after being initialized from a longer run 2D simulation.} with a secondary, the authors note that these effects are greatly diminished because the object now has a more realistic size in the azimuthal direction; in 2D the perturber was essentially a ring extending across the full 2$\rm \pi$ in azimuth.  

The black holes in our simulations have influence radii self-consistently set by the dynamical interaction between gravity and the MHD fluid, but we have roughly estimated them as $\sim$ 2--3 $\rg$, 4--5 $\rg$, 10 $\rg$, and 18--19 $\rg$ for ($\rbhi=20 \rg$, $i_0=90^\circ$), ($\rbhi=20 \rg$, $i_0=45^\circ$), ($\rbhi=80 \rg$, $i_0=90^\circ$), and ($\rbhi=80 \rg$, $i_0=45^\circ$), respectively.
These are all larger than the fiducial $1 \rg$ used in S21, yet we see almost no effect on the resulting primary accretion rates and the magnetic flux threading the black hole, and only marginal changes in the time and angle-averaged radial profiles of gas quantities.  
We do however, see quasi-periodic outflows caused by the secondaries similar to those of S21, though the peak outflows are only 2--4 times the ``background'' outflow rates provided by the disk.  
This is consistent with their 3D result.
Our peaks are also much more variable in magnitude and shape likely due to the increased turbulence and variability in the 3D disk compared with 2D.
 
The biggest differences between our simulations and S21 are 1) all of our simulations are fully 3D and 2) we focus specifically on black hole perturber and utilize a full treatment of the resulting binary metric.
1) is particularly important for a number of reasons.
First, there is no realistic way to treat a ballistic spherical object moving through an azimuthally symmetric accretion flow.  
The 2D perturber will act as ring with a substantially enhanced effect on the flow, as noted by S21.
The 3rd dimension also allows us to study inclined orbits in a more straight-forward way.
Furthermore, 2D, MRI-driven accretion flows are unrealistic when run for any significant length of time (more than a few thousand $M$) because the MRI is not sustainable in axisymmetry \citep{Cowling1933}.  
2) allows us to self-consistently study the effects of the black holes on the accretion flow and to investigate the effects of spin-orbit coupling on the primary black hole.
For example, the influence radii of the black holes are set by a combination of the orbital parameters, the accretion flow dynamics, and the secondary-to-primary mass ratio.
The precession of the primary accretion disk caused by spin-orbit coupling may also be the most significant observable effect of the secondary for certain parameters.

\subsection{Tilted Disk Simulations}
\label{sec:tilted_comp}
The interaction of the accretion disk in our $\rbhi=20\rg$ simulations with the precessing primary black hole spin axis (due to spin-orbit coupling with the orbit of the secondary black hole) is in some ways similar to the interaction of an incoming accretion disk tilted with respect to a fixed black hole spin axis.
For thick disks, such a situation has been studied by several authors in GRMHD (e.g., \citealt{Fragile2007,McKinney2013,Liska2018,White2019b,Ressler2023,Chatterjee2023}).
These simulations include both SANE and MAD disks, which were found to have different alignment properties.  
In particular, the strong magnetic fields rotating with the black hole in MAD disks are very efficient at aligning accretion flows and jets (called magneto-spin alignment in \citealt{McKinney2013}), at least for $a \gtrsim 0.9$ and misalignments $\lesssim 60^\circ$ (larger misalignments tend to inhibit the MAD state, \citealt{Ressler2023,Chatterjee2023}).
Alignment in MAD disks is seen in the simulations out to at least $\gtrsim 100 \rg$ \citep{Ressler2023,Chatterjee2023} and in reality could reach even larger distances on longer timescales.
The inner accretion flow ($r\lesssim$ 10--20$\rg$) tends to align on timescales of $\sim$ $10^4 M$ for misalignments of $\lesssim 60^\circ$ (see Figure 14 in \citealt{Ressler2023} and Figure 3 of \citealt{Chatterjee2023}).
The accretion flow at larger and larger radii takes progessively longer times to align (e.g., $\sim$  $2 \times 10^4 M$ at $r=50\rg$).
Jets in tilted MAD disks tend to align on even shorter timescales and out to several hundred $\rg$ (e.g., Figure 15 in \citealt{Ressler2023}).

Thick SANE disks, on the contrary, do not align efficiently but instead tend to form standing shocks as the gas accretes from the disk onto the black hole \citep{Fragile2007,White2019b} and perhaps  precess about the spin axis due to the \citet{LT} effect in the azimuthal direction (defined with respect to the black hole spin axis, \citealt{Liska2018}), though it is argued in \citet{Chatterjee2023} that this precession is short lived and the true steady state of tilted SANE flows is instead a warped disk without precession.

Our simulations all contain MAD disks.
The biggest difference between our study and the aforementioned works on tilted disks (apart from the presence of the secondary black hole) is that the misalignment between the angular momentum of the disk and the primary black hole spin axis is introduced gradually via the slowly changing spin axis instead of suddenly.
Despite this, the properties of the disk alignment agree very well with previously studied tilted MAD simulations. 
There is, however, an observable difference between the single black hole case and the binary case.  
Because thick MAD disks align so well, precession is not seen in single black hole simulations.
For binary simulations with spin-orbit coupling, however, precession would be observed even with strong alignment because the primary spin axis is changing with time.
Moreover, these disks would also be persistently warped (i.e., the angular momentum vector changes with radius) because the outer part of the accretion flow aligns with the time-averaged primary spin axis while the inner part of the accretion flow aligns with the instantaneous spin axis.  

\section{Limitations of Our Study}
\label{sec:limitations}
The simulations we have presented are formally applicable only to low luminosity supermassive accretion flows where radiative cooling is inefficient and the disk is thick due to thermal pressure support.
However, most observed AGN are in the radiative efficient regime where the disk is either thin from rapid cooling or thick from radiative pressure support.
We expect that the effect of a secondary on a thin disk to be more dramatic than a thick disc because it will impact a larger fraction of its volume (see our analytic argument in \S \ref{sec:analytic_general}).
At very small disc thicknesses, the process of accretion and ejection will no longer be well approximated by our Bondi-Hoyle-Lyttleton framework described in \S \ref{sec:analytic} because a spatially extended wind will no longer be a good approximation in the frame of the secondary.  
Orbits of low inclination may also be able to have more significant unbound mass outflows because they will fully enter and exit the polar regions (unlike our $i_0=45^\circ$ simulations).
On the other hand, in the radiative pressure dominated, thick disk regime, it is reasonable to expect that many of our qualitative conclusions may still hold due to the similar geometry of the system compared to thick, non-radiative disks. 
The biggest difference would likely be the significantly lower gas temperatures.  
These are particularly important for determining the emission associated with the ejection of the gas into the polar region, which is determined by a combination of geometry and the photospheric temperature of the ``blobs'' \citep{Franchini2023}.
For the highest accretion rates, self-gravity of the accreting gas and dynamical friction on the secondary may also become important, which would affect the orbit of the secondary.

In this work, we have only considered MAD accretion flows.  
While recent work has shown that MAD disks may be relatively common for supermassive black holes (e.g., \citealt{EHT5,EHT_SGRA_5,Ressler2020b,LiskaDynamo}), there are still likely many AGN either in less magnetized states or with more toroidally dominated magnetic fields given that most do not display obvious jet signatures (whereas MAD disks generally have strong jets).   
The powerful magnetic fields in MAD disks are known to be more efficient at aligning the accretion disk to tilted black hole spins than the weaker magnetic fields in less magnetized disks \citep{McKinney2013,White2019b,Ressler2023,Chatterjee2023}.
Thus the disk and jet in a SANE flow may not as closely mirror the black hole spin axis as our simulations.
The outflow and jet from MAD disks are also much stronger than SANE flows by factors of 10--100 \citep{Sasha2011,Ressler2023}, which could alter the behavior of the blobs ejected by the secondary black hole.  

We have also only considered rapidly rotating primary black holes.
In less rapidly rotating black hole systems, the accretion flow would be less affected by spin-orbit coupling because the alignment would be less efficient.
The jets would also be weaker than those in our $a=0.9375$ simulations (jet power is a strong function of $a$, $\tilde \propto$ $a^2$--$a^4$; \citealt{BZ1977,Mckinney2005,Tanabe2008,Sasha2010}) and perhaps be less efficient at blowing away gas brought into the polar regions by the secondary.  

Similarly, we have focused only on secondary black holes with no spin.  
More rapidly rotating secondaries would likely produce their own jets which could have a stronger effect on the primary accretion flow.
The strength of these jets would depend on the amount of magnetic flux accreted by the secondary; for secondaries accreting flux at the MAD level, their jets would inject $\sim |\dot M_{\rm s} c^2|$ worth of power into the primary accretion flow, which can be a sizable fraction of $\sim |\dot M_{\rm p} c^2|$ (i.e., comparable to the primary jet power, see Figures \ref{fig:secondary_accretion_rbh_80} and \ref{fig:secondary_accretion_rbh_20}).
The effect that such a deposition of energy would have on the primary flow is unclear, but it could be dramatic.
Depending on the spin orientation with respect to the orbit and the primary black hole spin, the secondary spin would change direction due to spin orbit coupling in a similar way to the primary black hole spin (though on a much shorter timescale due to the lower relative mass).
The jets from the primary and secondary could also in principle interact with each other (\citealt{Molnar2017,Volonteri2022,Gutierrez2023}; though this may only be significant for mass ratios closer to one where the jets are of comparable strength and size).
In addition, a rapidly spinning secondary could torque the surrounding accretion flow and alter the dynamics in that region more prominently than a non-spinning secondary.

Here we have presented only one choice of mass ratio, $q=0.1$.  
For mass ratios higher than this our assumption of a stationary primary black hole and steady state accretion disk would start to break down and the system would no longer be in the perturber regime.
We have performed simulations with significantly smaller mass ratios and found the effect of the secondary on outflows and accretion to be almost neglible and thus not particularly interesting to present.
Smaller mass ratios also make it more computationally demanding to resolve the secondary's influence region.  

Finally, we have only considered primary accretion flows that are aligned with the primary black hole spin axis.  
In reality, supermassive black hole accretion disks could be significantly tilted, which would reduce the amount of material in the polar regions, reduce the strength of the jet, and change the geometry of the accretion flow \citep{Fragile2007,White2019b,Chatterjee2020,Ressler2023,Chatterjee2023}.
The consequences of these combined effects could have on our results are not clear, though if the jet is weaker and more entrained with matter the outflows produced by the secondary black hole may be relatively weaker. 

Future work should explore a larger amount of parameter space by investigating disks of different thicknesses, tilts, magnetic field strengths (i.e., SANE vs.\ MAD flows), and black hole spins. 

\section{Discussion and Conclusions}

\label{sec:conc}

We have presented the results of 3D simulations of small mass ratio  ($q=0.1$) binary black holes in the inspiral regime at the centers of galaxies, considering four different scenarios: secondaries with orbits at {initial separations of} $20 \rg$ and $80 \rg$ {with} initial inclinations of 90$^\circ$ and $45^\circ$ from the midplane.  
We did this by implementing the time-dependent approximate metric from \citet{Combi2021} describing a superimposition of boosted Kerr black holes into {\tt Athena++}.
This metric, though approximate, fully captures {general relativistic} effects such as spin-orbit coupling on both the primary and secondary black holes, the inspiral of the secondary, the event horizon contractions when the black holes are moving close to the speed of light, and the \citet{BZ1977} process that can drive electromagnetically dominated jets. 
The secondaries are introduced into an accretion flow around a supermassive black hole that has already evolved for $10^5 M$ (Figure \ref{fig:init_torus}), long enough to come into inflow equilibrium out to $\sim$ $100\rg$ (Figure \ref{fig:1d_torus_plots}).  

We find that even for a relatively high secondary mass of 1/10 the mass of the primary, the overall effects on the primary accretion flow are small as measured by the accretion rate, dimensionless magnetic flux threading the black hole (Figure \ref{fig:mdot_phibh_rbh_80}), and the angle averaged radial quantities (Figure \ref{fig:radial_profiles}).   
This is because the area on a spherical shell swept up in the disk by the secondary during an orbit at the orbital radius is much smaller than the overall area of the disk at that radius.
In other words, the fraction of the disk affected by the secondary as determined by the orbit and influence radius is very small (Figure \ref{fig:rbh80_contour}). 

The secondary black holes in our simulations do, however, produce quasi-periodic outbursts with periods $\gtrsim$ the orbital period as seen in the unbound mass outflow rates (Figures \ref{fig:mdot_out_all}).
These outbursts are caused by the secondary bringing gas from the disk into the polar regions, where the electromagnetically dominated jet then blows this gas away (Figure \ref{fig:qpe_contour}).
When measured at distances from the primary of twice the secondary orbital radius, the peaks of the mass unbound outflow rates are $\sim$ 2--4 times the ``background'' unbound outflow rate provided by the accretion disk.  
The orbital frequency is clearly seen in the power spectra of the unbound outflow rates at these distances (Figure \ref{fig:psd_all}), especially for secondaries with $i_0=90^\circ$ orbits. 
However, there are also several other prominent frequencies, particularly lower frequencies and especially for $\rbhi=20\rg$. 
At distances farther away from the primary (e.g., $5\rbhi$), peaks in unbound outflow rates become less frequent and more spread out in time (Figure \ref{fig:mdot_out_all_further_out}), with power shifting to frequencies lower than half the orbital frequency. 
At even larger radii (e.g., $\gtrsim 10\rbhi$), the variability in the unbound outflow rates shifts to even lower frequencies and instead of having well-defined peaks is characterized by a more continuous distribution in time consistently $\sim$ 20--30\% larger than the single black hole unbound outflow rates.
This means that quasi-periodic features are only present in the unbound outflow rates for $\rbhi \lesssim r\lesssim 10\rbhi$.
We interpret this as a result of the mass ejected from the primary accretion disk by the secondary not propagating/accelerating as a coherent structure as it is propelled to larger radii by the jet.
Instead, it spreads out in space and can even catch up and merge with previously ejected matter.
Eventually, by about $r\approx10\rbhi$ the initially discrete chunks of outflow have diffused into a more continuous wind.

Because the accretion disk used in this work is so thick, orbits with $i_0=45^\circ$ do not fully escape the disk into the polar regions and so their ability to deposit matter into the jet is reduced.
At the same time, the secondaries on these orbits have larger influence radii because the relative velocity of the gas in the frame of the secondary is reduced.
These combined effects results in in similar outburst magnitudes as orbits with $i_0=90^\circ$ (Figure \ref{fig:mdot_out_all}).
Our simple analytic model predicts that for thinner disks in which the secondary fully passes into the polar regions, secondaries with $i_0=45^\circ$ orbital inclinations will have larger outbursts than secondaries with $i_0=90^\circ$ orbital inclinations (\S \ref{sec:analytic_general}). 

For smaller $\rbhi$ (e.g., $\rbhi=20\rg$), spin-orbit coupling between the primary black hole spin and the orbit of the secondary can cause the spin direction of the primary to change by up to $60^\circ$ on timescales we can simulate ($\sim 10^4 M_1$, Figure \ref{fig:spin_orbit}).
This results in both the disk and jet of the primary tilting along with the spin axis (Figures \ref {fig:jet_disk_3D}, \ref{fig:th_tilt}, and \ref{fig:th_jet}) in an approximately adiabatic way. 
As the spin axis returns to its original orientation, the jet and inner region of the accretion disk ($r \approx 5 \rg$) also return to their original orientation.
The larger radii accretion disk ($r \gtrsim 50 \rg$), however, remains tilted by $\sim$ half of the peak spin tilt, even as the primary spin axis continues to change.
This is because the dynamical timescales in the disk at these radii are longer than the timescale for the spin to change, resulting in the larger radii flow seeing an effective time-average of the primary spin.
Previous work in tilted accretion flows around single black holes has found that thick disks tend to either align or warp but not consistently precess (see \S \ref{sec:tilted_comp}).
Thus observations of precession in thick AGN disks would be strong evidence for the presence of a secondary black hole companion.

The secondary black holes themselves accrete a significant amount of mass from the primary accretion disk, from 0.1 times the primary accretion rate up to greater than the primary accretion rate (Figures \ref{fig:secondary_accretion_rbh_80} and \ref{fig:secondary_accretion_rbh_20}), meaning that the Eddington ratio for the secondaries is larger than the Eddington ratio for the primary.
Accretion rates vary in a quasi-periodic manner, with peak accretion occuring as the secondary passes through the midplane of the primary accretion disk.
Accretion rates generally increase with increasing $\rbhi$ (roughly as $\rbhi^{0.7}$), though there is significant overall time-variability in the magnitude of the individual peaks. 
Accretion rates onto secondaries with $i_0=45^\circ$ orbits are 2--3 times higher than those onto secondaries with $i_0=90^\circ$ orbits because the relative velocity of the gas in the accretion disk in the frame of the secondary is reduced.  
The black hole flux threading the secondaries is, on average, $\sim$ 15\% of the MAD limit for all orbital configurations.
We see no indication that the secondaries will ever become MAD because of the frequent dips in accretion rate during which the magnetic flux can leak out of the black hole.

Directly predicting the electromagnetic emission from our simulations is difficult without a detailed treatment of radiation, which we reserve for future work.  
in particular, there are several features of our simulations could imprint themselves on observations.  
For instance, the quasi-periodic ejection of matter from the disk to the jet could directly result in quasi-periodic emission if the hot gas radiates significantly between 1--10 $\rbhi$.  
This could happen via thermal emission from the gas soon after it is expelled, or by interactions between jet/outflow material and the disk or a surrounding medium.
Alternatively, emission could be generated via the shock front of the secondary and the associated accretion onto the secondary.
Note, however, that red noise contamination can also result in apparent periodic features in AGN light curves and thus make distinguishing true periodic signals more challenging.
The slow precession of the disk and jet from spin-orbit coupling between the orbit of the secondary and the primary black hole spin axis could be detectable through direct jet imaging or by variations in Doppler effects.
Finally, the matter brought into the polar regions by the secondary induces an interaction between the magnetic fields in the highly magnetized jet and in the mildly magnetized matter from the secondary. The turbulence resulting from this interaction may cause dissipation of strong magnetic field, e.g., through magnetic reconnection, that could potentially power bright and fast flares.
More accurately studying this possibility would require much higher resolution simulations.

Our results are valid for thick accretion disks associated with low luminosity AGN.  
They may also be qualitatively applicable to thick accretion disks associated with $\sim$ Eddington and super-Eddington accretion where the gas becomes radiation pressure dominated due to the similar geometry.
Future studies of the effect of secondary black holes on thin accretion disks will require a proper treatment of radiative cooling.    
Future work should also explore a larger range of parameter space in this binary black hole perturber scenario by investigating SANE disks, different black hole spins, tilted disks, and different mass ratios.

This work has implications for observed quasi-periodic outflows/outbursts/eruptions and the detection of companion supermassive or intermediate mass black holes.  
We have shown that the signatures of the latter may be subtle even when the mass ratio is relatively high ($q=0.1$) if the disk is thick.

\begin{acknowledgments}
We thank the referee for their careful reading of the manuscript and useful comments.
We acknowledge the support of the Natural Sciences and Engineering Research Council of Canada (NSERC), [funding reference number 568580]
Cette recherche a \'et\'e financ\'ee par le Conseil de recherches en sciences naturelles et en g\'enie du Canada (CRSNG), [num\'ero de r\'ef\'erence 568580].
B.R. is supported by the Natural Sciences \& Engineering Research Council of Canada (NSERC) and by a grant from the Simons Foundation (MP-SCMPS-00001470). Research at the Flatiron Institute is supported by the Simons Foundation.
L.C. and H.Y. are supported in part by
Perimeter Institute for Theoretical Physics.
Research at Perimeter Institute is supported in part by the Government of Canada through the Department of Innovation, Science and Economic Development Canada and by the Province of Ontario through the Ministry of Colleges and Universities. 

The computational resources and services used in this work were partially provided by facilities supported by the VSC (Flemish Supercomputer Center), funded by the Research Foundation Flanders (FWO) and the Flemish Government – department EWI. 
This research was supported in part by the NSF
through XSEDE computational time allocation TG-AST200005 on Stampede2 and Bridges-2.  This work was made possible by computing time granted by UCB on the Savio cluster.
\end{acknowledgments}

\software{{\tt Athena++}, \citep{White2016,Athenapp},
{\tt CBwaves}, \citep{cbwaves}}


\bibliographystyle{mn2efix}
\bibliography{perturber}

\appendix

\section{Tests of Metric Implementation}
\label{app:metric_tests}
\subsection{Transport in an Accelerating Reference Frame}
\label{app:1d_test}
Consider a gas at rest in Minkowski space $[u^\mu = (1,0,0,0)]$ with density profile $\rho = 0.5 \rho_0 [1 + \cos(2 {\rm \pi} x)] + \rho_0$, pressure profile $P = P_0 - 0.25 P_0 [1+ \cos(2 {\rm \pi} x)]$, and magnetic field $b^t = 0$ and $B^y = b^y = \sqrt{2(P_0 - P)}$ for $-0.5 \le x \le 0.5$, while $\rho = \rho_0$, $P=P_0$, and $B^y = b^y = 0$ otherwise.  
This state represents a static equilibrium and should thus be maintained for all time.

Our test is to take this solution and to boost into an accelerating reference frame:
\begin{equation}
  \begin{aligned}
    t^\prime & = \Gamma(t)  \left[ t - \beta(t) x \right] \\
    x^\prime & = \Gamma(t)  \left[ x - \beta(t) t \right],
  \end{aligned}
  \end{equation}
  where 
  \begin{equation}
  \beta(t) = v_0 \sin\left( \frac{2 {\rm \pi} t}{T} \right),
\end{equation}  
$\Gamma = [1-\beta(t)^2]^{-1/2}$, $v_0$ is a constant representing the maximum velocity of the frame, and $T$ is a constant representing the period of frame oscillation.
The associated transformation matrix is 
\begin{equation}
  \Lambda^\mu_\nu = \frac{d(x^\prime)^\mu}{dx^\nu} = 
  \begin{pmatrix}
\Gamma(t)  + \Gamma^3(t)  \frac{d \beta(t)}{dt} \left[\beta(t) t-x\right]&  - \Gamma(t) \beta(t)  \\
 - \Gamma(t) \beta(t) + \Gamma^3(t)  \frac{d \beta(t)}{dt}  \left[\beta(t) x -t\right]  &  \Gamma(t)\\ 
\end{pmatrix},
\\
\label{eq:Lambda_1d}
\end{equation}
resulting in a metric 
\begin{equation}
  g_{\mu \nu} = \left(\Lambda^{-1}\right)^\alpha_\mu \left(\Lambda^{-1}\right)^\beta_\nu \eta_{\alpha \beta}.
  \end{equation}
  Note that there is a coordinate singularity in this metric at $x^\prime = \Gamma(t) d\beta(t)/dt$.
  This metric/frame combination is not meant to be necessarily physically significant.  
  Our only goal in constructing it was to have a quasi-periodic time-dependent metric for testing purposes.
  
  \begin{figure}
\includegraphics[width=0.49\textwidth]{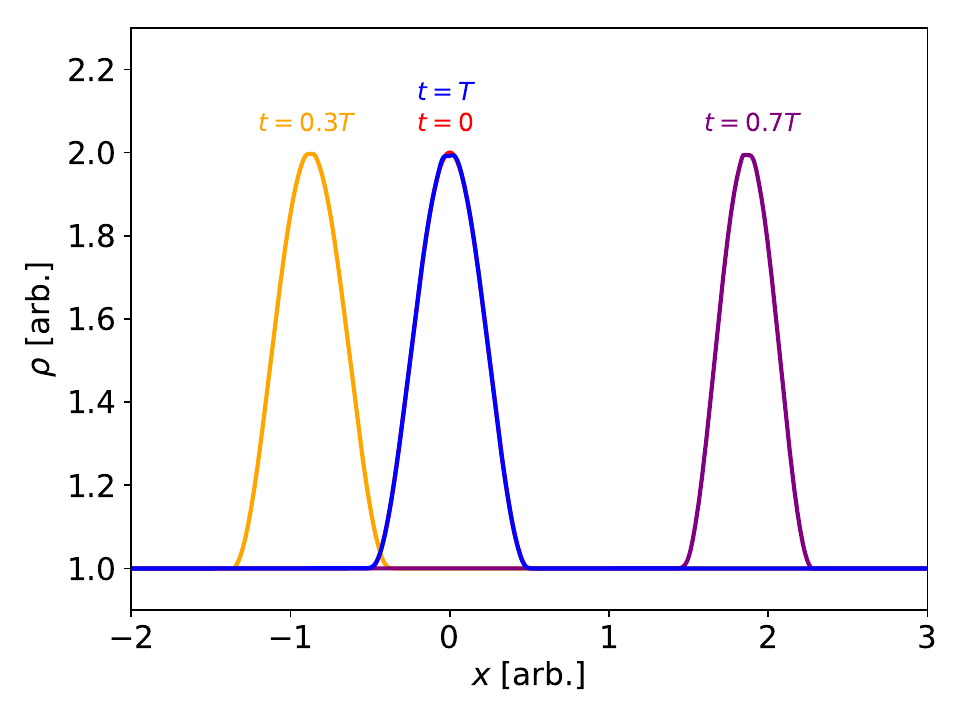}
\caption{Density vs.\ $x$ coordinate at four different times in our 1D accelerating reference frame test described in Appendix \ref{app:1d_test} for a resolution of $N=1024$.  The gas is in pressure equilibrium due to a combination of pressure and magnetic forces in its rest frame; the motion seen in this figure is due entirely to the reference frame. 
After one period ($t=T$) the gas should return to its initial state.
Our simulation reproduces this result very well for this resolution.   } 
\label{fig:1d_metric_test_profile}
\end{figure}
  
In this frame, the density profile should move from left to right in a quasi-periodic manner.  At $t=T$ the primed and unprimed coordinates are identical, so the fluid and magnetic field quantities should be identical to their initial condition at $t=0$ (when the primed and unprimed coordinates are also identical).  
We therefore initialize the gas in the primed frame with the density and pressure profiles described above using $\rho_0 = P_0 = 1$, as well as $(u^\prime)^\mu = \Lambda^\mu_\nu u^\nu$, and $(b^\prime)^\mu = \Lambda^\mu_\nu b^\nu$ using $v_0=0.3$ and $T=10$.
We use a domain of $-3.5 < x^\prime < 3.5$ and run until $t^\prime = T$ using piecewise-linear reconstruction and the HLLE Riemann solver.
Note that in order to keep the initial overdensity from passing through the coordinate singularity $v_0$ must be sufficiently small, $\lesssim 0.37$, so that $v_0 T \lesssim | \Gamma d\beta/dt |_{\rm max}$, where the maximum is taken over all time.

\begin{figure}
\includegraphics[width=0.47\textwidth]{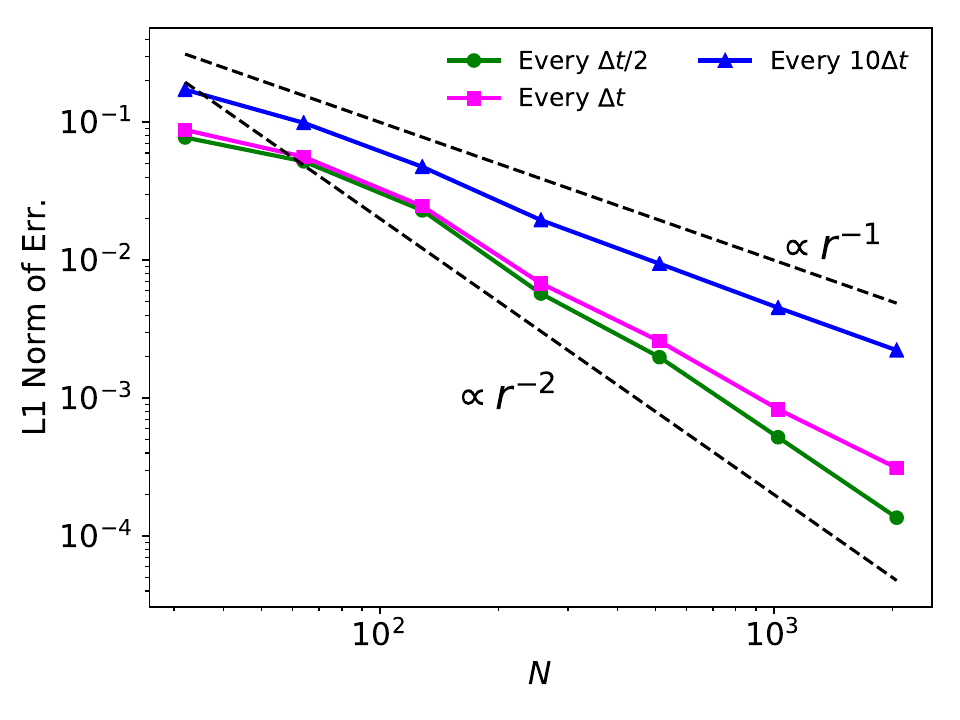}
\caption{L1 norm of the error in density, $\rho$, vs.\ resolution, $N$m in our 1D accelerating reference frame test described in Appendix  \ref{app:1d_test} when the metric is updated ever half time-step (green circles), every full time step (magenta triangles), and every 10 timesteps (blue squares).
Updating the metric every half timestep results in asymptotically 2nd order convergence while updating every 10 timesteps asymptotes to 1st order convergence.  
The curve for updating ever timestep is very close to the updating every half time step curve until relatively high resolution ($N \gtrsim 512$) at which point the convergence flattens out to slower than 2nd order but faster than 1st order.  
At higher resolution we expect this curve to continue to flatten until it reaches $\tilde\propto$ $r^{-1}$. 
The L1 norm of the error in pressure and magnetic field look similar.   
} 
\label{fig:1d_metric_test_error}
\end{figure}

Figure \ref{fig:1d_metric_test_profile} shows the resulting density profile at four different times for $N=2048$.  The initial overdensity feature moves to the left and then to the right (asymmetrically) before returning to the center, essentially overlapping with the initial condition.  

Plotted in Figure \ref{fig:1d_metric_test_error} is the L1 norm of the density error, 
\begin{equation}
  \sum \limits _i |\rho_i(t^\prime=T) - \rho_i(t^\prime =0)|
\end{equation}
vs.\ resolution for three different frequencies of updating the metric.  
The most accurate updates the metric every sub-timestep of the algorithm (i.e., every half time step), the next most accurate updates the metric every full timestep, and the least accurate updates the metric ever 10 full timesteps.  
Figure \ref{fig:1d_metric_test_error} demonstrates that all three of these methods converge, though at different orders.  
As expected, updating every substep of the algorithm results in asymptotically 2nd order convergence with resolution, reproducing the convergence properties of a static metric flow in {\tt Athena++} \citep{White2016}. 
Updating the metric only every full timestep formally reduces the order of the algorithm to 1st order, however, we find that the for this test problem the convergence is noticeably less than 2nd order but beats 1st order at least up until $N = 2048$, the highest resolution we simulate.
This is because spatial errors in the gradient of the gas can still dominate the temporal errors at low resolution.
Finally, updating the metric only once every 10 timesteps results in asymptotically 1st order convergence. 

The absolute values of the errors in Figure \ref{fig:1d_metric_test_error} are specific to this simple test problem and should not be generalized.  
Instead, our goal of this test is to demonstrate that the metric update described in \S \ref{sec:methods} converges properly (at 2nd or 1st order depending on whether the metric is updated every sub-timestep or just every full timestep or less).
In particular, we demonstrate that even updating the metric every 10 timesteps (the method employed in the main body of this work) results in 1st order convergence.

  \begin{figure}
\includegraphics[width=0.49\textwidth]{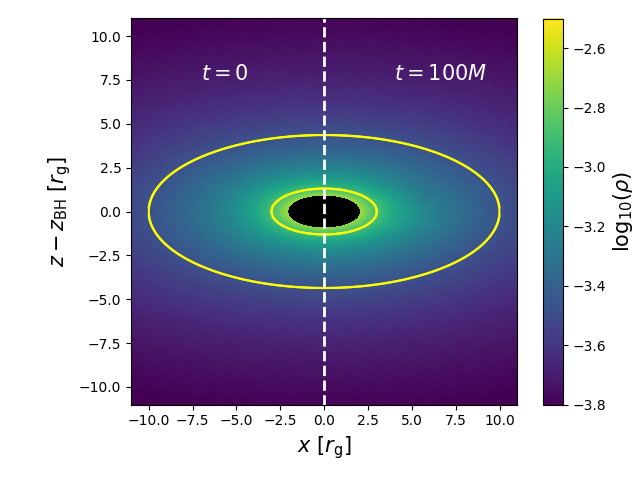}
\caption{2D slice of density of the 3D boosted Bondi test described in Appendix \ref{app:bondi_test} for a black hole moving at $0.9c$ and base resolution of $128^3$ (including 5 levels of adaptive mesh refinement).  Left: initial conditions.  Right: final solution at $t=100 M$. For plotting purposes the $z$-axis is shifted so that the black hole is located at the origin. 
Yellow lines demarcate the regions that are ``active'' in the simulation.  Regions outside the outermost ellipse and inside the innermost ellipse are fixed to the analytic solution.  The inner black surface covers the event horizon of the black hole.  Note that the boosted Bondi solution is no longer spherical, and regions of constant $r^\prime$ (radial distance from the black hole in its rest frame) are ellipses in the lab frame.  
The differences between the final and initial states are indistinguishable by eye.    
} 
\label{fig:bondi_contour}
\end{figure}

\subsection{Boosted Bondi Accretion}
\label{app:bondi_test}
Another test of our implementation is a single black hole accreting spherically in its rest frame moving through a grid at some velocity $v_{\rm BH}$. 
To do this, we take the spherically symmetric accretion solution for a nonspinning Schwarzschild black hole (\citealt{Hawley1984}, see also Section 4.4 of \citealt{White2016}) and then boost into the frame described by the trajectory $s^x = s^y = 0$, $s^z =  v_{\rm BH} t + z_0$.
This is a special case of the metric described in \S \ref{sec:approx_metric}, where $M_2=0$ and the transformation is exact since $v_{\rm BH}$ is constant in time.  
The resulting solution is length-contracted in the direction of motion ($z$) and thus no longer spherically symmetric in the lab frame but only trivially time-dependent in that it moves uniformly with a constant velocity.

  \begin{figure}
\includegraphics[width=0.47\textwidth]{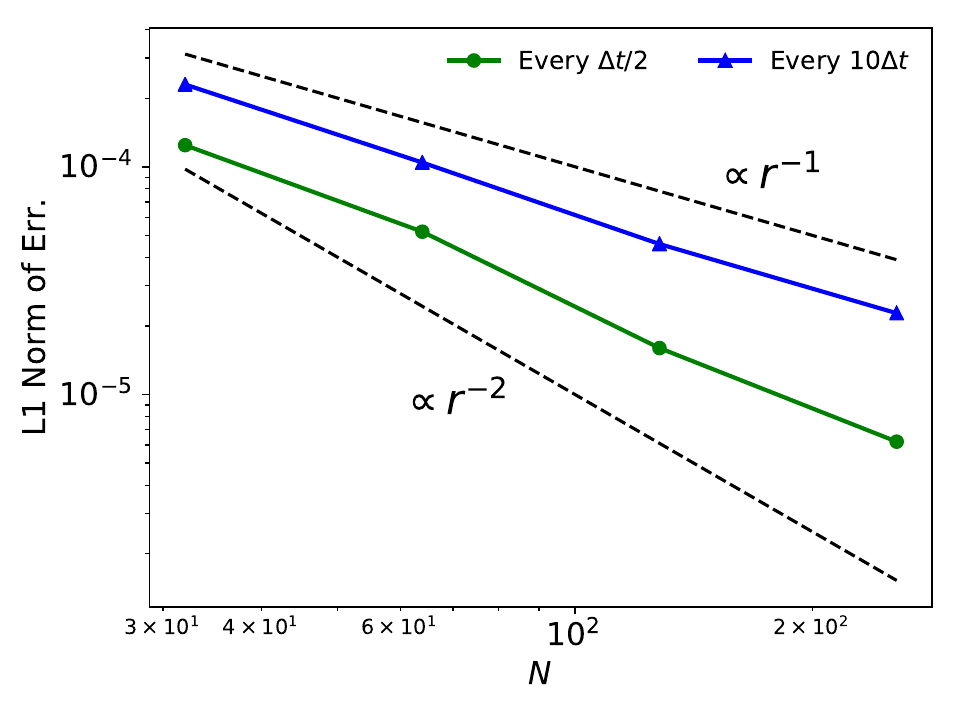}
\caption{ L1 norm of density error in the boosted Bondi accretion test described in Appendix \ref{app:bondi_test} vs.\ 3D resolution at $t=100 M$ for a simulation that updates the metric every half timestep (blue triangles) compared to a simulation that updates the metric ever $10$ timesteps (green circles).  
Note that the resolution $N$ on the $x$-axis is the base resolution in one direction (which is the same in every direction). On top of this, there are 5 additional levels of adaptive mesh refinement for each simulation.
At the highest base resolution tested ($N=256$), the simulation that updates the metric every half timestep converges at a rate steeper than $N^{-1}$ but shallower than $N^{-2}$.
The lower-than-expected order of convergence is likely due to the nature of the moving boundary condition.  
The simulation that updates the metric every 10 timestep does quite well in comparison, converging at approximately $r^{-1}$ and only a factor of $\sim$ 2 higher error than the simulation that updates the metric every half timestep.  
Both errors are quite low at all resolution ($\lesssim 3 \times 10^{-4}$).
 } 
\label{fig:bondi_error}
\end{figure}

We simulate this by initializing a $-100 r_{\rm g} < x,y,z < 100 r_{\rm g}$ box with the boosted Bondi solution for $z_0=80 r_{\rm g}$ and $v_{\rm BH} = 0.9$.  We use 5 levels of nested adaptive mesh refinement centered on the moving black hole.
The cells within $r^\prime \le 3$ and $r^\prime \ge 10$, where $r^\prime$ is the radial distance from the black hole in its rest frame, are fixed to the analytic solution and the simulation is run for $100$ $M$.  
The resulting mass density is shown in Figure \ref{fig:bondi_contour} plotted alongside the initial condition, showing that the simulation accurately reproduces the analytic solution.  

Quantitatively, we show the convergence properties of the L1 norm of the density error at $t=100$ $M$ in Figure \ref{fig:bondi_error}, comparing a simulation that updates the metric every half timestep to one that updates the metric every 10 timesteps.  
Both simulations do very well at maintaining the Bondi solution, even at relatively low resolution (e.g., L1 norms of density errors $\lesssim 3 \times 10^{-3}$ even at a base resolution of $32^3$).  
The errors in both simulations converge to zero and are within a factor of $\sim$ 2 of each other.  
The simulation that updates the metric every 10 timesteps converges roughly at the expected rate of $N^{-1}$.
The simulation that updates the metric every half timestep, however, converges at a slower rate than expected, just moderately steeper than $N^{-1}$ but shallower than $N^{-2}$.  
There are at least three possible causes of this.
1) The inner/outer boundaries at $r^\prime = 3 r_{\rm g}$ and $r^\prime = 10 r_{\rm g}$, which move along with the black hole so that cells are continuously entering and exiting the computational domain.
2) This same boundary is ellipsoidal on a Cartesian grid, so it is misaligned with the coordinate faces and is located at slightly different locations at different resolution.  
3) AMR, which can focus resolution at slightly different locations at different base resolution. 
The errors introduced by these three factors, of course, converge to zero with increasing resolution, but they converge at a rate $\tilde \propto$ $N^{-1}$ (at least for the latter two, since the errors are independent of time and $\propto \Delta x$) .
If these errors are dominant that would explain why the simulations that update every 10 timesteps and those that update every half timestep converge at a similar rate.
Given the relative complexity of this test problem compared to a stationary, face-aligned boundary and uniform grid test problem, however, we do not consider it worrisome that the convergence rate does not reach 2nd order, especially since 1) the errors in both simulations are small and still converge to zero and 2) we update the metric every 10 timesteps in our target problem (see \S \ref{sec:methods}) and thus would not expect to see $N^{-2}$ convergence anyway.  

\subsection{Boosted Bondi-Hoyle-Lyttleton Accretion}
\label{app:bhl_test}

  \begin{figure}
\includegraphics[width=0.49\textwidth]{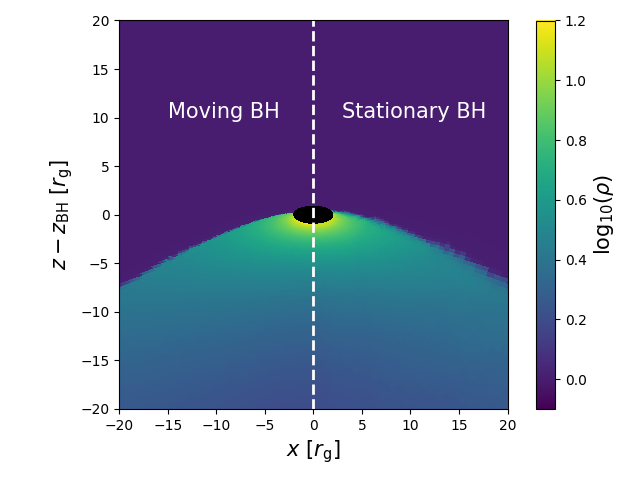}
\caption{2D slice of density in the 3D boosted Bondi-Hoyle-Lyttleton test described in Appendix \ref{app:bhl_test} for a black hole/gas moving at $0.9c$ and base resolution of $128^3$ (including 5 levels of adaptive mesh refinement for the stationary black hole simulations and 7 levels fo adaptive mesh refinement for the moving black hole simulation).  Left: simulation for a moving black hole through an initially stationary gas with the metric updated every 10 timesteps at $t=300M$.  Right: simulation for a stationary black hole with gas moving at an initially constant velocity at $t^\prime = 300M$. For plotting purposes the $z$-axis is shifted so that the black hole is located at the origin.
The stationary black hole simulation is also boosted into the frame of the moving black hole for ease of comparison.
The inner black surface covers the event horizon of the black hole (which is ellipsoidal in this frame). 
At these late times the flow has settled into a steady state solution of an extended shock front trailing the black hole.  
The differences between the solutions in the two simulations are almost indistinguishable by eye.    
} 
\label{fig:bhl_contour}
\end{figure}
For this test, we compare a simulation of a stationary black hole embedded in uniform gas moving with constant velocity to a simulation of a moving black hole through a stationary, uniform gas.  
The latter is simply a boosted version of the former, and so we can use the metric of \S \ref{sec:approx_metric} with a constant velocity and $M_2 = 0$. 
This problem, in general, doesn't have a known analytic solution, and is in fact still being actively researched (e.g., \citealt{Blakely2015,Gracia-Linares2023,Kaaz2023}). 
It is, however, a good test to see whether the time-dependency of the metric is accurately captured by our simulations in a complex problem.  

For the stationary black hole simulation, we use a grid of size $(100 r_{\rm g})^3$ with resolution of $128^3$ and 5 additional levels of static mesh refinement centered on the black hole. 
The gas is initialized with $\rho = 1$, $P = 5$, $u^x=u^y=0$, and $u^z/u^t = - v_{\rm BH} $ everywhere except for regions with $r^\prime \le5 r_{\rm g}$, where $\rho$ and $P$ are set to the floors and the three-velocities are set to zero.
These parameters describe a mildly supersonic flow, with Mach number $\mathcal{M}\approx 1.15$ \citep{Mignone2007}.

For the moving black hole simulation, we use a grid of size $(400 r_{\rm g})^3$ with resolution of $128^3$ and 7 additional levels of adaptive mesh refinement centered on the black hole (achieving the same effective resolution as the stationary black hole case). The gas is initialized with $\rho = 1$, $P = 5$ and three-velocities set to zero everywhere except for regions with $r\le5 r_{\rm g}$, where $\rho$ and $P$ are set to the floors.
The black hole is initially located at $x=0$, $y=0$, and $z=-380 r_{\rm g}$ moving at $v_{\rm BH}=0.9$ in the $+z$ direction.
The metric is updated every 10 timesteps.

Both simulations are run for $300$ M in their own reference frame, at which point the solutions reach a steady state. 
2D slices of the density in these solutions are shown in Figure \ref{fig:bhl_contour}.
When boosted for comparison, these two solutions agree very well, both showing a wide shock front extending behind the black hole as it moves through the uniform medium (or as the uniform wind impacts the black hole in the stationary black hole reference frame).
The agreement between the solutions is strong evidence that our treatment of the time-dependent, boosted metric is accurate for complex problems even when the metric is only updated every 10 timesteps.

\section{Nonlinear Coordinate Equation}
\label{app:nonlinear_expression}
 Equation \eqref{eq:coordinate_definition} in \S \ref{sec:approx_metric} results in the following nonlinear set of equations: 
     \begin{equation}
  \begin{aligned}
    t_\tau\left(\tau\right)  ={} & t - \beta^x\left[x-s^x( t_\tau)\right] -  \beta^y\left[y-s^y( t_\tau)\right] - \beta^z\left[z-s^z( t_\tau)\right] \\
    X =&  \left[1 + \left( \frac{1}{\Gamma}-1\right) \left(\frac{\beta^x}{\beta}\right)^2\right] \left[x-s^x(t_\tau)\right] + \left( \frac{1}{\Gamma}-1\right) \frac{\beta^x\beta^y}{\beta^2} \left[y-s^y(t_\tau)\right] \\ 
    &+\left(  \frac{1}{\Gamma}-1\right) \frac{\beta^x\beta^z}{\beta^2}\left[z-s^z(t_\tau)\right]\\
        Y =& \left(  \frac{1}{\Gamma}-1\right) \frac{\beta^x\beta^y}{\beta^2} \left[x-s^x(t_\tau)\right] +  \left[1 + \left( \frac{1}{\Gamma}-1\right) \left(\frac{\beta^y}{\beta}\right)^2\right] \left[y-s^y(t_\tau)\right]  \\ 
        &+ \left(  \frac{1}{\Gamma}-1\right) \frac{\beta^y\beta^z}{\beta^2}\left[z-s^z(t_\tau)\right] \\
        Z =&   \left(  \frac{1}{\Gamma}-1\right) \frac{\beta^x\beta^z}{\beta^2} \left[x-s^x(t_\tau)\right]  +  \left(  \frac{1}{\Gamma}-1\right) \frac{\beta^y\beta^z}{\beta^2} \left[y-s^y(t_\tau)\right] \\ 
        &+ \left[1 + \left(  \frac{1}{\Gamma}-1\right) \left(\frac{\beta^z}{\beta}\right)^2\right]\left[z-s^z(t_\tau)\right],
  \end{aligned}
  \label{eq:coordinates}
\end{equation}
and 
\begin{equation}
  \begin{aligned}
    t  ={} & t_\tau(\tau) + \Gamma \beta^x  X + \Gamma \beta^y  Y + \Gamma \beta^z  Z \\
    x =&  s^x\left(t_\tau\right) + \left[1 + \left( \Gamma-1\right) \left(\frac{\beta^x}{\beta}\right)^2\right] X + \left( \Gamma-1\right) \frac{\beta^x\beta^y}{\beta^2} Y \\ 
    &+\left( \Gamma-1\right) \frac{\beta^x\beta^z}{\beta^2}Z\\
        y =& s^y\left(t_\tau\right) + \left( \Gamma-1\right) \frac{\beta^x\beta^y}{\beta^2} X+  \left[1 + \left( \Gamma-1\right) \left(\frac{\beta^y}{\beta}\right)^2\right] Y  \\ 
        &+ \left( \Gamma-1\right) \frac{\beta^y\beta^z}{\beta^2}Z\\
        z =&  s^z\left(t_\tau\right)+  \left( \Gamma-1\right) \frac{\beta^x\beta^z}{\beta^2} X  +  \left( \Gamma-1\right) \frac{\beta^y\beta^z}{\beta^2} Y \\ 
        &+ \left[1 + \left( \Gamma-1\right) \left(\frac{\beta^z}{\beta}\right)^2\right]Z,
  \end{aligned}
  \label{eq:coordinates_inverse}
\end{equation}   
where $\beta^i$ and $\Gamma$ are evaluated at $t_\tau$.

\end{document}